\documentclass{article}
\usepackage[utf8]{inputenc}

\usepackage[margin=1in]{geometry}

\usepackage[section]{placeins}

\title{Parallel MCMC Algorithms:\\ Theoretical Foundations, Algorithm Design, Case Studies}

\date{}
\author{Nathan E. Glatt-Holtz, Andrew J. Holbrook, Justin A. Krometis, Cecilia F. Mondaini\\
\scriptsize{emails: negh@iu.edu, aholbroo@g.ucla.edu, jkrometi@vt.edu, cf823@drexel.edu}}

\usepackage{amsfonts, amssymb, amsmath, amsthm,esint,multicol}
\usepackage[colorlinks=true, pdfstartview=FitV, linkcolor=blue,
            citecolor=blue, urlcolor=blue]{hyperref}
\usepackage[usenames,dvipsnames,table]{xcolor}
\definecolor{Red}{rgb}{0.7,0,0.1}
\definecolor{Green}{rgb}{0,0.7,0}
\usepackage{accents}
\usepackage{comment}
\usepackage{graphicx}
\usepackage[capitalize,nameinlink,noabbrev]{cleveref}
\usepackage{enumerate}
\usepackage{enumitem}
\usepackage[numbers]{natbib}

\usepackage{bbm}
\usepackage{mathtools} %for coloneqq

\usepackage{tabularx}

\usepackage{listings}
\usepackage[textsize=tiny]{todonotes}
\usepackage{tikz}
\usetikzlibrary{shapes.misc}
\usepackage{etoolbox}
\usepackage{appendix}
\usepackage{subcaption}
\usepackage{wrapfig}
\usepackage{xr}
\usepackage{booktabs}

\numberwithin{equation}{section}

\newtheorem{Theorem}{Theorem}[section]
\newtheorem{Proposition}[Theorem]{Proposition}

\newtheorem{Corollary}[Theorem]{Corollary}

\newtheorem{Definition}[Theorem]{Definition}
\newtheorem{Remark}[Theorem]{Remark}

\usepackage[section]{algorithm}      %for algorithms
\usepackage[noend]{algpseudocode}    %for algorithms

\newcommand{\be}{\begin{equation}}
\newcommand{\ee}{\end{equation}}

\newcommand{\RR}{\mathbb{R}}
\newcommand{\NN}{\mathbb{N}}
\newcommand{\ZZ}{\mathbb{Z}}

\newcommand{\Exp}{\mathbb{E}}

\newcommand{\Prb}{\mathbb{P}}
\newcommand{\bfV}{\mathbf{v}}
\newcommand{\fm}{\phi}
\newcommand{\spq}{X}
\newcommand{\spv}{Y}
\newcommand{\dimx}{N}%{d}
\newcommand{\dimy}{D}
\newcommand{\Vker}{\mathcal{V}}
\newcommand{\Pker}{P}
\newcommand{\Qker}{Q}
\newcommand{\bQker}{\overline{Q}}
\newcommand{\Proj}{\Pi_1}
\newcommand{\cB}{\mathcal{B}}
\newcommand{\cC}{\mathcal{C}}
\newcommand{\tcC}{\tilde{\mathcal{C}}}
\newcommand{\cM}{\mathcal{M}}
\newcommand{\cX}{\mathcal{X}}
\newcommand{\cY}{\mathcal{Y}}

\newcommand{\bv}{\mathbf{v}}
\newcommand{\btv}{\tilde{\bv}}
\newcommand{\bbv}{\bar{\bv}}
\newcommand{\bq}{\mathbf{q}}
\newcommand{\btq}{\tilde{\bq}}
\newcommand{\bbq}{\bar{\mathbf{q}}}
\newcommand{\bz}{\mathbf{z}}
\newcommand{\bw}{\mathbf{w}}

\newcommand{\tQ}{\widetilde{Q}}
\newcommand{\dT}{\delta}
\newcommand{\dmu}{\pi} %pdf of target mu in the f-d case
\newcommand{\dQ}{f}%{q} %pdf of kernel \Qker
\newcommand{\dbQ}{\overline{f}}%{\overline{q}} %pdf of kernel \bQker
\newcommand{\dV}{g}%{v} %pdf of kernel \Vker
\newcommand{\dM}{h}
\newcommand{\bm}{\mathbf{m}}

\newcommand{\pmr}{\rho}
\newcommand{\bvv}{\mathbf{w}}

\newcommand{\Ham}{\mathcal{H}}
\newcommand{\Sol}{\hat{S}}
\newcommand{\Pot}{\Phi}
\newcommand{\VPot}{\Psi}
\newcommand{\pv}{\mathbf{y}} %position-velocity pair
\newcommand{\Sd}{\Xi}

\newcommand{\dimd}{\dimx}%{D}
\newcommand{\Od}{\mathcal{O}_{\dimd}}
\newcommand{\Hd}{\mathcal{H}_\dimd}
\newcommand{\vs}{\mathbf{w}}%{\mathbf{v}} %simplex vertices
\newcommand{\ed}{\nu}%edge length distribution
\newcommand{\rrt}{F}%rescaling-rotation-translation map
\newcommand{\bzero}{\mathbf{0}}

\newcommand{\diag}{\mbox{diag}}
\newcommand{\noise}{\eta}
\newcommand{\data}{y}
\newcommand{\Data}{\mathcal{Y}}
\newcommand{\G}{\mathcal{G}}
\newcommand{\param}{A_\bq}

\newcommand{\bg}{\mathbf{g}}
\newcommand{\bx}{\mathbf{x}}
\newcommand{\Obs}{\mathcal{O}}

\newcommand{\pd}{\partial}
\newcommand{\TT}{\mathbb{T}}

\newcommand{\bk}{\mathbf{k}}
\newcommand{\bee}{\mathbf{e}}
\newcommand{\bff}{\mathbf{f}}

\newcommand{\qS}{\mathfrak{Q}}
\newcommand{\bU}{\mathbf{u}}
\newcommand{\pS}{\theta}
\newcommand{\src}{f}
\newcommand{\bnh}{\hat{\mathbf{n}}}
\newcommand{\DD}{\mathcal{D}}
\newcommand{\rtrt}{\bar{\omega}}
\newcommand{\by}{\mathbf{y}}
\newcommand{\pth}{\phi}
\newcommand{\pr}{| \cdot |}
\newcommand{\ib}{\Gamma^i_{\bq}} %inner boundary
\newcommand{\ir}{r_{\text{min}}} %inner radius
\newcommand{\mr}{r_{\text{max}}} %middle radius
\newcommand{\oR}{R} %outer radius
\newcommand{\lnk}{c} %clamp or link function

\newcommand{\ExtSp}{\mathcal{Z}}
\newcommand{\bqv}{\mathbf{w}}

\newcommand{\br}{\mathbf{r}}
\newcommand{\btz}{\tilde{\mathbf{z}}}
\newcommand{\bhz}{\hat{\mathbf{z}}}
\newcommand{\bhq}{\hat{\mathbf{q}}}

\newcommand{\KR}{\mathcal{R}}%{\mathfrak{R}}
\newcommand{\KA}{\mathcal{A}}%{\mathfrak{A}}
\newcommand{\KP}{\mathcal{P}}%{\overline{P}}%{\mathfrak{P}}
\newcommand{\ProjES}{\mathcal{E}}%{\mathfrak{E}}
\newcommand{\MN}{\mathcal{N}}%{\overline{\mathcal{M}}}%{\mathfrak{N}}

\newcommand{\tk}{\tilde{k}}

\usetikzlibrary{matrix}
\usetikzlibrary{backgrounds}
\usetikzlibrary{calc}
\usetikzlibrary{arrows,shapes}
\usetikzlibrary{decorations}
\usetikzlibrary{decorations.pathmorphing}
\usetikzlibrary{fit}
\usetikzlibrary{decorations.pathreplacing}

\newtoggle{quickdraw}
\toggletrue{quickdraw} % Uncomment this to render more quickly (non-random)

\definecolor{darkgreen}{rgb}{0,0.3,0}
\definecolor{darkred}{rgb}{0.3,0,0}

\definecolorset{rgb}{}{}{darkred,0.8,0,0;darkgreen,0,0.5,0;darkblue,0,0,0.5}

\definecolor{trevorblue}{rgb}{0.330, 0.484, 0.828}
\definecolor{trevoryellow}{rgb}{0.829, 0.680, 0.306}

\newcommand{\arxiv}[1]{#1}

\begin{document}

\maketitle

\begin{abstract}
  Parallel Markov Chain Monte Carlo (pMCMC) algorithms generate clouds
  of proposals at each step to efficiently resolve a target
  probability distribution $\mu$. We build a rigorous foundational
  framework for pMCMC algorithms that situates these methods within a
  unified `extended phase space' measure-theoretic formalism. Drawing
  on our recent work that provides a comprehensive theory for
  reversible single proposal methods, we herein derive general
  criteria for multiproposal acceptance mechanisms which yield
  ergodic chains on general state spaces. Our formulation
  encompasses a variety of methodologies, including proposal cloud
  resampling and Hamiltonian methods, while providing a basis for the
  derivation of novel algorithms.  In particular, we obtain a top-down
  picture for a class of methods arising from `conditionally
  independent' proposal structures. As an immediate application of
  this formalism, we identify several new algorithms including a
  multiproposal version of the popular preconditioned Crank-Nicolson
  (pCN) sampler suitable for high- and infinite-dimensional target
  measures which are absolutely continuous with respect to a Gaussian
  base measure.
  
  To supplement the aforementioned theoretical results, we carry out a
  selection of numerical case studies that evaluate the
  efficacy of these novel algorithms. First, noting that the true
  potential of pMCMC algorithms arises from their natural parallelizability and
  the ease with which they map to modern high-performance computing
  architectures, we provide a limited parallelization study using
  \textsc{TensorFlow} and a graphics processing unit to scale pMCMC
  algorithms that leverage as many as 100k proposals at each step. Second, we use our multiproposal pCN algorithm (mpCN) to resolve a
  selection of problems in Bayesian statistical inversion for partial
  differential equations motivated by fluid measurement. These
  examples provide preliminary evidence of the efficacy of mpCN for
  high-dimensional target distributions featuring complex geometries and
  multimodal structures.
\end{abstract}

{\noindent \small {\it {\bf Keywords: } Parallel (Multiproposal)
    Markov Chain Monte Carlo (pMCMC), preconditioned Crank-Nicolson
    (pCN), Hamiltonian Monte Carlo (HMC), simplicial sampler, Bayesian statistical inversion, graphics processing units (GPU), high-performance computing, Metropolis-Hastings kernels.
  } } \\

\vspace{-0.3cm}
\begin{footnotesize}
\setcounter{tocdepth}{1}
\tableofcontents
\end{footnotesize}

\newpage

\section{Introduction}
\label{sec:Intro}

The efficient generation of random samples is a central task within
many of the quantitative sciences.  The workhorse of Bayesian
statistics and statistical physics, Markov chain Monte Carlo (MCMC)
comprises a large class of algorithms for sampling from arbitrarily
complex or high-dimensional probability distributions. The
Metropolis-Hastings method (MH) \cite{metropolis1953equation,
  hastings1970monte} stands as the seminal MCMC algorithm, and its
basic operation underlies most of the MCMC techniques developed to
this day. At each iteration, these algorithms choose the next Markov
chain state by (1) randomly generating a proposal state according to
an auxiliary proposal distribution and (2) accepting or rejecting the
proposed state according to a carefully derived threshold. Powerful,
modern algorithms such as Hamiltonian (hybrid) Monte Carlo (HMC)
\cite{duane1987hybrid,neal2011mcmc} share the same two-step approach
but use additional deterministic machinery to guide the random
proposal and obtain, e.g., lower autocorrelation between samples.
Recent independent works \cite{neklyudov2020involutive,
  glatt2020accept, andrieu2020general} develop an all-encompassing
mathematical framework that describes essentially any (reversible)
single proposal MH algorithm using three ingredients: a random
proposal, an involution on an extended phase space and an
accept-reject step.  This unified framework illuminates
under-appreciated relationships between a variety of known algorithms
while providing a means for deriving new methods.

Unfortunately, the overall structure of these MH extensions may fail
to fully exploit contemporary parallel computing resources such as
multi-core central processing units (CPUs) and many-core graphics
processing units (GPUs). While model-specific algorithmic subroutines
such as log-likelihood and log-likelihood gradient evaluations
sometimes admit parallelization \cite{massive, holbrook2021scalable,
  holbrook2022bayesian, holbrook2022viral}, the algorithms' generally
sequential nature can lead to under-utilization of increasingly
widespread hardware \cite{brockwell2006parallel}. Parallel MCMC
algorithms (pMCMC) offer a top-down approach to exploiting such
infrastructures through their use of multiple proposals at each
step. The multiproposal structure makes pMCMC algorithms amenable to
conventional parallelization resources such as GPUs (see
\cref{sec:ppMCMC}), and \cite{holbrook2023quantum} even demonstrates
advantages for implementations that leverage quantum computing.  While
general, efficient and user-friendly parallel implementations of pMCMC
algorithms remain an engineering challenge, there is reason to believe
that the pMCMC paradigm may prove useful when other, more prevalent
MCMC paradigms---such as, e.g., gradient-based sampling or
parallelization using multiple chains
\cite{gelman1992inference}---encounter their own challenges.

Whereas gradient-based MCMC methods such as HMC or Metropolis-Adjusted
Langevin Algorithm (MALA) can scale inference to extremely
high-dimensional settings, they are not always appropriate. Sometimes
the gradient is either not available, is computationally prohibitive
or involves difficult derivations (see \cref{sec:rot:Stokes}).  More
generally, these algorithms struggle to tackle distributions with
challenging geometries, even though there are sometimes solutions.  On
the one hand, if a target is roughly Gaussian---albeit with an
ill-conditioned covariance matrix---then one may effectively implement
HMC using an adaptive mass matrix \cite{holbrook2022viral}. On the
other hand, if a target is both ill-conditioned and strongly
non-Gaussian, then there are few gradient-based options beside Riemannian HMC
\cite{girolami2011riemann}, a user-intensive method that scales poorly
to high dimensions, .  Finally, even HMC can lose its competitive
advantage over vanilla MH in multimodal settings
\cite{mangoubi2018does}.

Parallel implementation of multiple MCMC chains is often a beneficial
strategy that adequately leverages hardware resources such as multiple
cluster nodes or CPU cores, but the paradigm is not without its
own difficulties. First, a lack of communication between chains means
that one chain's progress cannot aid that of another. This is a
problem because there is usually little benefit to combining multiple
poorly-mixing chains, where one may measure the quality of the
combined Monte Carlo estimator using, say, the potential scale
reduction factor (R-hat) of \cite{gelman1992inference}.  Thus, the
parallel chain strategy may fail when a target displays difficult
geometry that confounds individual chains. When such targets are also
high-dimensional, memory limitations may inhibit inference: $C$ chains
of length $S$ defined on a $D$-dimensional state space require
$\mathcal{O}(SDC)$ storage. Finally, the multi-chain strategy may also
fail when the target distribution is multimodal, as
\cite{neal1996sampling} notes:
\begin{quote}
  \emph{Unfortunately, multiple independent runs will not, in general,
    produce a sample in which each mode is fairly represented, because
    the probability of a run reaching a mode will depend more on the
    mode's `basin of attraction' than on the total probability in the
    vicinity of the mode.}
\end{quote}

Accordingly, we cast pMCMC as a paradigm that both complements other
MCMC paradigms and provides further opportunities for the use of
high-performance computing resources in MCMC.  Notwithstanding a
significant and growing recent literature
on pMCMC and related methods,  cf. \cite{neal2003markov, 
tjelmeland2004using, frenkel2004speed, delmas2009does, calderhead2014general, 
luo2019multiple, liu2000multiple, schwedes2021rao, holbrook2023generating},  
the subject remains
underdeveloped leaving a broad scope for the development and analysis
of novel pMCMC methods.  

This work provides a unified theoretical foundation for algorithm
design in this important direction in the sampling literature,
encompassing algorithms that use single (\cref{sec:framework}) or
multiple (\cref{sec:ext:Phase}) jumps within each proposal set.
Our approach then allows us to identify new algorithms while placing
existing methods in a broader context and rigorously justifying their
reversibility (\cref{sec:Ex:alg}). Finally, we provide a series of
numerical case studies (\cref{sec:case:Study}) which demonstrate the
efficacy of our new methods and explore the tuning of algorithmic
parameters.  We continue this introduction with a comprehensive
summary of our contributions (\cref{sec:cont}) before concluding with
an overview of the existing pMCMC literature (\cref{sec:back}).

\subsection{Contribution overview}
\label{sec:cont}

Our first contribution is to show that the unified framework developed
recently in \cite{glatt2020accept} can be fruitfully and nontrivially
extended to encompass a broad class of pMCMC methods. We develop an
involutive theory of pMCMC that (1) firmly places these multiproposal
algorithms in a broad measure-theoretic context and (2) makes clear
when a pMCMC algorithm is unbiased or reversible with respect to a
given target distribution $\mu$.  Note that our formalism includes
extensions that allow for proposal cloud resampling.

In \cref{sec:framework}, we begin by 
introducing a broad class of Markovian kernels of the form
\begin{align}
\label{eq:multi:prop:abs:intro}
  \Pker^{\alpha, S, \Vker}(\bq,d\btq) 
  = \sum_{j = 0}^p \int_\spv \alpha_j(\bq,\bv)
        \delta_{\Proj \circ S_j(\bq,\bv)}(d\btq) \Vker(\bq,d \bv),
\end{align}
where $\bq$ is the current state, the $\alpha_j$s determine the
acceptance probabilities, the $S_j$s are involutive operators acting
on an extended abstract phase space $\spq \times \spv$, $\Proj$ is the
projection onto the first component, namely $\Proj(\bq, \bv) = \bq$,
and $\Vker$ is the multiproposal mechanism.  Algorithmically, the
kernel $\Pker^{\alpha, S, \Vker}$ first draws a sample $\bv$ from
$\Vker(\bq,d\bv)$, then produces a cloud of proposals
\begin{align*}
  (\Proj \circ S_0( \bq,\bv), \ldots, \Proj \circ S_p( \bq,\bv)),
\end{align*}
and finally draws the next state from this cloud with probabilities
\begin{align*}
  (\alpha_0(\bq,\bv), \ldots, \alpha_p(\bq,\bv)).
\end{align*}
As we illustrate below, \eqref{eq:multi:prop:abs:intro} encompasses a
very broad class of pMCMC samplers inclusive of both random walk and
Hamiltonian type methods.

In a series of results (\cref{thm:main}, \cref{cor:ar:Barker},
\cref{cor:wedge:alpha}, \cref{thm:multi:ind:Tjmd} and
\cref{thm:cond:ind:mu:mu0}), we identify
various conditions on $\alpha_j$, $S_j$ and $\Vker$ such that, for a
given target measure $\mu$, we have
\begin{align}
  \label{eq:rev:intro}
  \Pker^{\alpha, S, \Vker}(\bq,d\btq) \mu(d\bq)
  =\Pker^{\alpha, S, \Vker}(\btq,d\bq) \mu(d\btq),
\end{align}
i.e., $\Pker^{\alpha, S, \Vker}$ is reversible and hence invariant
with respect to a given target $\mu$.  Notably, \cref{cor:wedge:alpha}
includes as a special case any single proposal MH
type algorithm which is reversible with respect to a given target.  As
such, the theory we develop here may be seen as a strict
generalization of \cite{glatt2020accept}.

On the other hand, it is worth highlighting some novel insights and
challenges to developing the multiproposal theory in contrast to
\cite{neklyudov2020involutive, glatt2020accept, andrieu2020general}.
The single proposal setting rightly emphasizes the traditional MH
acceptance mechanism, a choice which has shown to be optimal
\cite[Section 3]{tierney1998note}.  This mechanism frames
\cref{cor:wedge:alpha} herein.  By contrast, a Barker-type acceptance
mechanism \cite{barker1965monte} reveals its salience in the pMCMC
setting, and our results \cref{cor:ar:Barker},
\cref{thm:multi:ind:Tjmd} and \cref{thm:cond:ind:mu:mu0} lead to
\cref{alg:TJ:cor:FD} and \cref{alg:mpCN} later in \cref{sec:Ex:alg}.

To make this more explicit, we consider pMCMC algorithms that involve
acceptance probabilities for the $j$th proposed state taking the form
\begin{align}
    \alpha_j(\bq_0, \bq_1,\ldots,\bq_p) 
    = \frac{\pi(\bq_j)}{\sum_{k=0}^p \pi(\bq_k)},
    \label{eq:acp:intro}
\end{align}
where $\bq_0$ is the current state of the chain,
$(\bq_1,\ldots,\bq_p)$ is the cloud of proposed states made according
to some kernel $\Vker(\bq_0, d \bq_1,\ldots, d\bq_p)$, and the
involutions $S_j$ are the `flip' operators exchanging the $jth$ and
zeroth elements (see \eqref{def:Sj:cond:ind} below).  Here the
function $\pi$ may be the density of the target measure $\mu$ or taken to
be proportional to the likelihood in Bayesian settings.  The salience of
\eqref{eq:acp:intro} is immediately intuitive---it yields high
acceptance probabilities for points in the proposal cloud that inhabit
high probability regions of the target distribution.

Of course, we do not expect $\Vker$ in conjunction with
$(\alpha_0, \ldots, \alpha_p)$ defined according to
\eqref{eq:acp:intro} to yield an unbiased algorithm for a given target
$\mu$ in general.  But, in \cref{sec:cond:ind:prop}, we are able to
draw the following wide ranging conclusions by restricting attention
to a certain class of `conditionally independent' proposal kernels of
the form
\begin{align}
  \Vker(\bq_0, d \bq_1,\ldots, d\bq_p)
  =  \int
  \prod_{j=1}^p \Qker(\tilde{\bq}_0, d \bq_j)
  \bQker(\bq_0, d \tilde{\bq}_0),
	\label{eq:norwegian:wood}
\end{align}
where $\Qker, \bQker$ are single element Markov kernels on the
parameter space on which $\mu$ sits.  Note that, for such kernels
$\Vker$, one obtains samples from the current state $\bq_0$ by first
producing $\btq_0$ from $\bQker(\bq_0, \cdot)$ and then drawing
$(\bq_1, \ldots, \bq_p)$ independently from
$\Qker(\tilde{\bq}_0, \cdot)$.  For such proposal structures $\Vker$,
we are able to write down a Barker-like acceptance probability such
that the resulting Metropolized system is unbiased with respect to
essentially any target measure $\mu$. See \cref{thm:multi:ind:Tjmd},
for our precise formulation.

Interestingly, the naive choice
$\bQker(\bq_0, d \tilde{\bq}_0) = \delta_{\bq_0}( d \tilde{\bq}_0)$
that reduces $\Vker$ to $\prod_{j=1}^p Q(\bq_0, d \bq_j)$, i.e., $p$
independent proposals around the current state $\bq_0$, leads to a
complicated and potentially computationally onerous acceptance
probability $\alpha_j$. See, for example, \eqref{eq:non:TJ:cor:FD} in
\cref{sec:Ex:alg}.  On the other hand, inspired by
\cite{tjelmeland2004using}, if one carefully chooses $\bQker$ in
relation to $\Qker$ we show that one obtains an acceptance probability
of the form \eqref{eq:acp:intro}.  See \cref{thm:cond:ind:mu:mu0} for
precise formulations.  Note that in many cases $\bQker$ is readily
identifiable from $\Qker$.  Indeed, as observed previously in
\cite{tjelmeland2004using} in the specific case of Gaussian proposals,
we may take $\Qker = \bQker$ under a minimal symmetry condition on
$\Qker$.  Taken further this case leads us to novel `infinite
dimensional' algorithms as we describe presently (see
\cref{alg:mpCN}). Furthermore, in the finite dimensional setting we
show that by considering \emph{any} probability density
$r: \RR^N \to \RR^+$ we obtain suitable $\Qker, \bQker$ by taking
\begin{align}
  \Qker(\bq,d \btq) = r(\bq -  \btq) d \btq,
  \quad   \bQker(\bq,d \btq) = r(\btq -  \bq) d \btq,
\end{align}
thus accommodating non-symmetric proposal structures (see
\cref{alg:TJ:cor:FD}).

Another natural consideration in developing pMCMC methods is to
consider the possibility of performing multiple finite state space
jumps between the elements of a proposal cloud.  After all, when
target evaluations at proposal points represents the main
computational burden, there may be relatively little computational
overhead associated with additional inter-proposal jumps.  In
\cref{sec:ext:Phase}, we extend the formalism of \cref{sec:framework}
to support multiple jumps within the proposal set.  Here our framework
provide a rigorous justification for methods advanced in
\cite{calderhead2014general}.

In summary, the formalisms we develop in \cref{sec:framework},
\cref{sec:ext:Phase} provide a unified basis for concrete algorithm
design. In \cref{sec:Ex:alg}, we begin to explore this large space of
possible methods and to develop a selection of applications within our
paradigm.  Our most general results \cref{thm:main}, \cref{thm:tj} and
our observations around kernels of the form \eqref{eq:norwegian:wood},
\cref{thm:multi:ind:Tjmd}, \cref{thm:cond:ind:mu:mu0},
\cref{cor:simple:alpha}, yield the novel and immediately applicable
algorithms in \cref{alg:TJ:cor:FD} and in \cref{alg:mpCN},
\cref{alg:mpCN:TJ}. Meanwhile in \cref{sec:HMC}, we consider HMC
variants that use multiple integration times.  This study includes a
rigorous and more systematic framing of a method suggested previously
in \cite{calderhead2014general}. \cref{sec:simplicial} revisits the
simplicial sampler recently introduced in
\cite{holbrook2023generating}.

\cref{alg:mpCN}, \cref{alg:mpCN:TJ} provide multiproposal
generalizations of the so called preconditioned Crank-Nicolson (pCN)
algorithm \cite{beskos2008mcmc, cotter2013mcmc}. These methods address
a class of target measures of the form
\begin{align}
	\mu(d\bq) = Z^{-1} \exp(- \phi(\bq)) \mu_0(d\bq),
	\label{eq:stuart:form}
\end{align}
where $\mu_0$ is a centered Gaussian base measure.  Such measures
$\mu$ appear naturally in a variety of high- and infinite-dimensional
settings \cite{stuart2010inverse, dashti2017bayesian}.  
Note that while practical implementation of such schemes requires truncation to finite dimensions, deriving algorithms tailored to infinite-dimensional target measures can partially beat the curse of dimensionality in the sense that mixing rates for truncations of the full problem do not depend on the order of the truncation; see \cite{hairer2014spectral}, \cite{glatt2021mixing} for some rigorous results in this direction.

The pCN
algorithm---so named because it arises from a Crank-Nicolson
temporal discretization of an Ornstein–Uhlenbeck process
preconditioned by the covariance $\cC$ of $\mu_0$---exploits a
proposal kernel of the form
\begin{align}
	Q_{pCN}(\bq, d \tilde{\bq}) = \Prb( \rho \bq + \sqrt{1 - \rho^2} \bfV
       \in d \tilde{\bq}), \quad
	\text{ where } \bfV \sim \mu_0.
	\label{eq:pCN:prop:intro}
\end{align}
Here, $\rho \in [0,1]$ is an algorithmic parameter dictating the
degree of aggressiveness of the proposal.  In the context of
\eqref{eq:norwegian:wood}, we show that combining
$\Qker = \bQker := Q_{pCN}$ and an acceptance probability of the form
\eqref{eq:acp:intro} with $\pi(\bq)= \exp(- \phi(\bq))$ yields a pMCMC
algorithm which is reversible (and hence invariant) with respect to
measures of the form \eqref{eq:stuart:form}.  We dub this method the
Multiproprosal-pCN (mpCN) algorithm.  Notably, mpCN stands as a
gradient-free methodology that is easy to implement in comparison to
other `Hilbert space' methods used for certain parameter estimation
problems for partial differential equations (PDEs) considered in
statistical inversion.  Here, the existing alternatives to vanilla pCN
algorithm, $\infty$MALA and $\infty$HMC
\cite{cotter2013mcmc,Beskosetal2011}, typically require adjoint
methods and may be complicated or practically impossible to
implement. Following the discussion at the beginning of this
introduction, even properly implemented versions of these algorithms
can suffer under difficult target geometries.

Armed with \cref{alg:TJ:cor:FD} and \cref{alg:mpCN}, we conclude this
contribution by providing a series of numerical case studies in
\cref{sec:case:Study}.  These studies comprise preliminary
investigations into the efficacy and the tuning of algorithmic
parameters for these pMCMC algorithms.  We select these studies with an eye
towards effectively leveraging modern computer architectures while
identifying significant domains of immediate applicability.  In
\cref{sec:ppMCMC}, we demonstrate the natural parallelizability of
pMCMC with a limited high-performance computing study. Here, we use a
GPU to power \cref{alg:TJ:cor:FD} with up to 100,000 proposals at each
iteration. With a simple \textsc{TensorFlow} implementation, wall
times per iteration for a GPU are shown to be orders of magnitude
smaller than those for a comparable sequential implementation.
Furthermore, we show that these speedups increase with target
dimensionality. Under a fixed computational budget, GPU based
parallelization also confers greater sampling efficiency for massively
multimodal target distributions, with speedups again increasing with
mixture component counts. But the naturally parallel structure of
pMCMC makes it amenable to other parallel computing techniques as
well. In a companion paper \cite{holbrook2023quantum}, we combine
quantum optimization with the Gumbel-max trick and show that quantum
algorithms deliver significant theoretical speedups for pMCMC
algorithms similar to \cref{alg:TJ:cor:FD}.

To test efficacy of mpCN, namely \cref{alg:mpCN}, in
\cref{sec:B:Stat:Inv} we consider three stylized nonlinear statistical
inverse problems \cite{kaipio2006statistical, stuart2010inverse,
  dashti2017bayesian} motivated by fluid measurement, drawing on our
concurrent and recent work \cite{borggaard2020bayesian,
  borggaard2023statistical}.  For all of these problems
our aim is to provide a principled estimate for an unknown parameter
$\bq$ sitting in a Hilbert space $\spq$ from the observation of data
$\Data$ described by $\Data = \G(\bq) + \noise$.
Here $\G$ is a forward map determined from a physical model defined up
to the unknown parameter $\bq$.  The additive $\noise$ is a
probabilistic quantity representing measurement error.  We in effect
invert $\G$ by placing a (Gaussian) prior $\mu_0$ on $\bq$ and then invoking
Bayes theorem to obtain a posterior of the form \eqref{eq:stuart:form}
defined in terms of $\G$ and the distribution of $\noise$.  Note that,
in each of the problems we present, $\G$ is a nonlinear map from
higher dimensional unknown parameter space to lower dimensional
collections of observations so that we are addressing a severely
ill-posed inverse problem resulting in a non-Gaussian posterior target
measure $\mu$; see \cref{fig:ADToy_True_Hist} \arxiv{and also
  \cref{fig:ad:hist2dtrue}} below.  Furthermore, the latter two problems
presented in \cref{sec:AD:JUQ}, \cref{sec:rot:Stokes} involve a
naturally infinite-dimensional parameter space and feature forward
maps involving the numerically expensive resolution of a partial
differential equation.

As our first example we consider a problem that mimics the
mathematical form of PDE inverse problems of interest at a
smaller scale.  Our aim here is to provide `table-top lab' amenable to
computations which can be easily carried out on a personal computer in
a matter of minutes or hours for moderate dimensional problems and
thus allowing for more comprehensive parameter studies. This
problem estimates the coefficients of a $d\times d$ antisymmetric
matrix $\param$---parameterized by its $d(d-1)/2$ independent
components $\bq$---from the partial, noisy observations of solutions
$\Sol(\bq) = \bx(\param)$ of the corresponding problem
$(\param + \kappa I)\bx = \bg$, where here $\kappa> 0$ and
$\bg \in \RR^d$ are additional parameters assumed to be known a
priori.  We find that this class of inverse problems exhibits intricate
statistical structures reminiscent of far more complex PDE constrained
settings they mimic.  See \cref{fig:ADToy_True_Hist} below\arxiv{ and
  compare with, e.g., \cref{fig:ad:hist2dtrue}}.  We therefore think
this simple toy model holds independent interest outside of our
immediate considerations here.

Our other two inverse problems involve infinite-dimensional unknown
parameter spaces and a forward map $\G$ which involves resolution of a
partial differential equation or system of partial differential
equations.  One infinite-dimensional problem, considered previously in
\cite{borggaard2020bayesian}, involves the estimation of a divergence
free fluid flow field $\bfV$ through the sparse, noisy observation of
a solute which is passively advected and diffuses in the fluid medium.
Physically, we can think of estimating the motion of water filling a
glass by observing the concentration of a dye which has been
introduced.  Mathematically, this may be described as observing the
solution $\pS = \pS(\bq)$ of an advection diffusion equation
$\pd_t \pS + \bq \cdot \nabla \pS = \kappa \Delta \pS$,  which we
consider here on a periodic two-dimensional domain starting from a
given (known) initial condition $\pS_0$.  Here, as a second infinite-dimensional benchmark for mpCN, we revisit a particular stylized
example identified in \cite{borggaard2020bayesian} which exploits
natural symmetries in this model to produce complex high-dimensional
statistical correlation structures that are challenging to sample from
in an efficient manner.

Our final statistical inversion problem is a fluid domain shape
estimation problem which we develop in concurrent work
\cite{borggaard2023statistical}. For this problem the
unknown parameter $\bq$ specifies the shape of a domain upon which a
system of PDEs governing a time stationary Stokes flow $\bU$ and an
associated solute concentration $\theta$ are defined.  Our aim is
therefore to estimate the boundary shape from sparse or volumetrically
averaged observations of $\bU$ and $\theta$. The forward map $\G$
therefore requires solving a system of PDEs on an irregular domain, making the
derivation and implementation of an appropriate adjoint method a
difficult task. Moreover, the typical implementation will involve a
(typically third-party, black box) meshing algorithm to re-mesh the
domain at each iteration, ruling out use of automatic differentiation
algorithms. This problem is therefore a natural fit for exploration of
gradient-free algorithms.

\subsection{Literature review}\label{sec:back}

The pMCMC literature presents a patchwork of independent and sometimes
overlapping frameworks for using multiple proposals within a
generalized MH algorithm. Although these works come from different
disciplines (e.g., statistics \cite{neal2003markov}, physics
\cite{frenkel2004speed}, probability \cite{delmas2009does}, and machine
learning \cite{schwedes2021rao}), they all seek to answer a natural
question: ``Why not use more than one proposal within
Metropolis-Hastings?'' Perhaps due to the multi-disciplinary context,
the different answers to this question have often come with their own
terminology: \emph{waste-recycling} \cite{frenkel2004speed},
\emph{parallel MCMC} \cite{schwedes2021rao}, \emph{multiproposal
  methods} \cite{holbrook2023generating}, and \emph{generalized}
\cite{calderhead2014general} or \emph{multiple-try}
\cite{luo2019multiple} Metropolis-Hastings are a few. In this paper,
we explicitly do not consider the well-known multiple-try
Metropolis-Hastings algorithm of \cite{liu2000multiple}, which
randomly selects from a cloud of proposals \emph{before} an additional
Metropolis-Hastings step. Here, we are interested in methods that
subsume proposal selection into a single step.  Those who are interested in a thorough involutive treatment of multiple-try Metropolis-Hastings should consult \cite{andrieu2020general}, who successfully fit that particular algorithm into a single-proposal framework. Whereas the general multiple-proposal framework we construct in Section 2 admits the single-proposal framework of \cite{glatt2020accept} as a special case, it is difficult to say definitively that there is no theoretical mechanism by which our multiple-proposal extension may be viewed as a special case of the single-proposal theory.  We reserve this question for future work. 

\cite{neal2003markov}, \cite{tjelmeland2004using} and
\cite{frenkel2004speed} provide early contributions to the pMCMC
literature by investigating MCMC algorithms that generate multiple
proposals at each iteration. \cite{neal2003markov} considers the
special setting of non-linear state space models and develops an
algorithm for sampling from the distribution over hidden states
conditioned on an observed sequence. At each iteration, the method
generates a `pool' of candidate sequences, an element of which is the
current state, in an iterative manner and selects from among these
sequences with probability proportional to the target distribution
divided by a `pool' distribution that characterizes the probability of
obtaining the individual proposal conditioned on the other
proposals. \cite{neal2003markov} shows that this strategy with $P=10$
proposals greatly outperforms simple Metropolis steps on an iteration
by iteration basis and that computing the target probabilities with
the help of the linear-time forward-backward algorithm
\cite{scott2002bayesian} boosts implementation
speeds. \cite{tjelmeland2004using} works in a more general setting and
develops multiple proposal generation and multiple transition
strategies. Of the latter, the paper's `Transition alternative 1'
relates closely to that of \cite{neal2003markov}, randomly accepting a
proposed state with probability proportional to the target density
times the probability of generating the proposal conditioned on the
other proposed states. In addition to simply randomly accepting one of
the many proposals at each step, \cite{tjelmeland2004using} shows that
a carefully weighted average of all proposals provides an unbiased
estimator for an arbitrary estimand, assuming that the Markov chain
has converged to the target distribution. Among other results,
\cite{tjelmeland2004using} shows that empirically optimal proposal
scalings grow, and empirical variances of the weighted estimator
decrease, with the number of proposals.  \cref{sec:framework},
\cref{sec:ext:Phase} establishes a measure-theoretic framework that
reduces to the methods of \cite{tjelmeland2004using} as a special
case.

Working in the setting of statistical physics, \cite{frenkel2004speed}
proposes a method called `waste-recycling' that generates multiple
proposals at each iteration and constructs a running, weighted
estimator that is a special case of that of
\cite{tjelmeland2004using}, with weights that are products of target
probabilities and the probabilities of transitioning from all other
proposed states. \cite{frenkel2004speed} argues for the correctness of
this approach by viewing each iteration's contribution to the running
estimator as a `coarsening' of $P+1$ states that satisfies
`superdetailed balance' despite not satisfying detailed
balance. Finally, \cite{frenkel2004speed} applies waste-recycling to
estimating the probability distribution over total spin for a 2D,
discrete spin Ising model on a $32^2$ spin lattice and demonstrates an
over $10^{10}$-fold reduction in statistical error compared to
Swendsen-Wang. \cite{delmas2009does} prove consistency and asymptotic
normality of the waste-recycling estimator and show that the estimator
achieves a variance reduction over MH for certain selection
probabilities (alternatively, weights).  \cite{delmas2009does} calls
these the Boltzmann or Barker selection kernel, and they are
equivalent to those of \cite{neal2003markov, tjelmeland2004using,
  frenkel2004speed}.  We note that there is no fundamental need to combine the use of multiple proposals with the weighted average approach (as is done in waste-recycling): recent Markov chain importance sampling approaches demonstrate benefits of incorporating rejected proposals via a weighted estimator within the single proposal framework \cite{schuster2020markov,rudolf2020metropolis}.

Next, \cite{calderhead2014general} develops a distinct approach that
(1) generates multiple proposals and (2) simulates a \emph{finite}
state Markov chain, the states of which are the union of the current
state and proposed states, for a fixed number of iterations at each
step.  \cite{calderhead2014general} describes criteria necessary for
the finite state Markov chain transition probabilities to satisfy
detailed balance and argues that traditional MH is a special case of
this procedure that combines a single proposal with a single
transition. \cite{calderhead2014general} does not point out the fact
that the specific form of these transition probabilities admits as a
special case the form of the transition probabilities and weighting
schemes shared by \cite{neal2003markov, tjelmeland2004using,
  frenkel2004speed}.  Unlike these previous works,
\cite{calderhead2014general} incorporates the partially deterministic
proposal schemes of the Metropolis-adjusted Langevin algorithm (MALA)
\cite{roberts2002langevin} and HMC into his pMCMC algorithm.  Applying
these schemes to logistic regression and Bayesian inversion of an ODE,
\cite{calderhead2014general} shows, e.g.: that pMCMC using MALA or
adaptive transitions \cite{haario2001adaptive} and 1,000 proposals
enjoys 5-fold speedups over the algorithm using a single proposal; and
using multiple intermediate leapfrog steps from HMC transitions as
proposals leads to a roughly 60\% decrease in Monte Carlo
error. Unfortunately, \cite{calderhead2014general} stops short of
proving the correctness of such pMCMC algorithms that leverage
extended phase space strategies, and the development of such a
theoretical framework is one contribution of the present work
(\cref{sec:ext:Phase}).

More recently, \cite{yang2018parallelizable} extend the
waste-recycling method of \cite{frenkel2004speed} to include HMC
transitions and prove that the waste-recycling estimator is unbiased
and, as a Rao-Blackwellization, leads to variance reductions over the
estimator of
\cite{calderhead2014general}. \cite{yang2018parallelizable} accomplish
this for two general weighting schemes, one of which is
\cite{calderhead2014general}'s generalization of, and one of which is,
that of \cite{neal2003markov, tjelmeland2004using,
  frenkel2004speed}. \cite{schwedes2021rao} prove similar results to
\cite{yang2018parallelizable} for essentially the same algorithm with
the addition of an adaptive proposal mechanism extending
\cite{haario2001adaptive}. \cite{schwedes2021rao} conduct an empirical
study with important consequences for the present work and show that,
within the finite state space framework of
\cite{calderhead2014general}, jumping between proposed states for many
iterations leads to significantly lower Monte Carlo error. Remarkably,
this study uses $P$ finite state space iterations for $P$ proposals as
baseline and increases the number of finite state space iterations to
as many as $16P$. Extrapolating these results, one might expect that a
pMCMC algorithm that uses only a single finite state space iteration
at each step would generate unbiased estimators with astronomically
large variances.

Whereas \cite{frenkel2004speed, yang2018parallelizable,
  calderhead2014general, schwedes2021rao} generate multiple
independent and identically distributed proposals, one may generate
multiple proposals that share a more complex structure.  Earlier in
this section, we refer to the iterative strategy of
\cite{neal2003markov} in the context of non-linear state space models,
and \cite{tjelmeland2004using} advances two additional proposal
strategies designed to maintain exchangeability between proposals.
The first of these strategies, `Proposal alternative 1', (1) generates
a random center for the proposal distribution that itself follows a,
say, Gaussian distribution centered at the current state and (2)
generates $P$ proposals from the same distribution with updated
center.  The second strategy, `Proposal alternative 2', cleverly
enforces that the current state and all $P$ proposals be equidistant
from each other by iteratively generating and carefully manipulating
all proposals. Two recent works advance the two proposal alternatives
of \cite{tjelmeland2004using}. On the one hand, \cite{luo2019multiple}
extend the first proposal alternative to allow proposals to share a
general acyclic graphical structure. On the other hand,
\cite{holbrook2023generating} develops a proposal mechanism that
maintains equal distances between proposals; this \emph{simplicial
  sampler} initializes all proposals as the vertices of a
high-dimensional regular simplex and rotates these vertices according
to the Haar distribution on the orthogonal group. The formal
developments of \cref{sec:framework} include general, structured
proposal mechanisms, and \cref{sec:Ex:alg} applies this theory to the
structured proposals of \cite{tjelmeland2004using} and
\cite{holbrook2023generating}.

\section{An abstract framework for pMCMC algorithms}
\label{sec:framework}

In this section we introduce our abstract formulation of pMCMC
algorithms along with rigorous conditions which guarantee that the
generated Markov chains are ergodic or, even more, reversible.  The
section is organized as follows.  First, \cref{sec:Multi:MH:ker}
presents a generic definition of Metropolis-type Markov transition
kernels in this multiproposal setting. \cref{sec:Rev:Criteria} then
establishes general conditions leading to the invariance or
reversibility of such algorithms. Finally, \cref{sec:cond:ind:prop}
considers a special but still broad case of proposal mechanisms and
identifies conditions under which corresponding acceptance
probabilities assume a simplified proposal-independent expression.
We refer to \cref{subsec:meas:th} for various measure-theoretic elements used herein and postpone all mathematical proofs to \cref{appx:proofs}.

\subsection{Multiproposal Metropolis-Hastings Markov kernels}
\label{sec:Multi:MH:ker}

We formulate an abstract setting for multiproposal extended phase
space Metropolis-Hastings kernels on general state spaces as
follows. Throughout the manuscript, we denote by $\Pr(\mathcal{X})$
the set of all probability measures on a measurable space
$\mathcal{X}$.

\begin{Definition}
\label{def:gen:MMH:ker}
Let $(\spq, \Sigma_\spq)$ and $(\spv, \Sigma_\spv)$ be measurable
spaces and take $\Vker: \spq \times \Sigma_\spv \to [0,1]$ to be a
Markov kernel. Namely, we suppose that $\Vker(\bq,d \bv) \in \Pr(\spv)$
for each $\bq \in \spq$ and $\bq \to \Vker(\bq, E)$ is a measurable
map for each $E \in \Sigma_\spv$.  For each $j=0,1,\ldots,p$ consider
measurable mappings $S_j: \spq \times \spv \to \spq \times \spv$ and
$\alpha_j : \spq \times \spv \to [0,1]$ such that
\begin{align}\label{sum:alpha:1}
  \sum_{j = 0}^p \int_\spv \alpha_j(\bq,\bv)  \Vker(\bq,d \bv) = 1
  \quad \mbox{ for every } \bq \in \spq.
\end{align}
Then, denoting $\alpha = (\alpha_0, \ldots, \alpha_p)$ and
$S = (S_0, \ldots, S_p)$, we define the \emph{multiproposal
  Metropolis-Hastings Markov kernel given by $(\alpha, S, \Vker)$},
$\Pker^{\alpha, S, \Vker} : \spq \times \Sigma_\spq \to [0,1]$ as
\begin{align}
\label{eq:multi:prop:abs}
	\Pker^{\alpha, S, \Vker}(\bq,d\btq) 
	= \sum_{j = 0}^p \int_\spv \alpha_j(\bq,\bv) \delta_{\Proj  S_j(\bq,\bv)}(d\btq)
  \Vker(\bq,d \bv),
\end{align}
where $\delta_{\Proj S_j(\bq,\bv)}$ is the Dirac measure
concentrated at $\Proj S_j(\bq,\bv)$ and $\Proj$ denotes the
projection operator onto the $X$ component, i.e.,
$\Proj(\bq,\bv) = \bq$ for all $\bq \in \spq$, $\bv\in \spv$.
\end{Definition}
As described in \cref{sec:Intro}, this setup may be seen as a
multiproposal extension to the recent work of \cite{glatt2020accept}.
\cref{alg:main} makes the underlying algorithmic procedure explicit.
See also \cref{def:gen:MMH:ker:ext:phase} in \cref{sec:ext:Phase}
below for an extended abstract phase space pMCMC formalism that admits
multiple jumps within a proposal set and directly generalizes the
frameworks of \cite{tjelmeland2004using,calderhead2014general}.  Note
that the class of kernels identified in \cref{def:gen:MMH:ker}
encompasses a wide variety of algorithmic structures.  This includes
the multiproposal generalizations of random walk MH and pCN we describe in
\cref{sec:rw:alg:fd} and \cref{sec:mpCN}.  In these settings, the
$S_j$s represent coordinate exchanges as in \eqref{def:Sj:cond:ind},
and the proposal kernels $\Vker$ have the conditionally independent
structure \eqref{def:Vker:multi:ind}.
On the other hand, \cref{def:gen:MMH:ker} also accommodates a variety of
HMC-type algorithms where the $S_j$s are
related to the numerical integration of a Hamiltonian system
associated to the target measure (see \cref{sec:HMC}).

We also note that the general multiproposal algorithm specified by the transition kernel \eqref{eq:multi:prop:abs} clearly differs from so-called hybrid MCMC strategies given by mixtures of kernels, see e.g. \cite{tierney1994markov,RobertCasellabook}. In the latter, one specifies a set of probabilities $a_1,\ldots,a_m$ and Markov kernels $P_1,\ldots,P_m$. At each iteration, one selects one of such kernels according to the given probabilities (which may only depend on the current state) and then draws a proposal sample from the chosen kernel. In contrast, under \eqref{eq:multi:prop:abs} with $Y = X^p$, one first samples a cloud of proposals $\bv = (\bq_1,\ldots,\bq_p) \in X^p$ from the given proposal kernel $\Vker(\bq,\cdot)$, and then selects one of $\bq,\bq_1,\ldots,\bq_p$ according to probabilities $\alpha_0,\ldots,\alpha_p$ which depend on \emph{all} of $\bq,\bq_1,\ldots,\bq_p$. See e.g. \eqref{def:alphaj:pCN} and \eqref{def:alphaj:hmc} below for concrete examples.

\subsection{General criteria for invariance or reversibility}
\label{sec:Rev:Criteria}

Our first result, \cref{thm:main}, establishes a general set of
conditions on the kernel $\Vker$ and the mappings $S_j$, $\alpha_j$,
$j=0,1,\ldots,p$, under which the corresponding Markov kernel
$\Pker^{\alpha,S,\Vker}$ maintains a given target probability measure
$\mu$ on $\spq$ invariant, or additionally that it satisfies detailed
balance with respect to $\mu$.  See \cref{appx:thm:main} for the
proof.
\begin{Theorem}\label{thm:main}
  Let $(\spq, \Sigma_\spq)$ and $(\spv, \Sigma_\spv)$ be measurable
  spaces. Fix a probability measure $\mu$ on $\spq$ and a Markov
  kernel $\Vker: \spq \times \Sigma_\spv \to [0,1]$. Let $\cM$ be the
  probability measure on the product space $\spq \times \spv$ defined
  by
 \begin{align}\label{def:M}
  	\cM(d \bq, d \bv) = \Vker(\bq, d \bv) \mu(d \bq).
  \end{align}
  Fix any $p \geq 1$, and for each $j=0,1,\ldots,p$ consider the
  following statements for the given measurable mappings
  $S_j: \spq \times \spv \to \spq \times \spv$ and
  $\alpha_j: \spq \times \spv \to [0,1]$:
 \begin{enumerate}[label={(H\arabic*)}]
  \item\label{P1} $S_j$ is an involution, i.e. $S_j \circ S_j = I$, for $j = 0, 1,\ldots,p$;
  \item\label{P2} \eqref{sum:alpha:1} holds and
  \begin{align}\label{eq:inv}
  	\sum_{j=0}^p \int_\spv \alpha_j( S_j(\bq, \bv)) S^*_j \cM (d \bq, d \bv) = \mu(d\bq);
  \end{align}
  \item\label{P3} \eqref{sum:alpha:1} holds and, for every
    $j=0,1,\ldots, p$,\footnote{Equivalently,
      $\int_{\spq \times \spv} \varphi(\bq, \bv) \alpha_j( S_j(\bq,
      \bv)) S^*_j\cM (d \bq, d \bv) = \int_{\spq \times \spv}
      \varphi(\bq, \bv) \alpha_j(\bq, \bv) \cM (d \bq, d \bv)$ for
      every bounded and measurable function
      $\varphi: \spq \times \spv \to \RR$.}
  \begin{align}
    \label{eq:det:bal:cond}
    \alpha_j( S_j(\bq, \bv)) S^*_j\cM (d \bq, d \bv)
    = \alpha_j(\bq, \bv) \cM (d \bq, d \bv);
  \end{align}
 \end{enumerate}
 where in \eqref{eq:inv} and \eqref{eq:det:bal:cond} $S_j^* \cM$
 denotes the pushforward of $\cM$ under $S_j$ (see
 \eqref{def:pushfwd} below in \cref{subsec:meas:th}).
 
 Then, under \ref{P1} and \ref{P2}, it follows that the corresponding
 Markov kernel
 $\Pker^{\alpha, S, \Vker}: \spq \times \Sigma_\spq \to [0,1]$ defined
 in \eqref{eq:multi:prop:abs} maintains $\mu$ as an invariant measure,
 i.e. $\mu \Pker^{\alpha,S,\Vker} = \mu$. Moreover, \ref{P3} implies
 \ref{P2}, and under \ref{P1} and \ref{P3} the Markov kernel
 $\Pker^{\alpha, S, \Vker}$ additionally satisfies detailed balance
 with respect to $\mu$, i.e.
 \begin{align}\label{det:bal}
 P^{\alpha, S, \Vker} (\bq,d\btq) \mu(d\bq)  = P^{\alpha, S, \Vker} (\btq,d\bq) \mu(d\btq).
 \end{align}
\end{Theorem}

\cref{alg:main} describes the general 
sampling procedure justified by
\cref{thm:main}.

\begin{algorithm}[!t]
	\caption{}
	\begin{algorithmic}[1]\label{alg:main}
          \State Select the algorithm parameters:
          \begin{itemize}
          \item[(i)] The proposal kernel
            $\Vker(\bq,d \bv)$.
          \item[(ii)] The mappings $S_j$, $\alpha_j$,
            $j=0,1,\ldots,p$, satisfying \ref{P1} and \ref{P2}, or
            \ref{P1} and \ref{P3}.
          \end{itemize}
          \State Choose $\bq^{(0)} \in \spq$.  
          \For{$k \geq 0$}
          \State Sample $\bv^{(k+1)} \sim \Vker(\bq^{(k)}, \cdot)$.
          \State Compute $S_j(\bq^{(k)}, \bv^{(k+1)})$, for $j = 0,1,\ldots,p$.
          \State Set $\bq^{(k+1)}$ by drawing from $(\Pi_1 
          S_0(\bq^{(k)}, \bv^{(k+1)}), \ldots, \Pi_1 S_p(\bq^{(k)}, \bv^{(k+1)}))$
          according to the probabilities 
          \par
          $(\alpha_0(\bq^{(k)}, \bv^{(k+1)}), \ldots,$ $ \alpha_p(\bq^{(k)}, \bv^{(k+1)}))$.
          \State $k \to k + 1$.
          \EndFor
	\end{algorithmic}
\end{algorithm}

\begin{Remark}
  To allow the algorithm the possibility of staying at the current
  state at any given step, one simply has to choose one of the
  involution mappings $S_j$, $j = 0,1, \ldots, p$, to be the identity.
\end{Remark}

In the next two corollaries, we present two possible
definitions of sets of acceptance probabilities $(\alpha_0,\ldots,$
$\alpha_p)$ satisfying property \ref{P3}, and hence \ref{P2}, in
\cref{thm:main}.  Specifically, the $\alpha_j$s defined in
\cref{cor:ar:Barker} are of Barker-type \cite{barker1965monte},
whereas the ones in \cref{cor:wedge:alpha} correspond to the
Metropolis-Hastings type \cite{metropolis1953equation,
  hastings1970monte}, cf. \eqref{def:alphaj} and
\eqref{def:alphaj:metr:1}-\eqref{def:alphaj:metr:2}, respectively. See
\cref{appx:cor:ar:Barker} and \cref{appx:cor:wedge:alpha} for the
proofs.

\begin{Corollary}\label{cor:ar:Barker}
  Let $(\spq, \Sigma_\spq)$ and $(\spv, \Sigma_\spv)$ be measurable
  spaces. Fix any $\mu \in \Pr(\spq)$, a Markov kernel
  $\Vker: \spq \times \Sigma_\spv \to [0,1]$, and let
  $\cM \in \Pr(\spq \times \spv)$ be given by
  $\cM(d\bq, d\bv) = \Vker(\bq, d \bv) \mu(d\bq)$. Further, let
  $S_j: \spq \times \spv \to \spq \times \spv$, $j=0,1,\ldots,p$, be
  measurable mappings satisfying the involution assumption \ref{P1} of
  \cref{thm:main}. Assume additionally that, for every
  $j=0,1,\ldots, p$,
\begin{align}\label{sum:cond}
  \sum_{k=0}^p (S_j \circ S_k)^*\cM(E)
  = \sum_{k=0}^p S_k^*\cM(E)
  \quad \mbox{ for all } E \in \Sigma_{\spq \times \spv}.
\end{align}
Also, let $\alpha_j: \spq \times \spv \to [0,1]$, $j = 0,\ldots,p$, be
any measurable functions such that \eqref{sum:alpha:1} holds and
\begin{align}\label{def:alphaj}
    \alpha_j(\bq, \bv) =
    \frac{d S_j^*\cM}{d(S_0^*\cM + \cdots + S_p^*\cM)}(\bq, \bv)
\end{align}
for $\left(\sum_{j=0}^p S_j^* \cM\right)$-a.e.
$(\bq,\bv) \in \spq \times \spv$.  Then, under this setting, it
follows that condition \ref{P3} of \cref{thm:main} is satisfied (and
consequently also \ref{P2}). Therefore, the associated Markov kernel
$\Pker^{\alpha, S, \Vker}: \spq \times \Sigma_\spq \to [0,1]$ given in
\eqref{eq:multi:prop:abs}, with $\alpha_j$ as defined in
\eqref{def:alphaj}, satisfies detailed balance with respect to $\mu$
and thus maintains $\mu$ as an invariant measure.
\end{Corollary}

\begin{Remark}\label{rmk:alphaj:ac}
  Under the additional assumption that $S_j^* \cM \ll \cM$ for
  $j = 0, \ldots, p$, \eqref{def:alphaj} may be written as
  \begin{align}\label{def:alphaj:ac}
    \alpha_j(\bq, \bv) =
    \frac{\frac{d S_j^*\cM}{d \cM}(\bq, \bv)}
    {\frac{dS_0^*\cM}{d\cM}(\bq, \bv) + \cdots+\frac{dS_p^*\cM}{d\cM}(\bq, \bv)},
    \quad \text{ for } j = 0, \ldots, p.
  \end{align}
  On the other hand, the condition $S_j^* \cM \ll \cM$ is not required
  for \cref{cor:ar:Barker}.
\end{Remark}

\begin{Corollary}\label{cor:wedge:alpha}
  Let $(\spq, \Sigma_\spq)$ and $(\spv, \Sigma_\spv)$ be measurable
  spaces. Fix any $\mu \in \Pr(\spq)$, a Markov kernel
  $\Vker: \spq \times \Sigma_\spv \to [0,1]$, and let
  $\cM \in \Pr(\spq \times \spv)$ be given by
  $\cM(d\bq, d\bv) = \Vker(\bq, d \bv) \mu(d\bq)$. Further, set
  $S_0 \coloneqq I$, and let
  $S_j: \spq \times \spv \to \spq \times \spv$, $j=1,\ldots,p$, be
  measurable mappings satisfying assumption \ref{P1} of
  \cref{thm:main} and such that $S_j^*\cM \ll \cM$. Consider a
  collection of weights $\overline{\alpha}_j \in \RR^+$,
  $j = 1, \ldots,p$, satisfying
  $\sum_{j=1}^p \overline{\alpha}_j \leq 1$. Moreover, for each
  $j=0,1,\ldots,p$ define mappings
  $\alpha_j: \spq \times \spv \to [0,1]$ given by
\begin{gather}
  \alpha_j(\bq,\bv)
       := \overline{\alpha}_j \left[ 1 \wedge \frac{dS_j^*\cM}{d \cM}(\bq,\bv)\right],
    \quad j = 1,\ldots,p, \label{def:alphaj:metr:1}\\
    \alpha_0(\bq,\bv) := 1 - \sum_{j=1}^p \alpha_j(\bq,\bv),
    \label{def:alphaj:metr:2}
\end{gather}
for all $(\bq,\bv) \in \spq \times \spv$. Then, conditions \ref{P2}
and \ref{P3} of \cref{thm:main} hold. Consequently, the associated
Markov kernel
$\Pker^{\alpha, S, \Vker}: \spq \times \Sigma_\spq \to [0,1]$ given in
\eqref{eq:multi:prop:abs}, with $\alpha_j$ as defined in
\eqref{def:alphaj:metr:1}-\eqref{def:alphaj:metr:2}, satisfies
detailed balance with respect to $\mu$ and thus maintains $\mu$ as an
invariant measure.
\end{Corollary}

\begin{Remark}
  In the setting of \cref{cor:wedge:alpha}, a natural choice of
  weights would be $\overline{\alpha}_j = 1/p$, for all
  $j=1,\ldots,p$.  Under this specification, equations reminiscent of
  \eqref{def:alphaj:metr:1} and \eqref{def:alphaj:metr:2} appear in
  \cite{calderhead2014general}, albeit with no formal justification.
\end{Remark}

\begin{Remark}
  We notice that \cref{cor:wedge:alpha} encompasses Theorem 2.1 of
  \cite{glatt2020accept} for the case $p = 1$.
\end{Remark}

\subsection{The case of conditionally independent
  proposals}
\label{sec:cond:ind:prop}

Here, we specialize to a particular subclass of kernels
falling under the wider umbrella of \cref{def:gen:MMH:ker}.  This
setting encompasses and generalizes algorithms found in
\cite{tjelmeland2004using}.  Meanwhile, it permits the derivation of
new Hilbert space type algorithms. See \cref{sec:rw:alg:fd} and
\cref{sec:mpCN} respectively below.

We specify the elements $(\alpha, S, \Vker)$ composing the kernel
$P^{\alpha, S, \Vker}$ in \eqref{eq:multi:prop:abs} of
\cref{def:gen:MMH:ker} as follows.  Let $(\spq, \Sigma_\spq)$ be a
measurable space, and fixing $p \geq 1$, we take $Y = X^p$ to be the
$p$-fold product of $X$ which we endow with the standard product
$\sigma$-algebra.  Fixing any two Markov kernels
$\Qker, \bQker: \spq \times \Sigma_\spq \to [0,1]$, we take
\begin{align}\label{def:Vker:multi:ind}
    \Vker(\bq_0, d \bv) = \Vker(\bq_0, d \bq_1, \ldots, d \bq_p)
      = \int_\spq \prod_{i=1}^p \Qker(\bq, d \bq_i) \bQker(\bq_0, d\bq),
    \quad \mbox{ for } \bv = (\bq_1,\ldots, \bq_p) \in \spq^p,
\end{align}
and then consider the flip involutions $S_j: \spq^{p+1} \to \spq^{p+1}$, $j=0,\ldots,p$, given by
\begin{align}\label{def:Sj:cond:ind}
    S_0 \coloneqq I, \quad \mbox{ and }
    \quad S_j(\bq_0, \bv) = S_j(\bq_0, \bq_1, \ldots, \bq_p)
    \coloneqq (\bq_j, \bq_1, \ldots, \bq_{j-1}, \bq_0, \bq_{j+1},
  \ldots, \bq_p),
\end{align}
for all $(\bq_0, \bv) \in \spq \times \spq^p$ and $j = 1,\ldots,p$. We refer to this class
of proposal kernels defined by \eqref{def:Vker:multi:ind},
\eqref{def:Sj:cond:ind} as having a \emph{conditionally independent structure}. Viewed algorithmically, one makes such a proposal by
drawing from $\bQker$ from around the current state and then using
this new point we generate a cloud of $p$ proposal points according to $\Qker$ in a
conditionally independent fashion. The main advantage of this proposal
structure is that under suitable assumptions on the kernels
$\Qker, \bQker$ it can lead to simplified and more computationally
efficient expressions for the acceptance probabilities $\alpha_j$,
$j=0,1, \ldots ,p$ as we illustrate in \cref{thm:cond:ind:mu:mu0} and
\cref{cor:simple:alpha} below (see also \cref{sec:rw:alg:fd}).
\cref{alg:cond:ind} makes the associated sampling procedure precise.

\begin{algorithm}[!t]
	\caption{}
	\begin{algorithmic}[1]\label{alg:cond:ind}
		\State Select the algorithm parameters:
		\begin{itemize}
			\item[(i)] The  proposal kernels $\Qker,\bQker: \spq
			\times \Sigma_\spq \to [0,1]$.
                      \item[(ii)] The mappings $\alpha_j$,
                        $j=0,1,\ldots,p$, satisfying \ref{P2}
                        (invariance only) or \ref{P3}
                        (reversibility). Here, $\cM, \Vker$, and
                        $S_j$, $j=0,1,\ldots,p$, are as in
                        \eqref{def:M}, \eqref{def:Vker:multi:ind}, and
                        \eqref{def:Sj:cond:ind}, respectively. Note that
                        $\alpha_j$ may always be specified as in \eqref{def:alphaj}
                        (\cref{thm:multi:ind:Tjmd})
                        and see also \eqref{def:alphaj:bb}, \eqref{def:alpha:simple:MH}.
		\end{itemize}
		\State Choose $\bq_0^{(0)} \in \spq$.  
		\For{$k \geq 0$}
		\State Sample $\bbq \sim \bQker(\bq_0^{(k)}, \cdot)$.
		\State Sample $\bq_j^{(k+1)} \sim \Qker(\bbq, \cdot)$ independently,
		for $j =1, \ldots,p$. Set $\bv^{(k+1)}
		\coloneqq (\bq_1^{(k+1)}, \ldots, \bq_p^{(k+1)})$.
		\State Set $\bq_0^{(k+1)}$ by drawing from
		$(\bq_0^{(k)},\bq_1^{(k+1)},\ldots,\bq_p^{(k+1)})$
		according to the probabilities
                \par
		$(\alpha_0(\bq_0^{(k)},\bv^{(k+1)}), \ldots,$ $
		\alpha_p(\bq_0^{(k)},\bv^{(k+1)}))$. 
		\State $k \to k + 1$.
		\EndFor
	\end{algorithmic}
\end{algorithm}

\begin{Remark}      
  The proposal formulation in \eqref{def:Vker:multi:ind} allows for
  the particular case where $\bq_1, \ldots, \bq_p$ are directly and
  independently drawn from a probability distribution
  $\Qker(\bq_0,\cdot)$, by simply choosing
  $\bQker(\bq_0, d \bq) \coloneqq \delta_{\bq_0} (d\bq)$.  Note
  however that the condition \eqref{Q:mu0:det:bal} below
  typically leads
  to more computationally tractable acceptance probabilities.  Compare
  for example \eqref{eq:non:TJ:cor:FD} and \eqref{eq:AR:TJ:FD}
  below in \cref{sec:rw:alg:fd}.  In any case
  if we take $\Qker(\bq_0,\cdot)$ to be independent of $\bq_0$,
  \cref{alg:cond:ind} yields a multiproposal version of the standard
  independence sampler algorithm.
\end{Remark}

The following result states that under the choices of $\Vker$ and $S$
given in \eqref{def:Vker:multi:ind} and \eqref{def:Sj:cond:ind}, and
for a given target measure $\mu$, condition \eqref{sum:cond} of
\cref{cor:ar:Barker} is guaranteed to hold. Thus, by supplementing
such $\Vker$ and $S$ with Barker-like acceptance probabilities
$\alpha$ as \eqref{def:alphaj} one obtains a Markov transition kernel
$\Pker^{\alpha, S, \Vker}$ as in \eqref{eq:multi:prop:abs} resulting
from \cref{alg:cond:ind} that satisfies detailed balance with respect
to $\mu$.

\begin{Theorem}\label{thm:multi:ind:Tjmd}
  Fix any $p > 0$, $\mu \in \Pr(\spq)$ and Markov kernels
  $\Qker, \bQker: \spq \times \Sigma_\spq \to [0,1]$. Define another
  Markov kernel $\Vker: \spq \times \Sigma_{\spq^p} \to [0,1]$ as in
  \eqref{def:Vker:multi:ind} and let $\cM \in \Pr(\spq^{p+1})$ be given
  as $\cM(d\bq_0, d \bv) = \Vker(\bq_0, d \bv) \mu(d\bq_0)$. Further,
  let $S_j: \spq^{p+1} \to \spq^{p+1}$ be the involution mappings
  defined in \eqref{def:Sj:cond:ind}.  Then, for every
  $j=0, 1, \ldots ,p$, it holds that
  \begin{align}\label{eq:Sj:Sk:M:A}
	\sum_{k=0}^p (S_j \circ S_k)^*\cM(E) = \sum_{k=0}^p S_k^*\cM(E)
    \quad \mbox{ for all } E \in \Sigma_{\spq^{p+1}}.
  \end{align}
  Therefore, the associated Markov kernel
  $\Pker^{\alpha, S, \Vker}: \spq \times \Sigma_\spq \to [0,1]$ given
  in \eqref{eq:multi:prop:abs}, with $\alpha_j$ as defined in
  \eqref{def:alphaj}, satisfies detailed balance with respect to
  $\mu$.
\end{Theorem}

\noindent The proof of \cref{thm:multi:ind:Tjmd} is provided below in
\cref{sec:thm:multi:ind:Tjmd}.

The next theorem considers a particular setting in
\cref{thm:multi:ind:Tjmd}. Specifically, it identifies suitable
conditions on the target measure $\mu$ and the Markov kernels
$\Qker, \bQker$ which imply that $S_j^*\cM$ is absolutely continuous 
with respect to $\cM$ and for which the Radon-Nikodym derivative 
$d S_j^*\cM / d \cM$ assumes an explicit expression depending on 
$\mu$ but which is independent of the kernels $\Qker, \bQker$. 
This in turn yields in \cref{cor:simple:alpha} simplified
expressions for the acceptance probabilities $\alpha_j$,
$j=0,1,\ldots,p$ from \eqref{def:alphaj} and \eqref{def:alphaj:metr:1}-\eqref{def:alphaj:metr:2} 
that are independent of $\Qker, \bQker$, a crucial property for effective applications. 

\begin{Theorem}\label{thm:cond:ind:mu:mu0}
  Fix any $p > 0$, $\mu \in \Pr(\spq)$, and Markov kernels
  $\Qker, \bQker: \spq \times \Sigma_\spq \to [0,1]$. Define another
  Markov kernel $\Vker: \spq \times \Sigma_{\spq^p} \to [0,1]$ as in
  \eqref{def:Vker:multi:ind}. Take $\cM \in \Pr(\spq^{p+1})$ given by
  $\cM(d\bq_0, d \bv) = \Vker(\bq_0, d \bv) \mu(d\bq_0)$, and let
  $S_j: \spq^{p+1} \to \spq^{p+1}$, $j=0,1,\ldots,p$, be the
  involution mappings defined in \eqref{def:Sj:cond:ind}.
	
  Suppose $\mu \ll \mu_0$ for some $\sigma$-finite measure $\mu_0$ on
  $\spq$, with $\frac{d\mu}{d\mu_0}(\bq) > 0$ for a.e. $\bq \in
  \spq$. Additionally, assume that $\Qker$ and $\bQker$ satisfy the
  following balance-type condition
  \begin{align}\label{Q:mu0:det:bal}
		\Qker(\bq, d \btq) \mu_0(d\bq) = \bQker(\btq, d \bq) \mu_0(d \btq).
  \end{align}
  Then, for every $j=0,1,\ldots,p$, $S_j^*\cM \ll \cM$ and
  \begin{align}\label{RN:SjM:M}
    \frac{d S_j^*\cM}{d \cM}(\bq_0, \bv) = \frac{d \mu}{d\mu_0}(\bq_j)
    \left( \frac{d\mu}{d\mu_0}(\bq_0) \right)^{-1}
    \quad \mbox{ for $\cM$-a.e. } (\bq_0, \bv) = (\bq_0, \bq_1, \ldots, \bq_p)  \in \spq \times \spq^p.
  \end{align}
\end{Theorem}

\noindent See \cref{sec:thm:cond:ind:mu:mu0} for the proof of
\cref{thm:cond:ind:mu:mu0}. The next corollary follows immediately 
by plugging \eqref{RN:SjM:M} into the expressions \eqref{def:alphaj:ac} 
and \eqref{def:alphaj:metr:1}.

\begin{Corollary}\label{cor:simple:alpha}
Under the assumptions of \cref{thm:cond:ind:mu:mu0}, it follows that by
defining for any $(\bq_0,\bv) = (\bq_0, \bq_1, \ldots, \bq_p)
\in \spq \times \spq^p$
\begin{align}\label{def:alphaj:bb}
\alpha_j(\bq_0,\bv)
= \frac{\frac{d \mu}{d\mu_0}(\bq_j)}{\sum_{k=0}^p \frac{d\mu}{d\mu_0}(\bq_k)},
\quad j = 0,1,\ldots,p,
\end{align}
or
\begin{align}\label{def:alpha:simple:MH}
\alpha_j(\bq_0,\bv) = \overline{\alpha}_j 
\left[ 1 \wedge \frac{\frac{d \mu}{d\mu_0}(\bq_j)}{\frac{d \mu}{d\mu_0}(\bq_0)} \right], 
  \,\, j = 1, \ldots, p; \,\, \alpha_0(\bq_0,\bv) = 1 - \sum_{j=1}^p \alpha_j(\bq_0,\bv),
\end{align}
with $\overline{\alpha}_j \in \RR^+$ such that
$\sum_{j=1}^p \overline{\alpha}_j \leq 1$, then the associated kernel
$\Pker^{\alpha, S, \Vker}: \spq \times \Sigma_\spq \to [0,1]$, given
as in \eqref{eq:multi:prop:abs}, satisfies detailed balance with
respect to $\mu$.
\end{Corollary}

\begin{Remark}\label{rmk:non:sym:prop}
  One may obtain kernels $\Qker, \bQker$ satisfying
  \eqref{Q:mu0:det:bal} by selecting any measurable
  $\dQ: \spq \times \spq \to \RR^+$ such that
  \begin{align*}
    \int_\spq \dQ ( \bq, \btq) \mu_0(d\btq) =
    1 = \int_\spq \dQ ( \btq, \bq) \mu_0(d\btq),
  \end{align*}
  for any $\bq \in \spq$ and then defining $\Qker, \bQker: \spq
        \times \Sigma_\spq \to [0,1]$ according to
  \begin{align}
    \Qker(\bq, d \btq) =  \dQ ( \bq, \btq) \mu_0(d\btq),
    \quad
    \bQker(\bq, d \btq) =  \dQ ( \btq, \bq) \mu_0(d\btq).
    \label{eq:sym:gen:cond}
  \end{align}
  Of course, in the case that $\dQ$ is symmetric, namely
  that $\dQ (\bq, \btq) = \dQ (\btq, \bq)$ for every $\bq,\btq \in \spq$,
  then $\Qker = \bQker$.
\end{Remark}

\section{Incorporating multiple jumps between proposals}
\label{sec:ext:Phase}

Whereas the previous section develops a rigorous framework for pMCMC
methods that make a single jump from the current state to one of
multiple proposals, \cite{calderhead2014general} presents an
algorithm that allows for multiple resamples from a generated
proposal set. This section develops a rigorous `extended phase space'
multiple proposal, multiple jump formalism. In particular, we show
that the setup considered in \cite{tjelmeland2004using,
  calderhead2014general} permits an involutive, abstract state space
generalization inclusive of HMC and Hilbert space settings previously
unaddressed.  Indeed our formulation, culminating in \cref{thm:tj} 
and \cref{alg:tj:gen}, justifies drawing multiple samples from a given proposal 
cloud in a wide variety of
contexts as we sketch below in \cref{sec:Ex:alg}.  Note that, from a practical
standpoint, making multiple jumps within a proposal set may be
beneficial, insofar as the computational burden is not much greater
than the cost imposed for a single jump: one must always evaluate
the target density at all proposal points regardless of the number of
jumps one intends to make.

\subsection{An augmented extended phase space formulation}

Just as for \cref{def:gen:MMH:ker}, we let $(\spq, \Sigma_\spq)$ and
$(\spv, \Sigma_\spv)$ be measurable spaces and fix $p \geq 1$,
the number of proposals at each step.  We take
\begin{align}
   \bz = (\bqv, k) = (\bq, \bv, k)  \in \ExtSp
   := \spq \times \spv \times \{0, \ldots, p\},
\end{align}
and, as above, we define $\Proj : \spq \times \spv \to \spq$ 
as $\Proj ( \bq, \bv) = \bq$, 
the projection onto the $\spq$ coordinate.

We define transition kernels on $\ExtSp$ as follows:
\begin{Definition}
\label{def:gen:MMH:ker:ext:phase}
For $j=0,1,\ldots,p$, we consider 
\begin{align}
\Vker_j: \spq \times \Sigma_\spv \to [0,1] 
\quad \text{ to be Markov kernels}
\label{eq:mar:ker:ex}
\end{align}
and take
\begin{align}
  S_j: \spq \times \spv \to \spq \times \spv \quad
  \text{ to be measurable mappings}.
\label{eq:Sj:invol:ex}
\end{align}
\begin{itemize}
\item[(i)] Let $\KR: \ExtSp \times \Sigma_\ExtSp \to [0,1]$ be the
  Markov kernel defined by
\begin{align}
   \KR(\bq, \bv, k, d \btq, d\btv, d\tk)
   &= S_k^*(\Vker_k( \Proj S_k( \bq, \bv), d \btv)   \delta_{\Proj S_k( \bq,\bv)} (d \btq))
     \delta_{k}(d\tk).
     \label{eq:R:Trans}
\end{align}
\item[(ii)]
Let us furthermore suppose that for $k,j = 0, \ldots, p$ we have
$\alpha_{k,j}: \spq \times \spv \to [0,1]$ such that
\begin{align}
  \sum_{j = 0}^p \alpha_{k,j}(\bq, \bv) = 1,
  \quad \text{ for all $k =0, \ldots, p$ and $(\bq, \bv) \in \spq \times \spv$.}
  \label{eq:ext:sum:ac}
\end{align}
We then define $\KA: \ExtSp \times \Sigma_\ExtSp \to [0,1]$ as the
Markov kernel given by
\begin{align}     
   \KA(\bq, \bv, k, d \btq, d\btv, d\tk)
     &=   \delta_{(\bq,\bv)} (d \btq,d\btv)
       \sum_{j = 0}^p \alpha_{k,j}(\bq,\bv) \delta_{j}(d\tk).
            \label{eq:A:Trans}
\end{align}
\item[(iii)] We denote the composition kernel
\begin{align}
  \KP_1(\bz, d \btz) = \KP(\bz, d \btz) := \KR \KA(\bz, d \btz)
  =\int_{\ExtSp} \KA(\bhz, d\btz) \KR(\bz, d \bhz)
              \label{eq:P:Trans}
\end{align}
and furthermore iteratively define, for any $n \geq 1$,
\begin{align}
  \KP_n(\bz, d \btz) := \KP_{n-1} \KA (\bz, d \btz)
     =\int_{\ExtSp} \KA(\bhz, d\btz) \KP_{n-1}(\bz, d \bhz)
                \label{eq:P:Trans:n}
\end{align}
\item[(iv)] Finally we consider the projection operator
  $\ProjES: \ExtSp \to \spq$ as
\begin{align}
  \ProjES( \bq, \bv, k) =\Proj S_k( \bq, \bv).
  \label{eq:proj:ext}
\end{align}
\end{itemize}
\end{Definition}

We thus specify the kernels $\KR$, $\KA$, $\KP_n$ and the projection
$\ProjES$ by the triple $(\Vker^e, S, \alpha)$ where
$\Vker^e = (\Vker_0, \ldots, \Vker_p)$, $S = (S_0, \ldots, S_p)$ and
$\alpha = (\alpha_{j,k})_{j, k = 0, \ldots, p}$.  For a given
$(\Vker^e, S, \alpha)$ and any $n \geq 1$, \eqref{eq:P:Trans:n} and
\eqref{eq:proj:ext} yields a sampling procedure by taking, for any
$m \geq 1$,
$\br^{(m)} \sim \ProjES^* [(\KP_{n})^j \KP_{l}](\bz, d\bq)$ where
$m = nj +l$ for the appropriate $j,l \geq 0$, and $(\KP_n)^j$ denotes
the $j$-fold composition of $\KP_n$.  \cref{alg:tj:gen} describes this procedure.

\begin{algorithm}[!t]
	\caption{}
	\begin{algorithmic}[1]\label{alg:tj:gen}
          \State Select the algorithmic parameters
          \begin{itemize}
          \item[(i)] $p \geq 1$ the number of elements generated in
            each proposal cloud.
          \item[(ii)] The proposal kernels $\Vker^e = ( \Vker_0, \ldots,
            \Vker_p)$ as in \eqref{eq:mar:ker:ex}.
          \item[(iii)] The mappings $S = (S_0, \ldots, S_p)$ as in \eqref{eq:Sj:invol:ex}
          \item[(iv)] The transition probabilities $\alpha = (\alpha_{k,j})_{j,k = 0, \ldots, p}$ as in
            \eqref{eq:ext:sum:ac}.
          \item[(v)] The $n \geq 1$ number of samples drawn per
            generated proposal cloud.
          \end{itemize}
	\State Choose an initial $(\bq^{(0)}, \bv^{(0)}) \in \spq \times\spv$ and $k^{(0)} \in \{0, \ldots, p\}$.  
	\State Set $\br^{(0)} := \Pi_1 S_{k^{(0)}}(\bq^{(0)}, \bv^{(0)})$.
	\For{$j \geq 0$}
    	\State Set $\bbq^{(j+1)} := \br^{(nj)}$.
		\State Sample $\bbv^{(j+1)} \sim \Vker_{k^{(j)}}(\bbq^{(j+1)}, d\bv)$.
		\State Set $(\bq^{(j+1)}, \bv^{(j+1)}) := S_{k^{(j)}}(\bbq^{(j+1)}, \bbv^{(j+1)})$.
        \State Set $k_{cur} := {k^{(j)}}$.
        \For{$l = 1, \dots, n$}
              \State Draw $k_{nxt} \in \{0, \ldots, p\}$ with the probabilities
              $(\alpha_{k_{cur},0}(\bq^{(j+1)}, \bv^{(j+1)}),
              \ldots, \alpha_{k_{cur},p}(\bq^{(j+1)}, \bv^{(j+1)}))$.
              \State Set $k_{cur} := k_{nxt}$.
              \State Set $\br^{(nj + l)} := \Proj  S_{k_{cur}}(\bq^{(j+1)}, \bv^{(j+1)})$.
              \State $l \to l + 1$.
              \EndFor
              \State Set $k^{(j+1)}  = k_{cur}$.
	  	   \State $j \to j + 1$.
		\EndFor
              \end{algorithmic}
\end{algorithm}

\begin{Remark}
\label{rmk:stat:limit}
For each $(\bq, \bv) \in \spq \times \spv$, let $A(\bq,\bv)$ be the
$(p+1)\times (p+1)$ real matrix with entries
$A_{k,j}(\bq,\bv) := \alpha_{k,j}(\bq,\bv)$, $k,j=0,\ldots,p$. It is
not difficult to show the following alternative expression for the
kernel $\KP_n$, $n \geq 1$:
\begin{align}\label{KPn:An}
	\KP_n(\bz, d \btz) &= \KP_n (\bq,\bv,k, d \btq, d\btv, d \tk) \\
	&= \int_\spv \delta_{S_k(\Pi_1 S_k (\bq, \bv), \hat{\bv})} (d\btq, d \btv)
   \sum_{j=0}^p A_{k,j}^n(S_k(\Pi_1 S_k (\bq,\bv), \hat{\bv}))\delta_j(d \tk)
   \Vker_k(\Pi_1 S_k(\bq,\bv), d \hat{\bv}),
	\notag
\end{align}
where $A_{k,j}^n(\bq,\bv)$, $k,j = 0,\ldots,p$, denote the entries of
the matrix $A^n(\bq,\bv)$, i.e., the $n$-fold composition of
$A(\bq,\bv)$. Now denote by $A^\infty(\bq,\bv)$ the matrix with entries
$A^\infty_{k,j}(\bq,\bv) = \alpha^\infty_j(\bq,\bv)$ for all
$k,j = 0,\ldots,p$, where $(\alpha^\infty_j(\bq,\bv))_{j=0,\ldots,p}$
is the stationary vector of the finite state Markov chain with
transition matrix $A(\bq,\bv)$.  Notice that one obtains by formally taking
the limit $n \to \infty$ in \eqref{KPn:An} the following Markov kernel
\begin{align*}
	\KP_{\infty}(\bz, d \btz) := 
	 \int_\spv \delta_{S_k(\Pi_1 S_k (\bq, \bv), \hat{\bv})} (d\btq, d \btv)
  \sum_{j=0}^p \alpha^\infty_j(S_k(\Pi_1 S_k (\bq,\bv), \hat{\bv}))
  \delta_j(d \tk)
  \Vker_k(\Pi_1 S_k(\bq,\bv), d \hat{\bv}),
\end{align*}
which corresponds to the one-step transition kernel of the algorithm
introduced in \cite{calderhead2014general}. Moreover, observe that
when $\alpha_{k,j}$ is $k$-independent then it follows from condition
\eqref{eq:ext:sum:ac} that $A^n = A$ for all $n$, so that
$A = A^\infty$.
\end{Remark}

\subsection{Main result}

We turn to the main result of this section, \cref{thm:tj}. Consider a given target measure $\mu \in \Pr(\spq)$. 
Having fixed $\Vker^e = (\Vker_0, \ldots, \Vker_p)$ and
$S = (S_0, \ldots, S_p)$ as in \eqref{eq:mar:ker:ex} and
\eqref{eq:Sj:invol:ex}, we denote
\begin{align}\label{def:M:j}
  \cM_j(d \bq, d \bv) = \Vker_j(\bq, d \bv) \mu(d \bq),
  \quad j = 0,\ldots,p.
\end{align}
We then consider the following extended phase space measure
\begin{align}
  \MN(d \btq, d\btv, d\tk)
  &= \sum_{j = 0}^p \frac{1}{p+1} S_j^*\cM_j(d \btq, d \btv)
    \delta_{j}(d\tk).
    \label{eq:ext:target:msr}
\end{align}
\cref{thm:tj} establishes conditions on the elements
$(\Vker^e, S, \alpha)$ such that $\MN$ is invariant under $\KA$, $\KR$
and hence under $\KP_n$ for any $n \geq 0$.  This result furthermore
asserts that $\ProjES \MN = \mu$ so that we indeed have conditions
justifying \cref{alg:tj:gen} as an unbiased sampling procedure.

\begin{Theorem}\label{thm:tj}
  Let $\Vker^e = (\Vker_0, \ldots, \Vker_p)$ be a collection of Markov
  kernels as in \eqref{eq:mar:ker:ex}, $S = (S_0, \ldots, S_p)$
  measurable mappings as in \eqref{eq:Sj:invol:ex} and $\alpha
  = (\alpha_{k,j})_{k, j = 0, \ldots, p}$ acceptance probabilities
  as in \eqref{eq:ext:sum:ac}. Let also $\mu \in \Pr(\spq)$ and define probability measures $\cM_j$, $j=0,\ldots,p$, on $\spq \times \spv$ as in \eqref{def:M:j}. We assume that
  \begin{enumerate}[label={(H\arabic*)}]
  \item\label{P1:ext} $S_j$ is an involution, namely we suppose that
    $S_j \circ S_j = I$, for $j = 0, 1,\ldots,p$; and
  \item\label{P2:ext} \eqref{eq:ext:sum:ac} holds and 
\begin{align}
	S_j^* \cM_j( d \bq, d \bv)  
	= \sum_{k =  0}^p\alpha_{k,j}(\bq,\bv) S_k^*  \cM_k( d \bq, d\bv),
  \quad \text{ for each } j = 0, \ldots, p.
  \label{eq:ext:bal:ac}
\end{align}
 \end{enumerate}
 Under these conditions, define the Markov kernels $\KR$, $\KA$ on
 $\ExtSp$ as in \eqref{eq:R:Trans}, \eqref{eq:A:Trans}, respectively.
 Take $\KP_n$ as in \eqref{eq:P:Trans}, \eqref{eq:P:Trans:n} for any
 $n \geq 1$. Then:
\begin{itemize}
\item[(i)] $\MN$ is invariant under  $\KR$;
\item[(ii)] $\MN$ is invariant under $\KA$;
\item[(iii)] $\MN$ is invariant under $\KP_n$ for any $n \geq 1$;
\item[(iv)] $\ProjES^* \MN = \mu$, and therefore $\ProjES^* (\MN \, \KP_n)  = \mu$
for any $n \geq 1$.
\end{itemize}
\end{Theorem}
\noindent The proof of \cref{thm:tj} is found below in
\cref{sec:thm:tj}.

\begin{Remark}
  Under \eqref{eq:ext:sum:ac}, the condition on the acceptance
  probabilities \eqref{eq:ext:bal:ac} is implied by the following
  slightly stronger condition
  \begin{align}
	\alpha_{k,j}(\bq, \bv) S_k^* \cM_k( d \bq, d\bv)
    =\alpha_{j,k}(\bq, \bv) S_j^* \cM_j( d \bq, d\bv),
    \quad \text{ for every } k,j = 0, \ldots, p.
	  \label{eq:ext:d:bal:ac}
  \end{align}
\end{Remark}

Next, we identify two examples of sets of acceptance probabilities
$\alpha_{k,j}$, $k,j = 0,\ldots,p$, for which condition \ref{P2:ext}
of \cref{thm:tj} holds.  In fact, both cases satisfy the stronger
condition \eqref{eq:ext:d:bal:ac}. The proof follows similarly as in
\cref{cor:ar:Barker} and \cref{cor:wedge:alpha}, so we omit the
details.

\begin{Corollary}\label{cor:alphakj:res}
  Take $\mu$, $\Vker^e = (\Vker_0, \ldots, \Vker_p)$,
  $S = (S_0, \ldots, S_p)$, and $\cM_j$, $j=0,\ldots,p$, as in
  \cref{thm:tj}. Consider the following definitions:
\begin{enumerate}
\item[(i)]\label{i:alpha} Let
  $\alpha_{k,j}: \spq \times \spv \to [0,1]$, $k, j = 0, \ldots, p$,
  be any measurable mappings such that \eqref{eq:ext:sum:ac} holds and
  \begin{align}\label{def:alphaj:mult}
	\alpha_{k,j}(\bq, \bv) =
	\frac{d S_j^*\cM_j}{d (S_0^*\cM_0 + \cdots + S_p^*\cM_p)}(\bq, \bv),
  \end{align}
  for $(S_0^*\cM_0 + \ldots + S_p^* \cM_p)$-a.e.
  $(\bq,\bv) \in \spq \times \spv$, and for every $k=0,\ldots,p$.
\item[(ii)]\label{ii:alpha} Assume $S_j^*\cM_j \ll S_k^* \cM_k$ for
  all $k \neq j$. Take $\overline{\alpha}_{k,j} \in [0,1]$,
  $k,j = 0,\ldots,p$, such that
  $\sum_{j=0}^p \overline{\alpha}_{k,j} \leq 1$ for all
  $k = 0,\ldots, p$. Then, for each $k,j = 0,\ldots,p$ and
  $(\bq, \bv) \in \spq\times \spv$, define
  \begin{align}\label{def:alphakj:MH}
    \alpha_{k,j}(\bq, \bv) = 
		\begin{cases}
                  \overline{\alpha}_{k,j}
                  \left[ 1 \wedge \frac{d S_j^* \cM_j}{d S_k^* \cM_k}(\bq, \bv) \right]
                  &\mbox{ if } j \neq k, \\
                  1 - \sum_{\stackrel{l=0}{l \neq k}}^p
                  \alpha_{k,l}(\bq, \bv) &\mbox{ if } j = k .
		\end{cases}
  \end{align}
\end{enumerate}
Then, under the conditions in $(i)$ and $(ii)$, it follows that both
definitions \eqref{def:alphaj:mult} and \eqref{def:alphakj:MH} satisfy
properties \eqref{eq:ext:sum:ac} and \eqref{eq:ext:d:bal:ac}, and
consequently also assumption \ref{P2:ext} of \cref{thm:tj}.
\end{Corollary}

\begin{Remark}
  Note that any collection of $k$-independent $\alpha_{k,j}$s which
  satisfy condition \eqref{eq:ext:bal:ac} must coincide
  $(S_0^*\cM_0 + \ldots + S_p^* \cM_p)$-a.e. with the Barker-like
  expression in \eqref{def:alphaj:mult}.
\end{Remark}

\begin{Remark}\label{rmk:simple:alpha}
  Similarly as in \cref{cor:simple:alpha}, we obtain that under an
  analogous conditionally independent framework from
  \cref{thm:cond:ind:mu:mu0} the definitions in
  \eqref{def:alphaj:mult} and \eqref{def:alphakj:MH} assume a
  simplified expression. Indeed, take $\spv = \spq^p$ and suppose
  $\mu \ll \mu_0$ for some $\sigma$-finite measure $\mu_0$ on $\spq$,
  with $\frac{d\mu}{d\mu_0}(\bq) > 0$ for a.e. $\bq \in \spq$. Assume
  that $\Vker_0 = \ldots = \Vker_p =: \Vker$, with $\Vker$ as in
  \eqref{def:Vker:multi:ind} for fixed Markov kernels $\Qker, \bQker$
  satisfying \eqref{Q:mu0:det:bal}, and let
  $\cM(d\bq,d\bv) = \Vker(\bq,d\bv) \mu(d\bq)$. Moreover, take $S_j$,
  $j=0,\ldots,p$ to be the flip involutions defined in
  \eqref{def:Sj:cond:ind}. It thus follows from
  \cref{thm:cond:ind:mu:mu0} that $S_j^* \cM \ll \cM$ for
  $j=0,\ldots,p$ and \eqref{RN:SjM:M} holds, so that
  \eqref{def:alphaj:mult} and \eqref{def:alphakj:MH} reduce for any
  $(\bq_0,\bv) = (\bq_0, \bq_1,\ldots, \bq_p) \in \spq \times \spq^p$
  and $k,j = 0,\ldots, p$ to
\begin{align}\label{simpl:Bk}
	\alpha_{k,j}(\bq_0,\bv)
	= \frac{\frac{d \mu}{d\mu_0}(\bq_j)}{\sum_{l=0}^p \frac{d\mu}{d\mu_0}(\bq_l)},
\end{align}
and
\begin{align}\label{simpl:MH}
	\alpha_{k,j}(\bq_0, \bv) = 
	\begin{cases}
          \overline{\alpha}_{k,j}
          \left[ 1 \wedge \frac{\frac{d \mu}{d\mu_0}(\bq_j)}{\frac{d \mu}{d\mu_0}(\bq_k)} \right]
         &\mbox{ if } j \neq k, \\
        1 - \sum_{\stackrel{l=0}{l \neq k}}^p \alpha_{k,l}(\bq, \bv) &\mbox{ if } j = k,
	\end{cases}
\end{align}
respectively.
\end{Remark}

\begin{Remark}\label{rmk:overlap}
  One might expect that, by setting the number of jumps $n = 1$,
  \cref{alg:tj:gen} would essentially reduce to \cref{alg:main}. In
  fact, this seems to not be the case in general, owing crucially to
  step 7 in \cref{alg:tj:gen}. See \cref{sec:HMC},
  \cref{rmk:diff:multi:res:hmc}, below for a specific example of this
  non-equivalence.

  Nevertheless, there is a particular case where a
  suitable relationship between these two algorithms can be
  established. Indeed, assume the same setting from
  \cref{rmk:simple:alpha}, and let also $\alpha_j = \alpha_{k,j}$,
  $k,j=0,\ldots,p$, be the acceptance probabilities given as in
  \eqref{simpl:Bk}. Then, denoting $\Pker = \Pker^{\alpha,S,\Vker}$
  the corresponding kernel as defined in \eqref{eq:multi:prop:abs}, it
  is not difficult to show that
\begin{align}\label{eq:P:KP}
  \Pker^m (\Pi_1 S_k(\bq,\bv), d \btq)
  = \ProjES^* (\KP_n)^m(\bq, \bv, k, \cdot)(d\btq)
  \quad \mbox{ for all } (\bq, \bv) \in \spq \times \spq^p,
  \,\,\mbox{ and } m,n \in \NN,
\end{align}
so that, setting in particular $n=1$ and $k=0$, we have
$\Pker^m (\bq, d \btq) = \ProjES^* \KP^m (\bq, \bv, 0, \cdot)(d\btq)$
for all $(\bq,\bv) \in \spq \times \spq^p$ and $m \in \NN$. To show
\eqref{eq:P:KP}, one first crucially notices that due to
$\alpha_{k,j}$ given in \eqref{simpl:Bk} being $k$-independent and
also property \eqref{eq:ext:sum:ac} then it follows that the matrix
$A(\bq,\bv)$ with entries $\alpha_{k,j}(\bq,\bv)$, $k,j=0,\ldots,p$,
satisfies $A^n(\bq,\bv) = A(\bq,\bv)$ for all $n \in \NN$. Secondly,
from the definitions of the flip involutions $S_j$, $j=0,\ldots,p$, in
\eqref{def:Sj:cond:ind}, and $\alpha_j = \alpha_{k,j}$ given in
\eqref{simpl:Bk}, it follows immediately that, for any
$k,j = 0,\ldots,p$ and
$(\bq_0,\bv) = (\bq_0, \bq_1,\ldots,\bq_p) \in \spq \times \spq^p$,
$\alpha_j(S_k(\bq_0,\bv))$ is equal to: $\alpha_k(\bq_0,\bv)$ if
$j=0$; $\alpha_0(\bq_0,\bv)$ if $j=k$; and $\alpha_j(\bq_0,\bv)$ for
$j \in \{1,\ldots,p\}$ with $j \neq k$.  In essence, this nullifies
the effect of step 7 in \cref{alg:tj:gen}, and ultimately implies that
the samples $\{\bq^{(i)}\}_{i \in \NN}$ generated by \cref{alg:main}
are equivalent to the chain
$\{\Pi_1 S_{k^{(i)}}(\bq_0^{(i)},\bv^{(i)})\}_{i \in \NN}$ derived
from the samples $\{(\bq_0^{(i)},\bv^{(i)},k^{(i)})\}_{i \in \NN}$
generated by \cref{alg:tj:gen}.
\end{Remark}

\section{Applications for algorithm design}
\label{sec:Ex:alg}

This section leverages our abstract formalisms developed in
\cref{sec:framework}, \cref{sec:ext:Phase} in service of the design
and rigorous analysis of some concrete sampling algorithms. In
\cref{sec:rw:alg:fd} we provide a systematic treatment of
multiproposal methods for continuous probability distributions on
$\RR^N$.  This treatment generalizes previously observed finite
dimensional methods from e.g. \cite{tjelmeland2004using} to include
non-symmetric proposal kernels.  Next in \cref{sec:mpCN} we derive a
novel `Hilbert-space' method which we christen the multiproposal pCN
(mpCN) sampler in \cref{alg:mpCN}, \cref{alg:mpCN:TJ}.  As already
previewed above in the introduction, mpCN extends the preconditioned
Crank-Nicolson algorithm designed for infinite-dimensional target
measures which are absolutely continuous with respect to a Gaussian
base measure \cite{beskos2008mcmc, cotter2013mcmc} to a multiproposal
setting.  Elsewhere in \cref{sec:HMC} we address applications for
Hamiltonian type sampling methods while in \cref{sec:simplicial} we
consider simplicial methods developed recently in
\cite{holbrook2023generating}.

\subsection{Finite-dimensional multiproposal algorithms}
\label{sec:rw:alg:fd}

In this subsection, we consider the particular case of
finite-dimensional, continuously distributed measures in the
algorithms presented in \cref{sec:framework} and \cref{sec:ext:Phase}
above. We then rewrite the formulas for the acceptance probabilities
previously introduced in terms of the associated probability
densities, thus providing more directly applicable expressions. In all
such formulas presented below, namely \eqref{def:alphaj:Bk:fd},
\eqref{def:alphaj:MH:fd}, \eqref{eq:gen:conv:AR:FD},
\eqref{eq:AR:TJ:FD},
\eqref{def:alphakj:res:Bk}, \eqref{def:alphakj:res:MH}, and
\eqref{simpl:MH:fd}, the acceptance probability is defined by the
given expression at every point where the denominator is strictly
positive, and otherwise the probability is assumed to be zero.

Take $\spq = \RR^\dimx$ and $\spv = \RR^\dimy$ for some
$\dimx, \dimy > 0$, endowed with their corresponding Borel
$\sigma$-algebras. Consider a target distribution
\begin{align}
  \label{eq:fd:cont:target}
\mu(d\bq) = \dmu(\bq) d \bq \quad \text{ for some density function }
\dmu: \RR^\dimx \to \RR^+.
\end{align}
Following the framework from
\cref{sec:Rev:Criteria}, take a Markov kernel
$\Vker(\bq, d\bv) = \dV (\bq, \bv) d \bv$ for some measurable function
$\dV: \RR^{\dimx + \dimy} \to \RR^+$ with
$\int_{\RR^\dimy} \dV (\bq, \bv) d\bv =1$ for all $\bq \in
\RR^\dimx$. It follows that
\begin{align*}
\cM(d\bq, d \bv) = \Vker(\bq, d \bv) \mu(d\bq)
= \dM(\bq,\bv) \, d \bq \, d \bv, \mbox{ with } \dM(\bq, \bv) = \dV (\bq, \bv) \dmu(\bq).
\end{align*}
Then, given any $C^1$ involution mappings
$S_j: \RR^{\dimx + \dimy} \to \RR^{\dimx + \dimy}$, $j=0,1\ldots,p$,
it follows from \eqref{pfwd:diff:fd} that
\begin{align*}
S_j^*\cM(d\bq, d \bv)
= \dM(S_j(\bq,\bv)) |\det \nabla S_j(\bq,\bv)| d \bq d \bv,
\quad j = 0,1,\ldots,p.
\end{align*}

In this situation \cref{cor:ar:Barker} and \cref{cor:wedge:alpha}
yield the following general formulations.  In the case of
\cref{cor:ar:Barker} the condition \eqref{sum:cond} translates to
\begin{align*}
\sum_{k=0}^p \dM(S_k \circ S_j(\bq,\bv)) |\det \nabla (S_k \circ S_j)(\bq,\bv)| d \bq d \bv
  = \sum_{k=0}^p \dM(S_k(\bq,\bv)) |\det \nabla S_k(\bq,\bv)| \, d \bq \, d \bv,
\end{align*}
for all $j=0,\ldots,p$.  In this circumstance we obtain a reversible
sampling scheme from \cref{alg:main} by supplementing the input
parameters $\Vker$, $(S_0,\ldots,S_p)$ with acceptance probabilities
$\alpha_j$, $j=0,1,\ldots,p$, given as in \eqref{def:alphaj}, which
according to \eqref{RN:nu1:nu2} can be written here as
\begin{align}\label{def:alphaj:Bk:fd}
	\alpha_j(\bq, \bv)
	= \frac{ \dM(S_j(\bq,\bv)) |\det \nabla S_j(\bq,\bv)|}{\sum_{l=0}^p \dM(S_l(\bq,\bv)) |\det \nabla S_l(\bq,\bv)| },
  \quad (\bq, \bv) \in \RR^\dimx \times \RR^\dimy.
\end{align}
Alternatively, in the setting of \cref{cor:wedge:alpha} we assume that
$S_0 = I$ and consider the $\alpha_j$s in
\eqref{def:alphaj:metr:1}-\eqref{def:alphaj:metr:2}, given here by
\begin{align}\label{def:alphaj:MH:fd}
  \alpha_j(\bq,\bv) = \overline{\alpha}_j
  \left[ 1 \wedge \frac{\dM(S_j(\bq,\bv)) |\det \nabla S_j(\bq,\bv)|}{\dM(\bq,\bv)} \right], \,\, j =1, \ldots,p \,; \quad  
	\alpha_0(\bq,\bv) := 1 - \sum_{j=1}^p \alpha_j(\bq,\bv)
 \end{align}
 for $(\bq, \bv) \in \RR^\dimx \times \RR^\dimy$, where we recall that
 $\overline{\alpha}_j \in [0,1]$, $j=1,\ldots,p$, are any user defined weights
 satisfying $\sum_{j=1}^p\overline{\alpha}_j \leq 1$.

 Let us now turn to the `conditionally independent' setting of
 \cref{sec:cond:ind:prop}.  We will provide the details for the case
 of Barker type acceptance probabilities, \`a la
 \eqref{def:alphaj:Bk:fd}, which we anticipate as being the most
 relevant in this particular setting.  The reader will find the
 analogous reduction from \eqref{def:alphaj:MH:fd} for the
 Metropolis-Hastings case to be direct, following from the same considerations.

 In this conditionally independent situation we take
 $\spv = \RR^{p \dimx}$ and consider Markov kernels of the form
\begin{align}
  \label{eq:qker:fd:cont:form}
  \Qker(\bq, d \btq) = \dQ(\bq, \btq) d \bq d \btq \quad \text{ and } \quad
  \bQker(\bq, d \btq) = \dbQ(\bq, \btq) d \bq d \btq.
\end{align}
Here, $\dQ: \RR^{2\dimx} \to \RR^+$ and $\dbQ: \RR^{2\dimx} \to \RR^+$
are measurable functions such that
\begin{align}
  \label{eq:qker:fd:cont:form:int}
  \int_{\RR^\dimx} \dQ(\bq, \btq) d \btq = 1
  = \int_{\RR^\dimx}\dbQ(\bq, \btq) d \btq
  \text{ for every } \bq \in \RR^\dimx. 
\end{align}
In this
setting, the Markov kernel $\Vker$ from \eqref{def:Vker:multi:ind} can
be written as
\begin{align}\label{def:Vker:cond:ind:fd}
\Vker(\bq_0, d \bv) = \dV(\bq_0, \bv) d \bv, \quad \mbox{ with } \dV(\bq_0, \bv) = \int_{\RR^\dimx} \prod_{i=1}^p \dQ(\bq, \bq_i) \dbQ(\bq_0, \bq) d \bq
\end{align}
for all $\bq_0 \in \RR^\dimx$ and
$\bv = (\bq_1, \ldots, \bq_p) \in \RR^{p \dimx}$. We notice that the
assumed conditions on $\dQ, \dbQ$ imply that
$\int_{\RR^{\dimx p}} \dV(\bq_0,\bv) d \bv = 1$ for all
$\bq_0 \in \RR^\dimx$, so that $\Vker$ indeed defines a Markov kernel.
We also fix $S_j: \RR^{(p+1) \dimx} \to \RR^{(p+1) \dimx}$,
$j=0,1,\ldots, p$, to be the flip involutions defined in
\eqref{def:Sj:cond:ind}. Since these involutions are also linear, it
is not difficult to see that each $S_j$ is volume-preserving,
i.e. $|\det \nabla S_j(\bq, \bv)| = 1$ for all
$(\bq, \bv) \in \RR^{(p+1) \dimx}$.

We thus obtain the following particular expressions for the acceptance
probabilities $\alpha_j$, $j = 0,1,\ldots,p$, from
\eqref{def:alphaj:Bk:fd} in this setting,
which when input into \cref{alg:cond:ind} yield a reversible sampling
procedure:
\begin{align}
	\alpha_j(\bq_0, \bv) = \frac{ \dmu(\bq_j)
	\int_{\RR^\dimx} \prod_{\stackrel{i=0}{i\neq j}}^p \dQ(\bq, \bq_i)
	\dbQ(\bq_j, \bq) d \bq}{  \sum_{l=0}^p  \dmu(\bq_l)
	\int_{\RR^\dimx} \prod_{\stackrel{i=0}{i\neq l}}^p \dQ(\bq, \bq_i)
	\dbQ(\bq_l, \bq) d \bq},
\label{eq:gen:conv:AR:FD}
\end{align}
for
$(\bq_0, \bv) = (\bq_0, \bq_1,\ldots,\bq_p) \in \RR^\dimx \times
\RR^{p \dimx}$. Note that we can also consider the particular case
where $\bQker(\bq, \cdot) = \delta_\bq(\cdot)$. Of course, in this
case there is no density $\dbQ$ with respect to Lebesgue measure as
$\delta_\bq$ is not continuously distributed. Nevertheless, it follows
immediately from \eqref{def:Vker:multi:ind} that
$\Vker(\bq_0, d \bv) = \prod_{i=1}^p \dQ(\bq_0, \bq_i) d \bv$ for
$\bv = (\bq_1, \ldots, \bq_p)$, so that \eqref{def:alphaj:Bk:fd} now reduces 
to
\begin{align}
	\alpha_j(\bq_0, \bv) = \frac{\dmu(\bq_j)
	\prod_{\stackrel{i=0}{i\neq j}}^p \dQ(\bq_j, \bq_i)}{
	\sum_{k=0}^p \dmu(\bq_k) \prod_{\stackrel{i=0}{i\neq k}}^p
	\dQ(\bq_k, \bq_i)},
	\label{eq:non:TJ:cor:FD}
\end{align}
again for any $(\bq_0, \bv) = (\bq_0, \bq_1,\ldots,\bq_p)  \in \RR^\dimx \times
\RR^{p \dimx}$.

The acceptance probabilities \eqref{eq:gen:conv:AR:FD} and
\eqref{eq:non:TJ:cor:FD} are expensive or intractable to compute
in many situations which highlights the salience of
\cref{thm:cond:ind:mu:mu0} here.
Within this setting of \cref{thm:cond:ind:mu:mu0}, let us assume
for simplicity that $\mu_0(d\bq)$ is the Lebesgue measure $d\bq$, so
that $d\mu /d\mu_0(\bq) = \dmu(\bq)$ for a.e. $\bq \in \RR^\dimx$. Let
us also assume that $\dmu(\bq) > 0$ for a.e. $\bq \in \RR^\dimx$.
Moreover, we now suppose that the densities $\dQ$, $\dbQ$ associated to
$\Qker$, $\bQker$, respectively, satisfy \eqref{Q:mu0:det:bal}, or
equivalently in this setting
\begin{align}\label{cond:dQ:dbQ}
\int_A \int_B \dQ(\bq, \btq) \, d \btq \, d\bq
  = \int_B \int_A \dbQ(\btq, \bq) \, d \bq \, d\btq
  \quad \mbox{ for every Borel sets } A, B \subset \RR^\dimx.
\end{align}
Note that this condition, \eqref{cond:dQ:dbQ}, is satisfied if we find any $\dQ$ such that
\begin{align}\label{cond:dQ:dbQ:SC}
  \int_{\RR^\dimx} \dQ(\bq, \btq) d \btq =
  1 = \int_{\RR^\dimx} \dQ (\btq, \bq) d \btq
  \quad \text{ for any }\bq \in
  \RR^\dimx,
\end{align}
and then we set $\dbQ( \btq, \bq) := \dQ(\bq, \btq)$.  In particular, one may
select $\dQ (\bq, \btq) = r (\bq- \btq)$ for any probability density
$r: \RR^\dimx \to \RR^+$.  In this case, the simplified acceptance
probabilities $\alpha_j$, $j=0,1,\ldots,p$, from \eqref{def:alphaj:bb} can be written as
\begin{align}
\alpha_j(\bq_0, \bv) = \frac{\dmu(\bq_j)}{\sum_{l=0}^p
	\dmu (\bq_l)},
\label{eq:AR:TJ:FD}
\end{align}
for $(\bq_0, \bv) = (\bq_0, \bq_1, \ldots,\bq_p) \in \RR^{(p+1) \dimx}$.
Notice that in the special case $\dbQ( \btq, \bq) := \dQ(\bq, \btq)$, and under \eqref{cond:dQ:dbQ:SC}, the formula
\eqref{eq:AR:TJ:FD} is immediately obtained from 
\eqref{eq:gen:conv:AR:FD}.

We highlight this particular case of interest arising out of \cref{thm:cond:ind:mu:mu0}
from \eqref{cond:dQ:dbQ} and \eqref{eq:AR:TJ:FD} as \cref{alg:TJ:cor:FD}.  This
sampler is the basis for the numerical experiments carried out below
in \cref{sec:ppMCMC}.

\begin{algorithm}[!t]
	\caption{}
	\begin{algorithmic}[1]\label{alg:TJ:cor:FD}
		\State Select the algorithmic parameters:
		\begin{itemize}
                \item[(i)] Any measurable function
                  $\dbQ, \dQ: \RR^{2\dimx} \to \RR^+$ maintaining
                  \eqref{cond:dQ:dbQ} which we use to define the
                  proposal mechanism. In particular, one may consider any
                  probability density $r: \RR^\dimx \to \RR^+$ and set
                  $\dbQ (\bq, \btq) = r (\bq- \btq), \dQ (\bq, \btq) = r (\btq- \bq)$.
                \item[(ii)] the number of proposals $p \geq 1$.
		\end{itemize}
		\State Choose $\bq^{(0)}_0 \in \RR^\dimx$.  \For{$k
                  \geq 0$} \State Sample
                $\bar{\bq}^{(k)} \sim \dbQ (\bq^{(k)}_0, \btq) d\btq$
                \State Sample
                $\bq_j^{(k+1)} \sim \dQ (\bar{\bq}^{(k)}, \btq,) d\btq
                $, for $j =1, \ldots,p$. Set
                $\bv^{(k+1)}_0 \coloneqq (\bq_1^{(k+1)}, \ldots,
                \bq_p^{(k+1)})$.  \State Set $\bq^{(k+1)}_0$ by
                drawing from
                $(\bq^{(k)}_0,\bq_1^{(k+1)},\ldots,\bq_p^{(k+1)})$
                according to the probabilities
                \par
		$(\alpha_0(\bq^{(k)}_0,\bv^{(k+1)}), \ldots,$ $ \alpha_p(\bq^{(k)}_0,\bv^{(k+1)}))$ as defined in \eqref{eq:AR:TJ:FD}.
		\State $k \to k + 1$.
		\EndFor
	\end{algorithmic}
\end{algorithm}

\begin{Remark}
  \cref{alg:TJ:cor:FD} includes as a special case the proposal structure 
  considered in \cite[Section 3]{tjelmeland2004using}.  Specifically 
  the method in \cite{tjelmeland2004using} corresponds to the special case
  $\Qker = \bQker$, so that $\dQ = \dbQ$, and where $\dQ(\bq,\cdot)$
  is taken as the density of a multivariate Gaussian centered at
  $\bq$. Notice that since such an $\dQ = \dQ(\bq,\btq)$ is symmetric in
  the variables $\bq, \btq$ it is clear that \eqref{cond:dQ:dbQ:SC} holds
  trivially in this case.  Aside from \cref{thm:cond:ind:mu:mu0} being given in the
  broader context of general state spaces, yielding for example mpCN in \cref{alg:mpCN}
  below, it also shows that the
  simplified expression of acceptance probabilities in
  \eqref{eq:AR:TJ:FD} can be obtained from arbitrary, even
  \emph{non-symmetric} proposal densities $\dQ(\bq,\btq) := r( \bq - \btq)$, 
  by choosing
  $\dbQ(\bq, \btq) \coloneqq \dQ(\btq, \bq)$ as in
  \cref{alg:TJ:cor:FD}.
\end{Remark}

Finally, let us specialize the setting of \cref{sec:ext:Phase} to the case of finite-dimensional continuously distributed measures. We start by considering again the spaces $\spq = \RR^\dimx$, $\spv = \RR^\dimy$, for $\dimx, \dimy > 0$, a target distribution $\mu(d\bq) = \dmu(\bq) d\bq$, and generic involution mappings $S_j: \RR^{\dimx + \dimy} \to \RR^{\dimx + \dimy}$, $j=0,\ldots,p$. Then, consider Markov proposal kernels $\Vker_j(\bq, d \bv) = \dV_j(\bq,\bv) \, d \bv$, $j=0,\ldots,p$. Here, each $\dV_j: \RR^{\dimx + \dimy} \to \RR^+$ is a measurable function satisfying $\int_{\RR^\dimy} \dV_j(\bq, \bv) d\bv = 1$ for all $\bq \in \RR^\dimx$. We define
\begin{align*}
	\cM_j(d\bq, d \bv) = \Vker_j(\bq, d \bv) \mu(d\bq) = \dM_j (\bq, \bv) \, d\bq \, d \bv, \quad \mbox{ with } \dM_j (\bq, \bv) \coloneqq \dV_j(\bq, \bv) \dmu(\bq), \,\, j = 0,\ldots,p.
\end{align*}
We thus obtain from \eqref{def:alphaj:mult} and \eqref{def:alphakj:MH} the following examples of acceptance probabilities that yield an unbiased sampling scheme according to \cref{alg:tj:gen}. Namely, for all $k,j = 0,\ldots,p$:
\begin{align}\label{def:alphakj:res:Bk}
	\alpha_{k,j}(\bq, \bv) =  \frac{ \dM_j (S_j(\bq,\bv)) |\det \nabla S_j(\bq,\bv)|  }{\sum_{l=0}^p \dM_l (S_l(\bq,\bv)) |\det \nabla S_l(\bq,\bv)| },
\end{align}
and
\begin{align}\label{def:alphakj:res:MH}	
	\alpha_{k,j} (\bq,\bv) = 
	\begin{cases}
	\overline{\alpha}_{k,j} \left[ 1 \wedge \frac{\dM_j(S_j(\bq,\bv)) |\det \nabla S_j(\bq,\bv)|}{\dM_k (S_k(\bq,\bv)) |\det \nabla S_k(\bq,\bv)|} \right] &\mbox{ if } j \neq k, \\
	1 - \sum_{\stackrel{l=0}{l \neq k}}^p \alpha_{k,l}(\bq, \bv) &\mbox{ if } j = k 
	\end{cases}
\end{align}
for $(\bq, \bv) \in \RR^\dimx \times \RR^\dimy$, and where we recall that $\overline{\alpha}_{k,j} \in [0,1]$, $k,j = 0,\ldots,p$, are specified weights such that $\sum_{j=0}^p \overline{\alpha}_{k,j} \leq 1$ for all $k$.

Let us now assume the particular setting described in \cref{rmk:simple:alpha}. Namely, take $\spv = \RR^{p \dimx}$, assume for simplicity $\mu_0$ is the Lebesgue measure, so that $d \mu/d\mu_0(\bq) = \dmu(\bq)$ for a.e. $\bq \in \RR^\dimx$, and suppose $\dmu(\bq) > 0$ for a.e. $\bq \in \RR^\dimx$. Also, assume $\Vker_0= \ldots = \Vker_p  \eqqcolon \Vker$, with $\Vker$ as given in \eqref{def:Vker:cond:ind:fd}, and let $S_j: \RR^{(p+1) \dimx} \to \RR^{(p+1) \dimx}$, $j=0,\ldots, p$, be the flip involutions defined in \eqref{def:Sj:cond:ind}. Then, under this setting, \eqref{def:alphakj:res:Bk} reduces to the same expression as in \eqref{eq:AR:TJ:FD}, cf. \eqref{simpl:Bk}, whereas \eqref{def:alphakj:res:MH} reduces to
\begin{align}\label{simpl:MH:fd}
	\alpha_{k,j} (\bq_0,\bv) = 
	\begin{cases}
		\overline{\alpha}_{k,j} \left[ 1 \wedge \frac{\dmu(\bq_j)}{\dmu(\bq_k)} \right] &\mbox{ if } j \neq k, \\
		1 - \sum_{\stackrel{l=0}{l \neq k}}^p \alpha_{k,l}(\bq_0, \bv) &\mbox{ if } j = k 
	\end{cases}
\end{align}
for $(\bq_0, \bv) = (\bq_0, \bq_1, \ldots,\bq_p) \in \RR^{(p+1) \dimx}$, cf. \eqref{simpl:MH}.

\subsection{Multiproposal pCN algorithms}
\label{sec:mpCN}

We next develop a multiproposal version of the preconditioned Crank-Nicolson 
algorithm (pCN) \cite{beskos2008mcmc, cotter2013mcmc}. As in the previous 
subsection, we proceed by drawing on the conditionally independent formalism 
we developed above in \cref{sec:cond:ind:prop} to extend the standard pCN proposal mechanism.

Recall that the pCN algorithm is a methodology built to resolve
infinite-dimensional measures which are absolutely continuous with
respect to a Gaussian base measure $\mu_0$.  We therefore begin by
briefly reviewing this Gaussian formalism as suits our purposes here;
see e.g. \cite{bogachev1998gaussian, DPZ2014} for a systematic
treatment. Take $\spq$ to be a real separable Hilbert space, with
inner product and norm denoted by $\langle \cdot, \cdot \rangle$ and
$|\cdot|$, respectively. Consider any $\tcC: \spq \to \spq$ which is a
trace-class, symmetric and strictly positive definite linear
operator.\footnote{In other words $\tcC$ is bounded and linear such
  that
  $\langle \tcC \bq, \btq \rangle = \langle \tcC \btq, \bq \rangle$,
  for any $\btq, \bq \in X$, $\langle \tcC \bq, \bq \rangle > 0$
  whenever $\bq \in X \setminus \{0\}$ and
  $\sum_{k =1}^\infty \langle \tcC \bee_k, \bee_k \rangle < \infty$
  for any complete orthonormal system $\{\bee_k\}_{k \geq 0}$.}  Let us
recall that $\nu_0$ is a Gaussian measure on $\spq$ with mean
$\bm \in \spq$ and covariance operator $\tcC$, denoted
$\nu_0 = N(\bm,\tcC)$, if $\varphi^* \mu_0$ is normally distributed for any
bounded linear functional $\varphi : \spq \to \RR$ and
\begin{align*}
  \int_\spq \langle \bhq,  \bq \rangle  \mu_0(d\bhq) = \langle \bm, \bq  \rangle,
  \quad
  \int_\spq \langle \bhq - \bm, \bq \rangle
       \langle \bhq -\bm , \btq  \rangle \mu_0(d\bhq)
      = \langle \tcC \bq, \btq \rangle,
  \quad \text{ for any } \bq, \btq \in \spq.
\end{align*}
Note that, as one would expect extending the finite-dimensional case, the associated characteristic function of $\nu_0$ is
\begin{align}
  \xi \in \spq \mapsto \int_\spq \exp( i \langle \xi, \bq \rangle) \nu_0(d\bq)
  = \exp\biggl(i
  \langle \xi, \bm \rangle - \frac{1}{2} \langle \tcC \xi, \xi \rangle
  \biggr).
  \label{eq:nu:char}
\end{align}

A standard way to draw samples from such a $\nu_0$ is to use a
\emph{Karhunen–Lo\`eve expansion} as follows.  According to the
Hilbert-Schmidt theorem, we can find a complete orthonormal system
$\{\bee_k\}_{k \geq 0}$ of eigenfunction of $\tcC$ so that
$\tcC \bee_k = \mu_k \bee_k$ for any $k \geq 0$.  Drawing an i.i.d
sequence $\{\xi_k\}_{k \geq 1}$ of normal random variables in $\RR$ with mean
zero and variance one, we find that
\begin{align}
  \bw := \bm + \sum_{k=1}^\infty \sqrt{\mu_k} \bee_k \xi_k \sim \nu_0.
  \label{eq:KL:HGaussian}
\end{align}

The pCN algorithm is used to sample from a target probability measure
on $\spq$ of the form
\begin{align}\label{mu:mpCN}
  \mu(d \bq) = \frac{1}{Z} e^{-\Phi(\bq)} \mu_0(d\bq), \quad 
         Z  = \int_{\spq} e^{-\Phi(\bq)} \mu_0(d\bq).
\end{align}
Here $\mu_0 = N(0,\cC)$ with $\cC$ symmetric, positive and trace-class,
and we suppose that $\Phi:\spq \to \RR$ is a potential function such
that $e^{-\Phi(\bq)}$ is $\mu_0$-integrable.  The idea in
\cite{beskos2008mcmc, cotter2013mcmc} is to develop a proposal kernel
to sample from $\mu$ by taking a Crank-Nicolson discretization of the
following Ornstein-Uhlenbeck dynamics
\begin{align}
  d \bq = - \frac{1}{2} \bq dt + \sqrt{\cC} dW,
  \label{eq:OU:dynamics}
\end{align}
where $W$ is a cylindrical Brownian motion on $\spq$ so that, for any
$t > s \geq 0$, $\sqrt{\cC} (W(t) - W(s)) = N(0, (t -s) \cC)$ (see
e.g. \cite{DPZ2014}). Here note that the preconditioned dynamics
\eqref{eq:OU:dynamics} maintains $\mu_0$ as an invariant and indeed
the choice of a Crank-Nicolson scheme for \eqref{eq:OU:dynamics} is
selected precisely to preserve this invariance under numerical
discretization. Concretely, this yields a proposal kernel
$Q: \spq \times \cB(\spq) \to [0,1]$ given by
\begin{align}
  Q(\bq_0, d \btq_0) = F(\bq_0,\cdot)^* \mu_0(d\btq_0) \sim
  N(\rho \bq_0, (1- \rho^2) \cC)
  \label{eq:pCN:propdist}
\end{align}
where $F: \spq \times \spq \to \spq$ is defined as
\begin{align}
  F(\bq, \bvv) = \pmr \bq + \sqrt{1 - \pmr^2} \bvv,
  \label{eq:pCN:propFn}
\end{align}
for some tuning parameter $\rho \in [0,1]$.  Note that $\rho$ is given
in terms of the time step $\delta$ from the Crank-Nicolson
discretization of \eqref{eq:OU:dynamics} as
$\rho = (4 - \delta)/(4 + \delta)$. Note moreover that proposals can be
generated from \eqref{eq:pCN:propdist}, \eqref{eq:pCN:propFn} by
making use of an expansion of the form \eqref{eq:KL:HGaussian}.

We derive our multiproposal version of the pCN algorithm based on $Q$
in \eqref{eq:pCN:propdist} to sample from $\mu$ as in \eqref{mu:mpCN}
as a direct corollary of \cref{thm:cond:ind:mu:mu0}. Fix the number
of samples per step as $p > 0$ and take $\spv = \spq^{p}$, i.e. the
$p$-fold product of $\spq$. After \eqref{def:Vker:multi:ind}, we take $\Vker: \spq \times \cB(\spq^p) \to [0,1]$ as
\begin{align}\label{Vker:mpCN}
    \Vker(\bq_0, d \bv) = \Vker(\bq_0, d\bq_1, \ldots, d\bq_p) := \int_{\spq}
    \prod_{k=1}^p Q(\bq, d \bq_k) Q(\bq_0, d \bq),
\end{align}
where $\bv = (\bq_1, \ldots, \bq_p)$ and with $Q$ given by \eqref{eq:pCN:propdist}. We 
%supplement this proposal kernel $\Vker$ by selecting 
also consider involution mappings
$S_j: \spq \times \spq^p \to \spq \times \spq^p$, $j=0,\ldots,p$, of the form
\eqref{def:Sj:cond:ind}, namely the $S_j$s are the coordinate flip
operators: $S_0 := I$;
$S_j(\bq_0, \bv) := S_j(\bq_0, \bq_1, \ldots, \bq_p) \coloneqq
(\bq_j,\bq_1, \ldots, \bq_{j-1}, \bq_0, \bq_{j+1}, \ldots, \bq_p)$, $j=1,\ldots,p$.

Regarding the assumptions in \cref{thm:cond:ind:mu:mu0}, notice from
\eqref{mu:mpCN} that $\mu \ll \mu_0$ and
$d\mu/d\mu_0(\bq) = e^{-\Phi(\bq)} > 0$ for a.e. $\bq \in
\spq$. Moreover, \eqref{Q:mu0:det:bal} reduces in our situation to
showing that $\mu_0$ is in detailed balance with respect to $Q$, which
can easily be verified via e.g. equivalence of characteristic
functionals. Indeed, it follows from \eqref{eq:nu:char} that for any
$\xi, \tilde{\xi} \in \spq$
\begin{align*}
\int_{\spq\times \spq} \exp( i (\langle \xi, \bq \rangle
+\langle \tilde{\xi}, \btq \rangle)) \Qker(\bq, d \btq) \mu_0(d\bq)
=&  \int_{\spq\times \spq}\exp( i (\langle \xi, \bq \rangle
+ \langle \tilde{\xi},  \pmr \bq + \sqrt{1 - \pmr^2} \btq
\rangle)) \mu_0(d\btq) \mu_0(d\bq)\\
=& \exp\biggl( - \frac{1}{2}( \langle \cC \xi , \xi  \rangle +
2 \rho\langle \cC \xi , \tilde{\xi}  \rangle
+  \langle \cC \tilde{\xi}, \tilde{\xi}\rangle 
)\biggr).
\end{align*}
An analogous calculation produces the same result for
$\Qker(\btq, d \bq) \mu_0(d\btq)$, allowing us to conclude that indeed
$\Qker(\btq, d \bq) \mu_0(d\btq) = \Qker(\bq, d\btq)
\mu_0(d\bq)$. According to \cref{thm:multi:ind:Tjmd},
\cref{thm:cond:ind:mu:mu0} and \cref{cor:simple:alpha}, we may thus
obtain a reversible sampling scheme by supplementing the above choices
of $\Vker$ and $S = (S_0,\ldots,S_p)$ with acceptance probabilities
given e.g. as in \eqref{def:alphaj:bb}, written here as
\begin{align}\label{def:alphaj:pCN}
\alpha_j(\bq_0, \bv) 
= \frac{\exp(-\Phi(\bq_j))}{\sum\limits_{k=0}^p \exp(-\Phi(\bq_k))}, 
\quad (\bq_0,\bv) = (\bq_0,\bq_1,\ldots,\bq_p) \in \spq \times
\spq^p, \,\, j = 0, \ldots, p,
\end{align}
with $\Phi$ as appears in \eqref{mu:mpCN}. Alternatively, one could also 
consider $\alpha_j$ as defined in \eqref{def:alpha:simple:MH}, but here we restrict our attention to the definition \eqref{def:alphaj:pCN} in view of avoiding potentially poor behavior of the MH-type $\alpha_j$ in \eqref{def:alpha:simple:MH} for a large number $p$ of proposals. See however \cref{sec:outlook} below.

\cref{alg:mpCN} summarizes the reversible sampling procedure that follows from
\cref{alg:cond:ind} with these choices.

\begin{algorithm}[!t]
	\caption{(Multiproposal pCN (mpCN))}
	\begin{algorithmic}[1]\label{alg:mpCN}
          \State Select the algorithm parameters:
          \begin{itemize}
              \item[(i)] $\pmr \in [0,1]$.
              \item[(ii)] the number of proposals $p \geq 1$.
          \end{itemize}
          \State Choose $\bq^{(0)} \in \spq$.  
          \For{$k \geq 0$}
          \State Sample $\bvv^{(k+1)} \sim \mu_0$ (cf. \eqref{eq:KL:HGaussian}).
          \State Compute $\bq \coloneqq \pmr \bq^{(k)} + \sqrt{1 - \pmr^2} \bvv^{(k+1)}$.
          \State Sample $\bvv_j^{(k+1)} \sim \mu_0$ independently for $j=1,\ldots,p$.
          \State Compute $\bq_j^{(k+1)} = \pmr \bq + \sqrt{1 - \pmr^2} \bvv_j^{(k+1)}$, $j=1,\ldots,p$. Set $\bv^{(k+1)} := (\bq_1^{(k+1)},\ldots,\bq_p^{(k+1)})$.
          \State Draw $\bq^{(k+1)}$ from the set $(\bq^{(k)},\bq_1^{(k+1)},\ldots,\bq_p^{(k+1)})$ according to the probabilities $(\alpha_0(\bq^{(k)},\bv^{(k+1)}), \ldots,$ $ \alpha_p(\bq^{(k)},\bv^{(k+1)}))$ as defined in \eqref{def:alphaj:pCN}.
          \State $k \to k + 1$.
          \EndFor
	\end{algorithmic}
\end{algorithm}

Under the framework of \cref{sec:ext:Phase}, we may also consider an
extension of \cref{alg:mpCN} by allowing for resampling among the
cloud of proposals at each iteration, as in
\cref{alg:tj:gen}. Specifically, we take $\alpha_{k,j} = \alpha_j$ as
defined in \eqref{def:alphaj:pCN} for all $k,j = 0,\ldots,p$, together
with Markov kernels $\Vker_0 = \ldots = \Vker_p = \Vker$, with $\Vker$
as in \eqref{Vker:mpCN}, and again the flip involutions $S_j$,
$j=0,\ldots,p$, as in \eqref{def:Sj:cond:ind}. In this case, it
follows from \cref{rmk:overlap} that \eqref{eq:P:KP} holds, so that we
may write this particular case of \cref{alg:tj:gen} as in \cref{alg:mpCN:TJ}. Notice
that by setting the number of resamples as $n =1$ the procedure
indeed coincides with \cref{alg:mpCN}.

\begin{algorithm}[!t]
	\caption{(Multiproposal pCN with Proposal Resampling)}
	\begin{algorithmic}[1]\label{alg:mpCN:TJ}
          \State Select the algorithm parameters:
          \begin{itemize}
              \item[(i)] $\pmr \in [0,1]$.
              \item[(ii)] the number of proposals $p \geq 1$.
              \item[(iii)] the number of resamples per proposal cloud $n \geq 1$.
          \end{itemize}
          \State Choose $\bq^{(0)} \in \spq$.  
          \For{$k \geq 0$}
          \State Sample $\bvv^{(k+1)} \sim \mu_0$ (cf. \eqref{eq:KL:HGaussian}).
          \State Compute $\bq \coloneqq \pmr \bq^{(k)} + \sqrt{1 - \pmr^2} \bvv^{(k+1)}$.
          \State Sample $\bvv_j^{(k+1)} \sim \mu_0$ independently for $j=1,\ldots,p$.
          \State Compute $\bq_j^{(k+1)} = \pmr \bq + \sqrt{1 - \pmr^2} \bvv_j^{(k+1)}$, $j=1,\ldots,p$. Set $\bv^{(k+1)} := (\bq_1^{(k+1)},\ldots,\bq_p^{(k+1)})$.
           \For{$l = 1, \dots, n$}
            \State Draw $\br^{(nk + l)}$ from the set $(\bq^{(k)},\bq_1^{(k+1)},\ldots,\bq_p^{(k+1)})$ 
            with the probabilities
            \par
            $\quad \; (\alpha_0(\bq^{(k)},\bv^{(k+1)}), \ldots,\alpha_p(\bq^{(k)},\bv^{(k+1)}))$
            defined in \eqref{def:alphaj:pCN}.
            \State $l \to l + 1$.
           \EndFor
          \State Set $\bq^{(k+1)} := \br^{(n(k + 1))}$.
          \State $k \to k + 1$.
          \EndFor
	\end{algorithmic}
\end{algorithm}

\subsection{Multiproposal HMC algorithms}
\label{sec:HMC}

In this section, we present a few instances of Hamiltonian Monte Carlo
(HMC)-like algorithms based on multiple proposals, restricting our
attention to the finite-dimensional case for simplicity. We emphasize
that the general scope of \cref{alg:main} and \cref{alg:tj:gen} allow
for the possibility of a variety of other HMC-like sampling schemes
under different choices of algorithmic parameters and acceptance
probabilities, which we will explore in future work.
 
We proceed by briefly recalling some generalities regarding HMC
algorithms, following a similar presentation as in our recent
contribution \cite{glatt2020accept}. For more complete details, we
refer to e.g. the pioneering works \cite{duane1987hybrid,neal1993} and
also to \cite{neal2011mcmc, LeimkuhlerReich2004, Hairer2006book,
  bou2018geometric}.

Let us consider spaces $\spq = \spv = \RR^\dimx$ and fix a target
measure of the form
\begin{align}\label{def:mu:HMC}
  \mu(d\bq) = \frac{1}{Z} e^{-\Pot(\bq)} d\bq, \quad Z
  = \int_{\RR^\dimx} e^{-\Pot(\bq)} d \bq,
\end{align}
for some potential function $\Pot:\RR^\dimx \to \RR$ which we assume
to be in $C^1(\RR^\dimx)$ and such that
$e^{-\Pot(\bq)} \in L^1(\RR^\dimx)$. An HMC algorithm samples from
this measure by first selecting a Hamiltonian function
$\Ham: \RR^{2 \dimx} \to \RR$ such that the marginal of the associated
Gibbs measure
\begin{align}\label{def:M:HMC:0}
	\cM(d\bq,d\bv) = \frac{1}{Z_\Ham} e^{-\Ham(\bq,\bv)} \, d \bq\, d\bv,
  \quad Z_\Ham 
	= \int_{\RR^{2\dimx}} e^{-\Ham(\bq, \bv)} \, d \bq \, d\bv
\end{align}
with respect to the ``position'' variable $\bq$ coincides with the
target measure $\mu$ in \eqref{def:mu:HMC}.  We may thus write such
Hamiltonian function in a general form as
\begin{align*}
	\Ham(\bq, \bv) = \Pot(\bq) + \VPot(\bq, \bv), \quad (\bq,\bv) \in \RR^{2\dimx},
\end{align*}
for some $C^1$ function $\VPot: \RR^{2\dimx} \to \RR$ with
$\int_{\RR^\dimx} e^{-\VPot(\bq, \bv)} d\bv = 1$ for all
$\bq \in \RR^\dimx$. Under this definition, $\cM$ can be written as
\begin{align}\label{def:M:HMC}
  \cM(d\bq, d \bv) = \Vker(\bq,d\bv) \mu(d\bq), \,\,
  \mbox{ for }\,\, \Vker(\bq,d\bv) = e^{-\VPot(\bq,\bv)} d\bv,
\end{align}
where it follows by construction that $\Vker$ is a Markov kernel.

In association with such $\Ham$, one considers the following
Hamiltonian dynamic for the pair $\pv = (\bq,\bv) \in \RR^{2 \dimx}$
\begin{align}\label{Ham:dyn}
	\frac{d\pv}{dt} = J^{-1} \nabla \Ham (\pv), \quad \pv(0) = (\bq_0,\bv_0),
\end{align}
for some $2\dimx \times 2 \dimx$ real matrix $J$ which is invertible
and antisymmetric.\footnote{Typical choices for $J$ and $\VPot$ are
  given by $J = \begin{pmatrix} 0 & - I \\ I & 0 \end{pmatrix}$ and
  $\VPot(\bq,\bv) = \frac{1}{2} \langle M(\bq)^{-1} \bv, \bv \rangle +
  \frac{1}{2} \ln ((2\pi)^\dimx \det(M(\bq)))$, where
  $\langle \cdot, \cdot \rangle$ denotes the Euclidean inner product,
  for some ``mass'' matrix $M$. In this case, $\VPot(\bq,\bv)$
  corresponds to the negative log-density of the Gaussian
  $N(0,M (\bq))$ in $\RR^\dimx$.} This dynamic leaves $\Ham$ invariant and preserves volume elements\footnote{The solution map is
  actually symplectic, a stronger property which implies the
  preservation of volumes.  See \cite[Definition 4.1]{glatt2020accept}
  in our context.}. As such, $\cM$ is an invariant measure under the
associated flow. This in turn implies that the projected dynamic of
the $\bq$-variable leaves invariant the $\bq$-marginal of $\cM$,
namely $\mu$.

Of course, it is typically intractable to exactly solve
\eqref{Ham:dyn}.  One must instead resort to a suitable numerical
integrator maintaining certain indispensable geometric properties. For
a chosen time step size $\dT > 0$, such a scheme
$\{\Sol_{j,\dT}(\bq_0,\bv_0)\}_{ j \in \NN}$ yields an approximation
of the solution of \eqref{Ham:dyn} at times $j\dT$, $j \in \NN$.

For the given step size $\dT > 0$, one builds
$\{\Sol_{j,\dT}\}_{ j \in \NN}$ starting from the single integration
step $\Sd_\dT: \RR^{2\dimx} \to \RR^{2\dimx}$, which is in general
supposed to be an invertible $C^1$ mapping.  We require that $\Sd_\dT$ satisfies the following standard geometric properties (see
e.g. \cite{Hairer2006book, LeimkuhlerReich2004}):
\begin{enumerate}[label={(P\arabic*)}]
\item\label{prop:i} $\Sd_\dT$ is \emph{volume-preserving}, i.e.
  \begin{align}\label{eq:step:vol}
  |\det \nabla \Sd_\dT (\bq,\bv)| = 1 \quad \text{ for every }
    (\bq,\bv) \in \RR^{2\dimx};
    \end{align}
  \item\label{prop:ii} $\Sd_\dT$ is \emph{reversible} with respect to
    the ``momentum''-flip involution
    \begin{align}\label{eq:mom:flip}
      R(\bq, \bv) := (\bq, -\bv),
      \end{align}
  i.e.
	\begin{align}\label{eq:Ham:nm:inv}
          R \circ \Sd_\dT (\bq, \bv) = \Sd_\dT^{-1} \circ R (\bq, \bv)
          \quad \mbox{ for all } (\bq, \bv) \in \RR^{2\dimx}.
	\end{align}
	Or, equivalently, $ (R \circ \Sd_\dT)^2 = I$, i.e. $R \circ
        \Sol$ is an involution.
\end{enumerate}
We now define
\begin{align}\label{eq:nm:sol:def}
  \Sol_{j,\dT} = \Sd_\dT^j \text{ for }  j \in \ZZ,
 \end{align}
 where $\Sd_\dT^j$ denotes the $j$-fold composition of $\Sd_\dT$ for
 $j \geq 1$, the identity map for $j = 0$ and the $j$-fold composition
 of $\Sd_\dT^{-1}$ for $j \leq -1$.  Note that it follows immediately
 that for $j \in \NN$, $\Sol_{j,\dT}$ maintains the same volume
 preservation and reversibility properties \`a la \eqref{eq:step:vol},
 \eqref{eq:Ham:nm:inv}.

As a concrete example, we recall that when the Hamiltonian is written
in the separable form $\Ham(\bq,\bv) = \Pot(\bq) + \VPot(\bv)$ and
$J = \begin{pmatrix} 0 & - I \\ I & 0 \end{pmatrix}$, the associated
Hamiltonian dynamic is $d\bq/dt = \nabla \VPot(\bv)$,
$d\bv/dt = - \nabla \Pot(\bq)$. In this case, the classical leapfrog
integrator is defined as
\begin{align}\label{eq:leap:ex:1}
\Sd_\dT := \Xi^{(1)}_{\dT/2} \circ \Xi^{(2)}_\dT \circ
  \Xi^{(1)}_{\dT/2},
\end{align}
  where $\Xi^{(1)}$ and $\Xi^{(2)}$ are the exact
solution operators of $d\bq/dt = 0$, $d\bv/dt = - \nabla \Pot(\bq)$,
and $d\bq/dt = \nabla \VPot(\bv)$, $d\bv/dt = 0$,
respectively. Namely,
\begin{align}\label{eq:leap:ex:2}
  \Xi^{(1)}_t (\bq,\bv) = (\bq, \bv + t \nabla \Pot(\bq))
  \quad \text{ and } \quad
  \Xi^{(2)}_t (\bq,\bv) = (\bq + t \nabla \VPot(\bv), \bv)
\end{align}  
for all $t \in \RR$.  This classical leapfrog scheme is
the basis for many of the common HMC implementations.

Fixing now $\delta > 0$, $p \geq 1$ and an integration scheme
defined by $\Sd_\dT$ maintaining \ref{prop:i},\ref{prop:ii}, we now
construct a multiproposal sampling procedure starting from the setting of
\cref{sec:Multi:MH:ker}, \cref{alg:main}. Our approach makes use of
a sequence of integration steps as the multiple proposals at each
iteration. For this, we define the mappings, cf. \eqref{eq:mom:flip}, \eqref{eq:nm:sol:def},
\begin{align}\label{eq:HMC:inv}
S_0 \coloneqq I, \quad 
  S_j \coloneqq R \circ \Sol_{j,\dT}, \text{ for } j=1,\ldots,p.
\end{align}
Each of these maps are involutions according to property \ref{prop:ii}
above.  Property \ref{prop:i} furthermore implies that each of these
maps $S_j$ are volume preserving so that,
cf. \eqref{def:alphaj:MH:fd},
\begin{align}
  \frac{d S_j^* \cM}{d \cM}(\bq,\bv) = \exp(\Ham(\bq,\bv) - \Ham(S_j(\bq,\bv))).
\end{align}
Therefore, it follows from \cref{cor:wedge:alpha} that we may obtain a
reversible sampling algorithm by supplementing $\Vker$ from
\eqref{def:M:HMC} and these mappings $S_j$ with the following
acceptance probabilities
\begin{align}\label{def:alphaj:hmc}
	\alpha_j(\bq, \bv) = \overline{\alpha}_j \left[ 1 \wedge
    e^{\Ham(\bq,\bv) - \Ham(S_j(\bq,\bv))} \right],
  \,\, j = 1,\ldots,p; \quad \alpha_0(\bq,\bv) = 1 - \sum_{j=1}^p \alpha_j(\bq,\bv),
\end{align}
for all $(\bq,\bv) \in \RR^{2\dimx}$, where $\overline{\alpha}_j$ are
given weights such that $\sum_{j=1}^p \overline{\alpha}_j \leq 1$. \cref{alg:mHMC} summarizes the resulting reversible procedure that follows as a
particular case of \cref{alg:main} with these choices.

\begin{algorithm}[!t]
	\caption{(Multiproposal HMC (mHMC))}
	\begin{algorithmic}[1]\label{alg:mHMC}
		\State Select the algorithm parameters:
		\begin{itemize}
			\item[(i)] The ``momentum'' kernel $\Vker(\bq, d\bv) = e^{-\VPot(\bq,\bv)} d\bv$.
			\item[(ii)] The time step size $\dT > 0$ and the
                          corresponding integration step 
                          $\Sd_\dT: \RR^{2\dimx} \to
                          \RR^{2 \dimx}$ satisfying \ref{prop:i},
                          \ref{prop:ii}  used for approximating the
                          Hamiltonian dynamic \eqref{Ham:dyn} (see
                          \eqref{eq:leap:ex:1}, \eqref{eq:leap:ex:2}).
			\item[(iii)] The number of integration steps
                          $p \geq 1$.
                        \item[(iv)] The acceptance weights
                          $\overline{\alpha}_j \in [0,1]$ such that
                          $\sum_{j=1}^p \overline{\alpha}_j \leq 1$.
		\end{itemize}
		\State Choose $\bq^{(0)} \in \RR^N$.  
		\For{$k \geq 0$}
		\State Sample $\bv^{(k+1)} \sim \Vker(\bq^{(k)}, \cdot)$.
		\State Set $(\bbq^{(k,0)}, \bbv^{(k,0)}) := (\bq^{(k)}, \bv^{(k+1)})$.
		\For{$j = 0,\ldots,p-1$}
		\State Compute $(\bbq^{(k,j+1)}, \bbv^{(k,j+1)}) := \Xi_\dT (\bbq^{(k,j)}, \bbv^{(k,j)})$.
		\State $j \to j + 1$.
		\EndFor
		\State Set $\bq^{(k+1)}$ by drawing from 
		$(\bq^{(k)}, \bbq^{(k,1)},\ldots, \bbq^{(k,p)})$
		according to the probabilities $(\alpha_0^{(k)}
                \ldots,\alpha_p^{(k)})$ where
                \begin{align*}
                 \quad \,\,\,\, \alpha_j^{(k)} := \overline{\alpha}_j \left[ 1 \wedge \exp(\Ham(\bbq^{(k,0)},
                  \bbv^{(k,0)}) - \Ham(\bbq^{(k,j)},- \bbv^{(k,j)})) \right]
                  \text{ for } j = 1, \ldots, p,
                  \text{ and }
                  \alpha_0^{(k)} := 1 - \sum_{j=1}^p \alpha_j^{(k)}.
                \end{align*}
		\State $k \to k + 1$.
		\EndFor
	\end{algorithmic}
\end{algorithm}

\begin{Remark}
  Note that one could also consider in \cref{alg:mHMC} acceptance
  probabilities $\alpha_j$, $j=0,\ldots,p$, in the corresponding
  Barker form \eqref{def:alphaj}. However, to guarantee reversibility
  or even just ergodicity in this case according to
  \cref{cor:ar:Barker}, the summation condition \eqref{sum:cond} would
  be imposed, thus introducing a seemingly restrictive assumption on
  the numerical integrator.
\end{Remark}

Furthermore, under the framework of \cref{sec:ext:Phase}, we can also
construct a multiproposal HMC algorithm that allows for proposal
resampling. In this case, we choose in \cref{alg:tj:gen} Markov
kernels $\Vker_0 = \cdots = \Vker_p \eqqcolon \Vker$, with $\Vker$ as
in \eqref{def:M:HMC}, and again consider the involution mappings $S_j$
from the integration of \eqref{Ham:dyn} given in
\eqref{eq:HMC:inv}. From \cref{cor:alphakj:res}, we may supplement
these choices with e.g. either of the following definitions of
acceptance probabilities to yield an unbiased sampling algorithm,
cf. \eqref{def:alphakj:res:Bk}-\eqref{def:alphakj:res:MH}:
\begin{align}\label{def:alphakj:HMC:Bk}
	\alpha_{k,j}(\bq,\bv) = \frac{e^{-\Ham(S_j(\bq,\bv))}}{\sum_{l=0}^p e^{-\Ham(S_l(\bq,\bv))}},
\end{align}
or
\begin{align}\label{def:alphakj:HMC:MH}
	\alpha_{k,j} (\bq,\bv) = 
	\begin{cases}
	\overline{\alpha}_{k,j} \left[ 1 \wedge e^{\Ham(S_k(\bq,\bv)) - \Ham(S_j(\bq,\bv))} \right] &\mbox{ if } j \neq k, \\
	1 - \sum_{\stackrel{l=0}{l \neq k}}^p \alpha_{k,l}(\bq, \bv) &\mbox{ if } j = k 
	\end{cases}
\end{align}
for all $k,j = 0,\ldots,p$ and $(\bq,\bv) \in \RR^{2\dimx}$, where in
\eqref{def:alphakj:HMC:MH} $\overline{\alpha}_{k,j} \in [0,1]$,
$k,j = 0,\ldots,p$, are chosen weights satisfying
$\sum_{j=0}^p \overline{\alpha}_{k,j} \leq 1$ for all $k$.

In fact, due to step 7 in \cref{alg:tj:gen}, at each iteration the
procedure requires the calculation of
$\alpha_{k,j}(S_m(\bq,\bv)) = \alpha_{k,j}(R \circ
\Sol_{m,\dT}(\bq,\bv))$ for some $(\bq,\bv) \in \RR^{2\dimx}$ and some $m \in \NN$. Note
that, according to \ref{prop:ii} above, we have
\begin{align*}
	S_j \circ S_m = R \circ \Sol_{j,\dT} \circ R \circ \Sol_{m,\dT} = \Sol_{j,\dT}^{-1} \circ \Sol_{m,\dT}
	= (\Sd_\dT^j)^{-1} \circ \Sd_\dT^m
	= (\Sd_{\dT}^{-1})^j \circ \Sd_\dT^m
	= \Sd_\dT^{m-j},
\end{align*}
for all $m,j=0,\ldots,p$, where recall that we denote
$\Sd_\dT^{-l}:= (\Sd_\dT^{-1})^l$ for all $l \in \NN$. Therefore,
it follows respectively under \eqref{def:alphakj:HMC:Bk} and
\eqref{def:alphakj:HMC:MH} that
\begin{align}\label{def:alphakj:HMC:Bk:Sk}
  \alpha_{k,j}(S_m(\bq,\bv)) = \alpha_{k,j}(R \circ \Sol_{m,\dT} (\bq,\bv))
  = \frac{\exp(-\Ham(\Sd_\dT^{m-j}(\bq,\bv)))}{\sum_{l=0}^p \exp(-\Ham( \Sd_\dT^{m-l}(\bq,\bv)))},
\end{align}
or
\begin{align}\label{def:alphakj:HMC:MH:Sk}
\alpha_{k,j}(S_m(\bq,\bv)) = \alpha_{k,j} (R \circ \Sol_{m,\dT} (\bq,\bv)) = 
\begin{cases}
  \overline{\alpha}_{k,j} \left[ 1 \wedge e^{\Ham(\Xi_\dT^{m-k} (\bq,\bv)) -\Ham(\Sd_\dT^{m-j}(\bq,\bv))} \right]
&\mbox{ if } j \neq k, \\
1 - \sum_{\stackrel{l=0}{l \neq k}}^p \alpha_{k,l}(\bq, \bv)
&\mbox{ if } j = k 
\end{cases}
\end{align}
for all $k,j,m = 0,\ldots,p$ and $(\bq,\bv) \in \RR^{2\dimx}$.  \cref{alg:mhmcpr} details the sampling procedure resulting from these observations.

\begin{algorithm}[!t]
	\caption{(Multiproposal HMC with Proposal Resampling)}\label{alg:mhmcpr}
	\begin{algorithmic}[1]\label{alg:mHMC:res}
		\State Select the algorithmic parameters
		\begin{itemize}
                \item[(i)] The ``momentum'' kernel
                  $\Vker(\bq, d\bv) = e^{-\VPot(\bq,\bv)} d\bv$.
                \item[(ii)] The time step size $\dT > 0$ and the
                  corresponding integration step
                  $\Sd_\dT: \RR^{2\dimx} \to \RR^{2 \dimx}$ satisfying
                  \ref{prop:i}, \ref{prop:ii} used in the scheme
                  $\Sol_{k,\dT} = \Sd_\dT^k$ approximating the
                  Hamiltonian dynamic \eqref{Ham:dyn} (see
                  \eqref{eq:leap:ex:1}, \eqref{eq:leap:ex:2}).
                \item[(iii)] The number of integration steps
                  $p \geq 1$.
                \item[(iv)] If using the mechanism
                  \eqref{def:alphakj:HMC:MH:Sk}, specify the
                  weights $\overline{\alpha}_j \in [0,1]$, $j=1,\ldots,p$, such that
                  $\sum_{j=1}^p \overline{\alpha}_j \leq 1$.
                \item[(v)] The $n \geq 1$ number of samples drawn per
                  generated proposal cloud.
		\end{itemize}
		\State Choose an initial $(\bq^{(0)}, \bv^{(0)}) \in
                \RR^{2\dimx}$ and $k^{(0)} \in \{0, \ldots, p\}$.
                \State Set
                $\br^{(0)} := \Pi_1 \Sol_{k^{(0)},\dT}(\bq^{(0)},
                \bv^{(0)})$.  \For{$i \geq 0$} \State Set
                $\bq^{(i+1)} := \br^{(ni)}$.  \State Sample
                $\bv^{(i+1)} \sim \Vker_{k^{(i)}}(\bq^{(i+1)}, d\bv)$.
                \State Set
                $(\bbq^{(k^{(i)},k^{(i)})}, \bbv^{(k^{(i)},k^{(i)})})
                := (\bq^{(i+1)}, \bv^{(i+1)})$.
		\For{$j=0,\ldots,k^{(i)}-1$}
			\State Compute $(\bbq^{(k^{(i)}, k^{(i)} - j - 1))}, \bbv^{(k^{(i)}, k^{(i)} - j - 1))}) := \Sd_\dT(\bbq^{(k^{(i)}, k^{(i)} - j )}, \bbv^{(k^{(i)}, k^{(i)} - j )}) $.
			\State $j \to j+1$.
		\EndFor
		\For{$j = k^{(i)}, \ldots, p-1$}
			\State Compute $(\bbq^{(k^{(i)}, j + 1))}, \bbv^{(k^{(i)}, j + 1))}) := \Sd_\dT^{-1}(\bbq^{(k^{(i)},  j )}, \bbv^{(k^{(i)},  j )}) $.
			\State $j \to j+1$.
		\EndFor
		\State Set $k_{cur} := {k^{(i)}}$.
		\For{$m = 1, \dots, n$}
		\State Draw $k_{nxt} \in \{0, \ldots, p\}$ with the
		probabilities  $(\alpha_{k_{cur},0}^i,
		\ldots, \alpha_{k_{cur},p}^i)$
		where 
		\begin{align*}
		\alpha_{k_{cur},j}^i := \frac{\exp(-\Ham(\bbq^{(k^{(i)},j)}, \bbv^{(k^{(i)},j)}))}{\sum_{l=0}^p \exp(-\Ham( \bbq^{(k^{(i)},l)}, \bbv^{(k^{(i)},l)})))}, \quad j =0,\ldots,p,
		\end{align*}
		\par 
		\quad or
		\begin{align*}
		\qquad \qquad \alpha_{k_{cur},j}^i =
		\begin{cases}
		\overline{\alpha}_{k_{cur},j} \left[ 1 \wedge \exp \left(\Ham(\bbq^{(k^{(i)}, k_{cur})}, \bbv^{(k^{(i)}, k_{cur})}) -\Ham(\bbq^{(k^{(i)}, j)}, \bbv^{(k^{(i)}, j)}) \right) \right]
		&\mbox{ if } j \neq k_{cur}, \\
		1 - \sum_{\stackrel{l=0}{l \neq k_{cur}}}^p \alpha_{k_{cur},l}(\bq, \bv)
		&\mbox{ if } j = k_{cur} .
		\end{cases}
		\end{align*}
		\State Set $k_{cur} := k_{nxt}$.
		\State Set $\br^{(ni + m)} := 
		\bbq^{(k^{(i)},k_{cur})}$. 
		\State $m \to m + 1$.
		\EndFor
		\State Set $k^{(i+1)}  = k_{cur}$.
		\State $i \to i + 1$.
		\EndFor
	\end{algorithmic}
\end{algorithm}
 
\begin{Remark}\label{rmk:diff:multi:res:hmc}
  In contrast to the situation for the multiproposal pCN algorithms in
  \cref{sec:mpCN}, cf. \cref{rmk:overlap}, it is clear by comparing
  the previous two algorithms that \cref{alg:mHMC:res} with $n=1$ and
  under the choice of $\alpha_{k,j}$ in \eqref{def:alphakj:HMC:MH:Sk}
  yields a different procedure than \cref{alg:mHMC}.
\end{Remark}

\subsection{Simplicial samplers}
\label{sec:simplicial}

In this section, we cast into the framework of \cref{sec:framework} and \cref{sec:ext:Phase} a class of algorithms recently introduced in \cite{holbrook2023generating} (see also \cite{tjelmeland2004using}), where proposals are generated as the vertices of a fixed regular simplex after a random rescaling of edge lengths, random rotation, and translation by the current state. We make the algorithmic setting more precise as follows. 

Fix $\dimx \geq 1$ and a continuously distributed target measure $\mu \in \Pr(\RR^\dimx)$, namely $\mu(d\bq) = \dmu(\bq) d \bq$ for some density function $\dmu: \RR^\dimx \to \RR^+$. Fix also 
$p \leq \dimx$ and a $p$-simplex in $\RR^\dimx$ with equidistant vertices $\vs_1,\ldots,\vs_p,\bzero \in \RR^\dimx$, where we set $\bzero$ as one of the vertices for convenience. Specifically, $\vs_1,\ldots,\vs_p$ are linearly independent vectors which we assume for simplicity to have unit norm and pairwise distance, namely
\begin{align}\label{equidist}
 \|\vs_j\| = 1 \,\, \mbox{ for } j = 1,\ldots, p; \,\, \mbox{ and }
\| \vs_j - \vs_k \| = 1, \,\, \mbox{ for all } j, k = 1,\ldots, p,\,\, j \neq k.
\end{align}
The corresponding $p$-simplex is defined as the convex hull of $\vs_1,\ldots,\vs_p,\bzero$, so that for, e.g., $p=1,2,3$ it is given respectively by a line, a triangle, and a tetrahedron.

To perform random rotations and rescalings of this fixed simplex, we set the following additional framework. Denote by $\Od$ the $\dimx$-dimensional orthogonal group, i.e., the group formed by all real $\dimd \times \dimd$ orthogonal matrices $Q$, namely $Q Q^T = I$. We endow $\Od$ with the Borel $\sigma$-algebra of $\RR^{\dimd \times \dimd} \simeq \RR^{\dimd^2}$. Since $\Od$ is a locally compact topological group, it admits a unique normalized left Haar measure $\Hd$ on $\Od$ \cite{stewart1980,Folland1999,halmos2013}. Here by normalized we mean precisely $\Hd(\Od) = 1$, and we recall that a left Haar measure is defined as a nonzero left-invariant Radon measure, where the left-invariance property means that 
\begin{align}\label{left:inv:Haar}
\Hd(Q(E)) = \Hd(E) \quad \mbox{ for all } Q\in \Od \mbox{ and Borel set } E \subset \Od,
\end{align}
or, equivalently, $(Q^T)^*\Hd = \Hd$ for every $Q \in \Od$.

Let us additionally fix a probability distribution of edge lengths $\lambda > 0$ given by a measure $\ed(d\lambda)$ in $\RR^+$. We define a mapping $\rrt: \RR^\dimx \times \RR^+ \times \Od \to \RR^{\dimx p}$ that for each $(\bq,\lambda, Q) \in \RR^\dimx \times \RR^+ \times \Od$ yields a rescaling of the vertices $\vs_1,\ldots,\vs_p$ by $\lambda$, applies the orthogonal transformation $Q$, and then shifts the output vertices $\lambda Q \vs_1,\ldots, \lambda Q\vs_p \in \RR^\dimx$ by $\bq$. Namely,
\begin{align*}
	\rrt(\bq,\lambda, Q) = \bv := (\bq + \lambda Q \vs_1, \ldots, \bq + \lambda Q \vs_p) \quad \mbox{ for all } (\bq,\lambda, Q) \in \RR^\dimx \times \RR^+ \times \Od.
\end{align*}

From a current state $\bq \in \RR^\dimx$, one then draws independently $\lambda \sim \ed$, $Q \sim \Hd$, and proposes $p$ new states given by $\rrt(\bq,\lambda,Q)$. This proposal procedure can be written explicitly in terms of the following Markov kernel
\begin{align}\label{def:Vker:simpl}
	\Vker(\bq, d \bv) = \rrt(\bq, \cdot)^*(\ed \otimes \Hd)(d\bv) \quad \mbox{ for } \bq \in \RR^\dimx, \,\, \bv \in \RR^{\dimx p}.
\end{align}

Next, similarly as in \eqref{def:Sj:cond:ind}, we define involution mappings $S_j: \RR^{\dimx(p + 1)} \to \RR^{\dimx(p + 1)}$, $j=0,\ldots,p$, given as the coordinate flip operators 
\begin{align}\label{def:Sj:simpl}
	S_0 := I; \quad S_j(\bq_0,\bv) := S_j(\bq_0,\bq_1,\ldots,\bq_p) := (\bq_j, \bq_1, \ldots, \bq_{j-1},\bq_0,\bq_{j+1},\ldots, \bq_p)
\end{align}
for all $(\bq_0,\bv) \in \RR^{\dimx(p+1)}$ and $j=1,\ldots,p$. 

To complete the parameter setup from \cref{alg:main}, we consider acceptance probabilities $\alpha_j$, $j=0,\ldots,p$, given as either the Barker type \eqref{def:alphaj} or the Metropolis-Hastings type \eqref{def:alphaj:metr:1}-\eqref{def:alphaj:metr:2}. Before providing the explicit interpretation of these formulas within this special setting, we justify these choices of $\alpha_j$ by noticing that condition \eqref{sum:cond} in \cref{cor:ar:Barker} and also the absolute continuity condition $S_j^*\cM \ll \cM$ from \cref{cor:wedge:alpha} are indeed satisfied. These statements follow from the next two propositions, whose proofs are presented in \cref{sec:prop:sum:simpl} and \cref{sec:prop:SjM:M:simpl}, respectively. 

\begin{Proposition}\label{prop:sum:cond:simpl}
Fix any $\mu \in \Pr(\RR^\dimx)$. Let $\Vker$ be the Markov kernel in \eqref{def:Vker:simpl}, and define the measure $\cM(d\bq, d\bv) = \Vker(\bq, d \bv) \mu(d\bq)$ in $\RR^{\dimx(p+1)}$. Let also $S_j: \RR^{\dimx(p+1)} \to \RR^{\dimx(p+1)}$, $j=0,\ldots,p$, be the flip involution mappings defined in \eqref{def:Sj:simpl}. Then, we have the following equivalence of measures:
\begin{align}\label{sum:cond:simpl}
	\sum_{k=0}^p (S_j \circ S_k)^* \cM = \sum_{k=0}^p S_k^* \cM 
\end{align}
for all $j=0,\ldots,p$.
\end{Proposition}

\begin{Proposition}\label{prop:SjM:M:simpl}
Let $\mu \in \Pr(\RR^\dimx)$ be given as $\mu(d\bq) = \dmu(\bq) d \bq$ for some density function $\dmu: \RR^\dimx \to \RR^+$. Set $\Vker$ to be the Markov kernel in \eqref{def:Vker:simpl}, and define the measure $\cM(d\bq, d\bv) = \Vker(\bq, d \bv) \dmu(\bq) d \bq$ in $\RR^{\dimx(p+1)}$. Let also $S_j: \RR^{\dimx(p+1)} \to \RR^{\dimx(p+1)}$, $j=0,\ldots,p$, be the flip involution mappings defined in \eqref{def:Sj:simpl}. Then, $S_j^* \cM \ll \cM$ for all $j=0,\ldots,p$, and 
\begin{align}\label{RN:SjM:M:simpl}
	\frac{d S_j^* \cM}{d \cM} (\bq_0, \bv) = \frac{\dmu(\bq_j)}{\dmu(\bq_0)} \quad \mbox{ for $\cM$-a.e. } (\bq_0,\bv) = (\bq_0,\bq_1,\ldots,\bq_p) \in \RR^{\dimx} \times \RR^{p\dimx}.
\end{align}
\end{Proposition}

It thus follows from \cref{cor:ar:Barker} and \cref{cor:wedge:alpha} that we can obtain a reversible sampling algorithm by combining the above choices of $\Vker$ and $S_j$, $j=0,\ldots,p$, in \eqref{def:Vker:simpl}-\eqref{def:Sj:simpl} with acceptance probabilities $\alpha_j$, $j=0,\ldots,p$, given either by \eqref{def:alphaj:ac} or \eqref{def:alphaj:metr:1}-\eqref{def:alphaj:metr:2}, which according to \eqref{RN:SjM:M:simpl} are written here respectively as
\begin{align}\label{def:alphaj:Bk:simpl}
	\alpha_j(\bq_0,\bv) = \frac{\dmu(\bq_j)}{\sum_{l=0}^p \dmu(\bq_l)}, \quad j =0,\ldots,p,
\end{align}
and
\begin{align}\label{def:alphaj:MH:simpl}
	\alpha_j(\bq_0,\bv) = \overline{\alpha}_j \left[ 1 \wedge \frac{\dmu(\bq_j)}{\dmu(\bq_0)}\right], \quad j=1,\ldots,p; \quad
	\alpha_0(\bq,\bv) = 1 - \sum_{j=1}^p \alpha_j(\bq_0,\bv),
\end{align}
for all $(\bq_0,\bv) = (\bq_0,\bq_1,\ldots,\bq_p) \in \RR^{\dimx(p+1)}$ such that the expression in the denominator in each case is strictly positive, and zero otherwise. As before, $\overline{\alpha}_j \in [0,1]$, $j=1,\ldots,p$, in \eqref{def:alphaj:MH:simpl} are user specified weights such that $\sum_{j=1}^p \overline{\alpha}_j \leq 1$.
Under this setting, \cref{alg:simpl} describes the reversible procedure for sampling from $\mu(d\bq) = \dmu(\bq) d\bq$ in $\RR^\dimx$ as a special case of \cref{alg:main}.

\begin{algorithm}[!t]
	\caption{}
	\begin{algorithmic}[1]\label{alg:simpl}
		\State Select the algorithm parameters:
		\begin{itemize}
			\item[(i)] Linearly independent vectors $\vs_1,\ldots,\vs_p \in \RR^\dimx$, for some $p \leq \dimx$, satisfying \eqref{equidist} to form a $p$-simplex with vertices $(\vs_1,\ldots,\vs_p,\bzero)$.
			\item[(ii)] The edge length distribution $\ed(d\lambda)$.
			\item[(iii)] If using the mechanism \eqref{def:alphaj:MH:simpl}, select the weights $\overline{\alpha}_j \in [0,1]$ for $j=1,\ldots,p$ such that $\sum_{j=1}^p \overline{\alpha}_j \leq 1$.
		\end{itemize}
		\State Choose $\bq_0^{(0)} \in \spq$.  
		\For{$k \geq 0$}
		\State Sample $\lambda^{(k+1)} \sim \nu$.
		\State Sample $Q \sim \Hd$.
		\State Compute $\bq_j^{(k+1)} = \bq_0^{(k)} + \lambda Q \vs_j$ for $j=1,\ldots,p$. Set $\bv^{(k+1)} := (\bq_1^{(k+1)}, \ldots, \bq_p^{(k+1)})$.
		\State Set $\bq_0^{(k+1)}$ by drawing from $(\bq_0^{(k)}, \bq_1^{(k+1)}, \ldots, \bq_p^{(k+1)})$ according to the probabilities 
		\par
		$(\alpha_0(\bq^{(k)}_0,\bv^{(k+1)}), \ldots,$ $ \alpha_p(\bq^{(k)}_0,\bv^{(k+1)}))$ as defined in \eqref{def:alphaj:Bk:simpl} or \eqref{def:alphaj:MH:simpl}.
		\State $k \to k + 1$.
		\EndFor
	\end{algorithmic}
\end{algorithm}

Furthermore, we can also obtain a simplicial sampler algorithm allowing for multiple selected states from the cloud of proposals at each iteration as in \cref{alg:tj:gen}. For this, we set $\Vker_0 = \ldots = \Vker_p := \Vker$, with $\Vker$ as in \eqref{def:Vker:simpl}, and $(S_0,\ldots, S_p)$ as in \eqref{def:Sj:simpl}, along with acceptance probabilities $\alpha_{k,j}$, $k,j = 0,\ldots,p$, that can be defined for example as in \eqref{def:alphaj:mult} or \eqref{def:alphakj:MH}. Note that, similarly as in \eqref{def:alphaj:Bk:simpl} and \eqref{def:alphaj:MH:simpl}, these formulas can be written explicitly in terms of the density $\dmu$ from the target $\mu$ by invoking \eqref{RN:SjM:M:simpl}. Moreover, it follows from the same arguments as in \cref{rmk:overlap} that by setting the number of jumps $n=1$ in the resulting algorithm would reduce it to \cref{alg:simpl}. We omit further details.

\section{Numerical case studies}
\label{sec:case:Study}

This section collects various case studies which illustrate some of
the potential scope of the pMCMC algorithms we derive above in
\cref{sec:Ex:alg}.  \cref{sec:ppMCMC} considers on two case studies
for \cref{alg:TJ:cor:FD} focusing on the case when $p$, the number of
proposals per step, is taken to be very large.  These examples
illustrate the promise of pMCMC methods in the context of modern
computational architectures while illuminating its efficacy in
addressing highly multimodal problems. In \cref{sec:B:Stat:Inv} we
present a series of examples in the framework of Bayesian statistical
inverse problems.  In order to provide some preliminary studies of
\cref{alg:mpCN} we focus on the case of Gaussian priors for our
unknown parameter so that we are resolving target measures of the form
\eqref{mu:mpCN}.  Here two of our examples confront PDE constrained
problems. These problems are naturally high (infinite) dimensional
featuring complex correlation geometry and multimodal structure and
where, particularly in our second example, gradient based methods may
be impractical or costly to employ.

\subsection{High-performance computing}\label{sec:ppMCMC}

\begin{figure}[ht!]
	\centering
	\includegraphics[width=\linewidth]{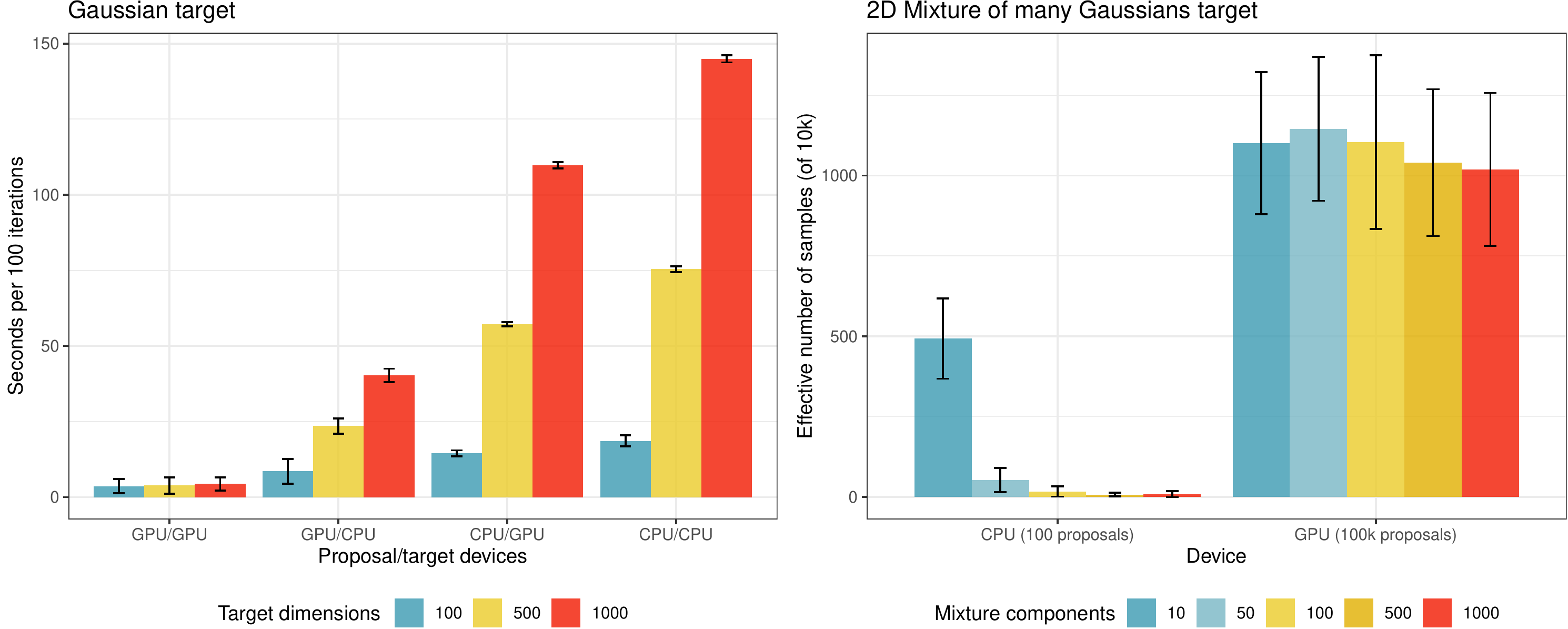}
	\caption{Parallelization study displaying means and 95\%
          confidence intervals. Left: we use a graphics processing
          unit (GPU) to accelerate embarrassingly parallel proposal
          and/or target evaluation steps for \cref{alg:TJ:cor:FD}
          and compare to conventional central processing unit (CPU)
          implementation. Here, we fix proposal count to
          100,000. Right: we demonstrate how parallelization across
          many target components can allow for more proposals when
          working with a fixed timing budget.  Here, GPU uses 100,000
          proposals while CPU uses 100.  Crucially, both take roughly
          the same amount of time per iteration.}\label{fig:par1}
\end{figure}

The general structure of parallel MCMC algorithms leverages parallel
computing resources that are becoming increasingly available and easy
to use. Whereas \cite{holbrook2023quantum} considers implementation of
parallel MCMC algorithms using quantum computers, here we consider
parallelization using conventional parallel computing resources. We
first compare the efficiency of sequential and massively parallel
implementations of \cref{alg:TJ:cor:FD} using an Intel Xeon CPU with 8
cores and 26GB RAM and an Nvidia Tesla P100 GPU with 3,584 cores, 16GB
RAM and 730 GB/s memory bandwidth. Since \cref{alg:TJ:cor:FD} enables
simplified acceptance probabilities \cite{holbrook2023generating}, its
two main computational bottlenecks are the generation of $p$ proposals
$\bq_j$ and the calculation of $p$ target values $\pi(\bq_j)$. Within
the \textsc{TensorFlow} \cite{abadi2016tensorflow} framework, the
\textsc{random} module accelerates the former task with GPU powered
implementations of the parallel PRNG algorithms of
\cite{salmon2011parallel}.  For the latter task, the
\textsc{Distributions} library \cite{dillon2017tensorflow} enables GPU
powered implementations of popular probability density functions that
parallelize across evaluation arguments and density components when
appropriate. All experiments use the isotropic Gaussian proposal
kernel $Q(\bq, d \btq)$ with variance tuned to obtain a 50\% acceptance
rate.

\cref{fig:par1} shows results from two parallelization experiments
that demonstrate the speedups associated with such GPU implementations
compared to sequential implementations on a CPU. In the first
experiment, we record the number of seconds required for 100
iterations of \cref{alg:TJ:cor:FD} using 100,000 proposals with
parallelized proposal and/or acceptance steps for centered/isotropic
multivariate Gaussian targets of 100, 500 and 1000 dimensions. In
general, higher-dimensional targets require more computations, but the
burden of dimensionality is significantly greater for CPU
implementations. We find that both proposal and acceptance steps
benefit from parallelization but that parallelizing proposals is more
beneficial at these scales. In the second experiment, we maintain a
fixed time budget for parallel MCMC implementations that use CPU (with
100 proposals) or GPU (with 100,000 proposals) for both proposal and
acceptance steps.  The target distributions are 2-dimensional mixtures
of 10, 50, 100, 500 and 1000 standard Gaussians centered at
$\boldsymbol{0}$, $\boldsymbol{10}$, $\boldsymbol{20}$,
etc.~(\cref{fig:deathMixture}). We monitor effective sample sizes
after 10,000 MCMC iterations.  In this context, the target evaluations
scale linearly in the number of mixture components, but the
\textsc{Distributions} library enables parallelization across
proposals and mixture components.  The upshot is a thousand-fold
increase in the number of proposals when operating under a fixed time
budget, and this increase in proposals leads to significantly better
exploration of the heavily multimodal targets.

\begin{figure}[ht!]
	\centering
	\includegraphics[width=0.9\linewidth]{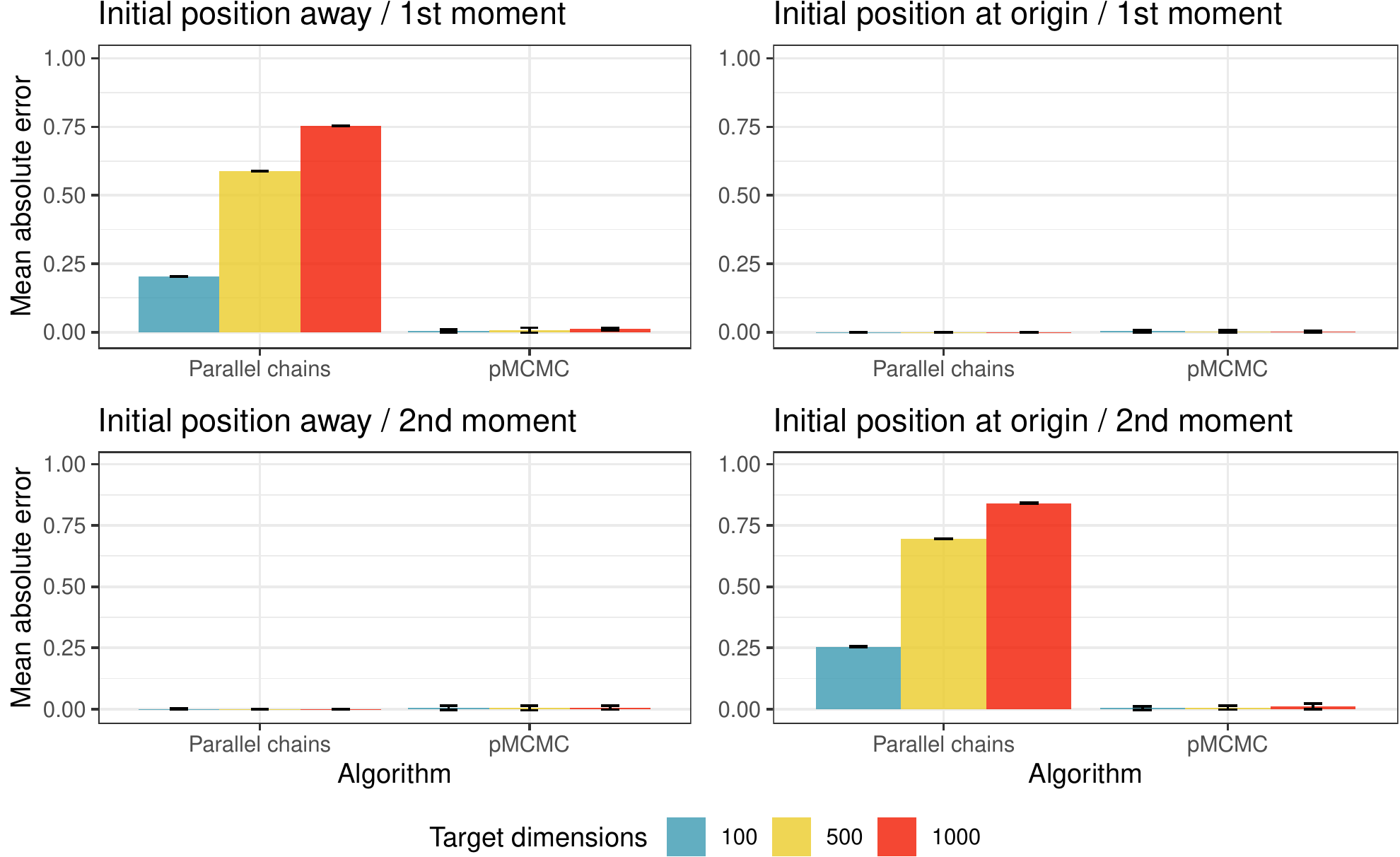}
	\caption{Means and 95\% confidence intervals for mean absolute
          estimation error from 10k parallel Metropolis-Hastings
          chains versus a single parallel MCMC chain produced by \cref{alg:TJ:cor:FD} using 100k
          proposals. Both target isotropic Gaussians of 100, 500 and
          1000 dimensions using 10k MCMC iterations. Estimator from
          10k parallel chains demonstrates adverse dependence on
          starting point: starting at distribution mean
          $\boldsymbol{0}$ harms 2nd moment estimator, but starting at
          $\boldsymbol{1}$ harms 1st moment estimator. The difference
          between chain and proposal counts arises from memory
          constraints on the multi-chain approach and does not
          appreciably change results.}\label{fig:par2}
\end{figure}
 
 \begin{figure}[ht!]
	\centering
	\includegraphics[width=0.9\linewidth]{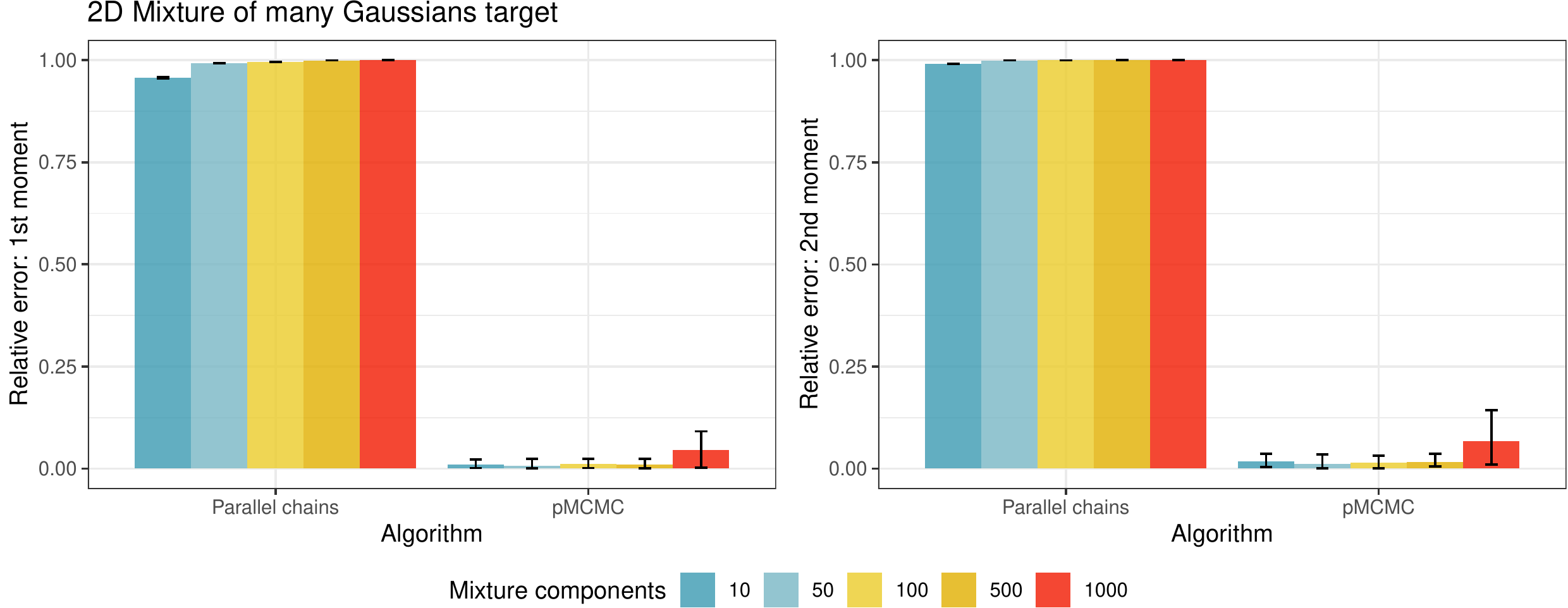}
	\caption{Means and 95\% empirical intervals for mean absolute
          estimation error from 10k parallel Metropolis-Hastings
          chains versus a single parallel MCMC chain produced by \cref{alg:TJ:cor:FD} using 10k
          proposals. Both algorithms target mixtures of isotropic
          Gaussians using 10k MCMC iterations. To facilitate
          comparison between errors associated with the estimation of
          first and second moments, we divide mean absolute errors by
          true moment values.}\label{fig:par3}
\end{figure}
 
Next, we compare the same parallel MCMC algorithm to the massively
parallel implementation of independent random walk Metropolis Markov
chains to see if the latter, embarrassingly parallel approach is
sufficient. Importantly, we implement both strategies within
\textsc{TensorFlow} using the same GPU. An immediate difference
between the two approaches is that the memory burden for independent
chains quickly accrues. For example, collecting $S$ states each from
$C$ chains with $D$ dimensional targets requires $O(SCD)$ memory. In
high-dimensional settings with thousands of chains, we therefore find
it necessary to severely thin the chains, saving a relatively small
number of states.  \cref{fig:par2} shows results from just such a
comparison, targeting centered, isotropic and high-dimensional
Gaussians. We run 10,000 parallel chains but find it necessary to
remove 99 states for every 100.  An adverse dependence of Monte Carlo
estimator accuracy on initial positions demonstrates that the act of
combining such a large number of chains fails to make up for the
weaknesses of individual chains despite their being tuned for optimal
scaling \cite{gelman1997weak}. On the other hand, the parallel MCMC
algorithm achieves accuracy when run for the same number of
iterations.  \cref{fig:par3} shows similar results using the
2-dimensional mixture of many Gaussians targets from the previous
experiment (\cref{fig:par1}).  Thanks to the low-dimensional nature of
the problem, we are able to save all 10,000 states of 10,000 parallel
chains, but this approach fails nonetheless. A harbinger of this
failure is that the parallel chains achieve a potential scale
reduction factor of 2.30 \cite{gelman1992inference}. Again, the
parallel MCMC algorithm performs well for the same number of MCMC
iterations and, free from memory constraints, is able to use as many
as 100,000 proposals within each iteration.

\subsection{Statistical inversion with infinite-dimensional unknowns}
\label{sec:B:Stat:Inv}

We next present three case studies involving target measures arising
from ill-posed inverse problems framed in the setting of Bayesian
statistical inversion, \cite{kaipio2006statistical, stuart2010inverse,
  dashti2017bayesian}.  Here we are particularly focused on 
  PDE constrained
problems with infinite-dimensional unknowns with a particular emphasis
on fluid measurement problems in two different settings we advocated
recently in \cite{borggaard2020bayesian,
  borggaard2023statistical}.
As noted above, while such problems must be truncated to fit on a computer, study in the full infinite-dimensional setting has yielded MCMC algorithms with beneficial convergence results in the high-dimensional limit \cite{beskos2008mcmc,cotter2013mcmc,Beskosetal2011,hairer2014spectral}.

Before delving into the specific details of each of the three problems
let us first recall some generalities of the statistical inversion
framework.  Suppose we are trying to recover an unknown parameter
$\bq$ sitting in a parameter space $\spq$, typically a separable
Hilbert space.  Our parameter is observed through a forward model
$\G : \spq \to \RR^k$ and is subjected to an additive observational noise
$\noise$.  Namely we would like to estimate $\bq \in \spq$ from the
observation model
 \begin{align}
 	\Data = \G(\bq) + \noise.
	\label{eq:stat:inv}
 \end{align} 
 By placing a prior probability $\mu_0$ on $\spq$ for our unknown
 $\bq$ and assuming that $\noise$ is continuously distributed with
 density $p_\noise$, Bayes theorem (see \cite{dashti2017bayesian},
 \cite{borggaard2020bayesian}) uniquely determines the posterior
 probability measure for $\bq$ given observed data $\Data$ as
\begin{align}
  \mu(d \bq)
  := \frac{1}{Z} p_\noise( \Data - \G(\bq)) \mu_0(d \bq),
  \quad \text{ where }
  Z = \int_\spq p_\noise( \Data - \G(\btq)) \mu_0(d \btq).
	\label{eq:stat:inv:sol}
\end{align}
We view this posterior $\mu$ as an optimal solution of the ill-posed
inverse problem \eqref{eq:stat:inv}.

As noted above, we are interested in estimating unknown parameters
appearing in a partial differential equation.  Typically we consider
the situation where
\begin{align}
  \G = \Obs \circ \Sol,
  \text{ where $\Sol$ is a `PDE parameter to solution map' and $\Obs$ is an
  `observation operator'.}
  \label{eq:forward:map:decomp}
\end{align}
This second operator $\Obs$ projects the PDE solution down to a
finite-dimensional space.  Notice that in the natural case of a
Gaussian prior $\mu_0$ and e.g. Gaussian observational errors
$\noise_j$, \eqref{eq:stat:inv:sol} takes the form
\eqref{eq:stuart:form} which falls in the scope of \cref{alg:mpCN}

\subsubsection{A finite-dimensional toy model}
\label{sec:AD:Toy:Model}

Our first statistical inverse problem has a finite-dimensional
unknown parameter space, but was designed as a toy model for much
more complex and computationally involved PDE inverse problems we
consider below in \cref{sec:AD:JUQ} and \cref{sec:rot:Stokes}.  Our goal
here is to have a `table-top' statistical experiment which mimics some
of the structure (and indeed produces statistics reminiscent of) these
more involved problems but which can be fully resolved on a laptop in a
matter of hours.

\smallskip
\noindent\textbf{Problem specification}
\smallskip

We let $\spq$, our unknown parameter space, be the collection of skew
symmetric matrices acting on $\RR^d$ so that
$m :=\dim(\spq) = d(d -1)/2$.  We then take
$\Sol = \Sol_{\bg, \kappa}: \RR^m \to \RR^d$ to be the solution of
 \begin{align}
 	(\param + \kappa I) \bx = \bg,
	\label{eq:AD:toy:gen}
 \end{align}
 namely $\bx(\param) = \Sol(\bq) := (\param + \kappa I)^{-1} \bg$.  Here
 $\kappa >0$ and $\bg = (g_1, \ldots, g_d) \in \RR^d$ are fixed and
 known parameters in our model.  Regarding the observation procedure
 $\Obs: \RR^d \to \RR^k$ we consider resolving some of the components of
 $\Sol(\bq) = (\Sol(\bq)_1, \ldots , \Sol(\bq)_d)$, namely
 \begin{align*}
   \Obs(\Sol(\bq))
   = (\langle\Sol(\bq), \bz_1 \rangle, \ldots, \langle\Sol(\bq), \bz_k \rangle)
 \end{align*}
 for some directions $\bz_1, \ldots, \bz_k \in \RR^d$.  $\Obs$ could
 be taken to be nonlinear in general, for example in the case of the
 norm $\Obs(\Sol(\param)) = \| \Sol(\param)\|$.  Note that
 \eqref{eq:AD:toy:gen} explicitly mimics some of the structure of a
 steady state version of the advection-diffusion problem,
 \eqref{eq:ad:eqn}, we consider further on below.

Turning now to the specifics of the numerical study carried out here
we let $d = 4$ so that the dimension of the unknown
parameter space is $\mbox{dim}(\spq) = 6$.  Regarding the
observational noise we assume $\noise \sim N(0, \sigma^2_\noise I_k)$.  We
observe the first two components of $\Sol(\param)$ (i.e $k =2$) so
that our target measure from \eqref{eq:stat:inv:sol} takes the form
\begin{align}
  \mu(d \bq)
  = \frac{e^{-\Phi(\bq)}}{Z} \mu_0(\bq), \quad \text{ where } \quad
  \Phi(\bq) := \frac{1}{2 \sigma^2_\noise}
  \left( (y_1 - \Sol(\bq)_1)^2
        +  (y_2 - \Sol(\bq)_2)^2\right)
	\label{eq:AD:toy:Ex}
\end{align}
where the prior $\mu_0$ is a centered Gaussian with a diagonal
covariance specified by two parameters $\sigma \in \RR$ and
$\gamma > 0$ as
\begin{align*}
  \mu_0 = N(0,
  \diag(\sigma^2, \sigma^2 2^{-\gamma}, \sigma^2 3^{-\gamma},
      \sigma^2 4^{-\gamma},  \sigma^2 5^{-\gamma}, \sigma^2 6^{-\gamma})).
\end{align*}
Regarding the choice of parameters defining \eqref{eq:AD:toy:Ex} we
selected
\begin{align}
  \sigma^2_\noise = 2, \quad
  y_1 = 4.601,\quad 
  y_2 = 18.021, \quad
  \sigma^2 = 5, \quad
  \gamma = 1.5,
  \label{eq:AD:toy:Ex:Prm1}
\end{align}
and to specify $\Sol$ from \eqref{eq:AD:toy:gen} we considered
\begin{align}
	\kappa =  .1, \quad
	\bg = (0,0,5,2).
	  \label{eq:AD:toy:Ex:Prm2}
\end{align} 
Note that, regarding the selection of $y_1$ and $y_2$, we found a
value of $\bar{A}$ through a Monte Carlo search such that
$\bx(-\bar{A})_1 \approx \bx(\bar{A})_1 = y_1$ and
$\bx(-\bar{A})_2 \approx \bx(\bar{A})_2 = y_2$ up to an error of
approximately $.09$.  However, from a preliminary, non-systematic,
exploration of the parameter space, we experimentally found covariance
structures which were broadly similar topologically with regard to the
covariance structures appearing in the resulting target measures
across a variety of different parameter values in comparison to those
chosen above.

\smallskip
\noindent\textbf{Numerical results}
\smallskip

\begin{figure}[!ht]
	\centering
	\includegraphics[width=.8\linewidth]{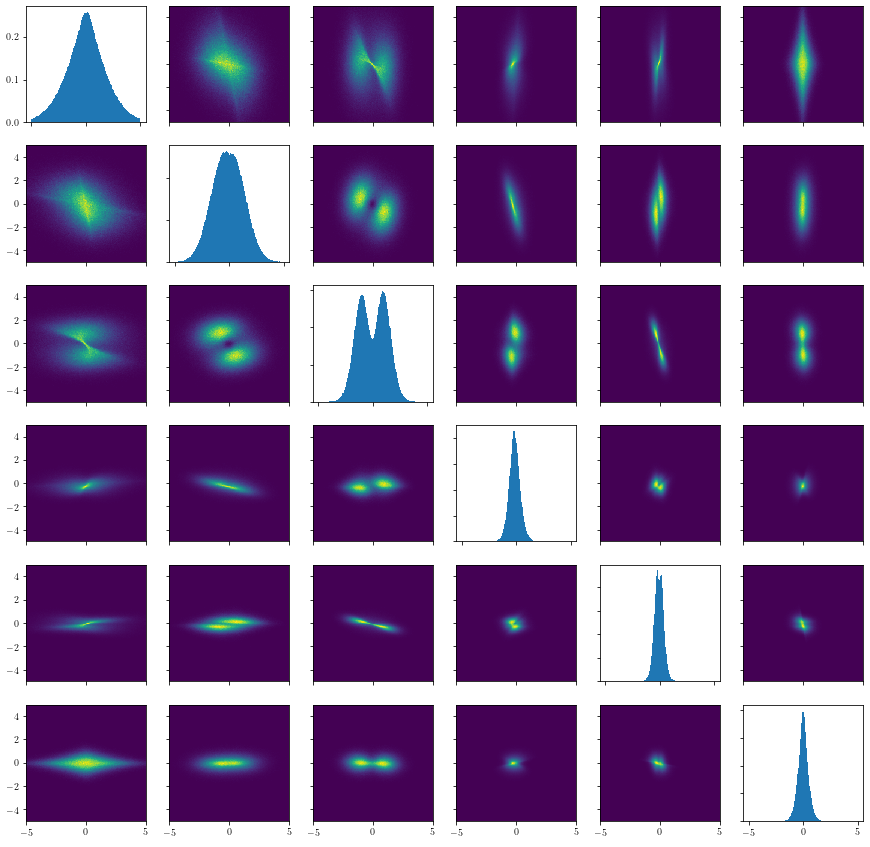}
	\caption{Histogram of the one and two dimensional marginals of
          $\mu$ defined by \eqref{eq:AD:toy:Ex} relative to the
          parameters \eqref{eq:AD:toy:Ex:Prm1},
          \eqref{eq:AD:toy:Ex:Prm2}. While relatively low-dimensional,
          this target measure provides an interesting test case
          exhibiting both ill-conditioned geometry and
          multimodality. In fact, two different flavors of
          multimodality are exhibited: the dimensions 2-3 marginal
          exhibits energetic multimodality while the dimensions 1-3
          marginal exhibits entropic
          multimodality.}\label{fig:ADToy_True_Hist}
\end{figure}

As a benchmark we fully resolved \eqref{eq:AD:toy:Ex} using a
traditional pCN sampler with $10^8$ samples taking the algorithmic
parameter $\rho = .99$ which yielded an approximate acceptance
ratio of $.22$.  As a sanity check we also ran mpCN out to $10^7$
samples under the algorithmic parameter $\rho = .6$, $p =100$.  These
two benchmark runs produced very similar statistics.  A histogram
representing the one and two dimensional marginals of the fully
resolved target are pictured in \cref{fig:ADToy_True_Hist}.

\begin{figure}[!htb]
	\centering
	\includegraphics[width=\linewidth]{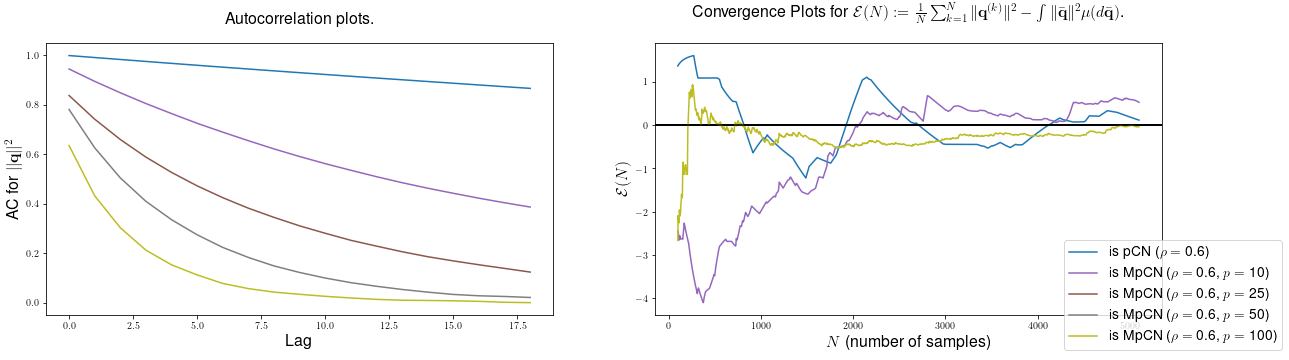}
	\caption{Two different illustrations of improving per-sample
          performance for multiproposal preconditioned Crank-Nicolson (mpCN, \cref{alg:mpCN}) for the observable
          $\Psi(\bq) = \|\bq\|^2$.  The left panel provides
          autocorrelation plots while the right panel measures the
          convergence of ergodic averages.}\label{fig:btr:p:perfm}
\end{figure}

\begin{figure}[!htb]
	\centering
	\includegraphics[width=\linewidth]{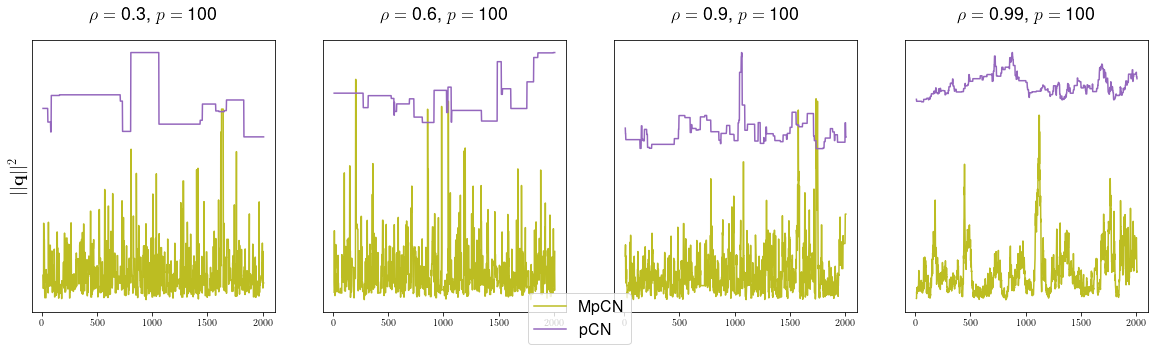}
	\caption{Time series plots comparing multiproposal preconditioned Crank-Nicolson  (mpCN, \cref{alg:mpCN}) with $p = 100$ and pCN for various values of $\rho$ for the observable $\Psi(\bq) = \|\bq\|^2$.}\label{fig:AD:TS:Plots}
\end{figure}

\begin{figure}[!htb]
	\centering
	\includegraphics[width=\linewidth]{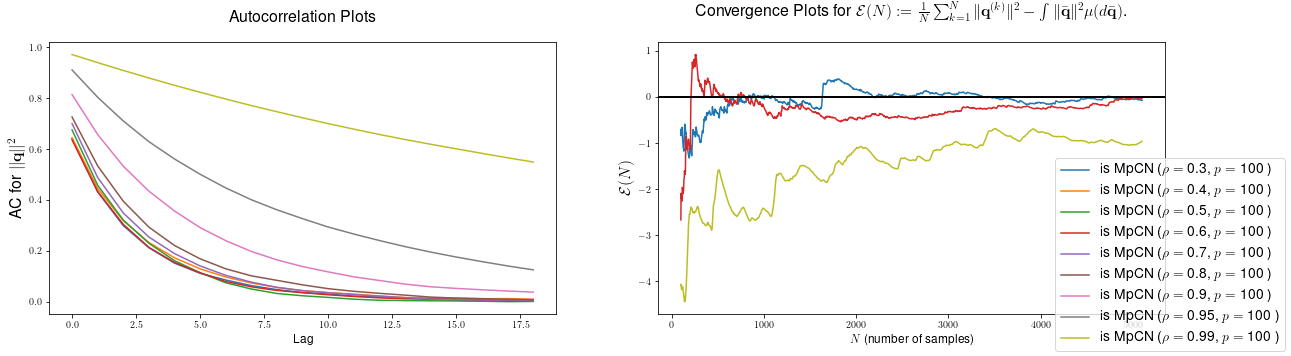}
	\caption{Convergence plots comparing multiproposal preconditioned Crank-Nicolson  (mpCN, \cref{alg:mpCN}) with $p = 100$ and pCN for various values of $\rho$ over different observables.}\label{fig:AD:conv:plt}
\end{figure}

To provide an assessment of the efficacy of mpCN (\cref{alg:mpCN}) for this model
\eqref{eq:AD:toy:Ex} we ran mpCN out to $10^5$ samples over 
a range of algorithmic parameter values $\rho \in [0,1]$ and $p \geq
1$ namely:
\begin{align*}
	\rho \in \{ .3, .4, .5, .6, .7, .8, .9, .95, .99\}
	\quad
	p  \in \{10, 25, 50, 100\}.
\end{align*}
Unsurprisingly performance improved on a per sample basis for fixed $\rho$
as $p$ increases as illustrated in \cref{fig:btr:p:perfm}.  
An unambiguous heuristic for the selection of an optimal choice for
$\rho$ for a given value of $p$ remains unclear.  What is evident from
our numerical experiments is that, firstly, $\rho$ should be chosen to
be rather more aggressive in comparison to its single proposal pCN
counterpart. Secondly, it appears that effective chains are produced
over a fairly robust range of values for $\rho$.  These observations
are illustrated in \cref{fig:AD:TS:Plots}, \cref{fig:AD:conv:plt}.  \arxiv{For further
  illustrations of these points see \cref{fig:AD:conv:plt:EXT},
  \cref{fig:AD:TS:Plots:EXT} and \cref{fig:AC:Conv:rho:for:p100}  in
  \cref{sec:appndx:sup:mat} below.}

\subsubsection{Estimation of velocity fields from a passive solute}
\label{sec:AD:JUQ}

Our second model problem, developed previously in
\cite{borggaard2020bayesian}, involves the estimation of a time
independent, divergence free vector field $\bq$ from the sparse
measurement of a solute passively advected and diffusing in the fluid
medium. By choosing observations that leverage symmetries in the
problem, we generate a posterior measure with complex correlation and
multimodal structures that make efficient sampling difficult.

\smallskip
\noindent\textbf{Problem specification}
\smallskip

The basic physical model for this situation is the
advection-diffusion equation
\begin{align}
  \pd_t \theta + \bq \cdot \nabla \theta = \kappa \Delta \theta,
  \label{eq:ad:eqn}
\end{align}
where $\theta$ represents the concentration of the solute.  We
consider \eqref{eq:ad:eqn} in the case of a periodic two dimensional
domain $\TT^2 = [0,1]^2$ with a known initial condition
$\theta(0, x) = \theta_0(x)$ (i.e. initial solute concentration) and
diffusivity parameter $\kappa > 0$.

To develop concrete examples in this setting we need to specify the
prior for the velocity field $\bq$, the solute observation procedure
$\Obs$ in \eqref{eq:forward:map:decomp} and the observation noise
structure $\noise$ in \eqref{eq:stat:inv}.  We consider the unknown
parameter space as the Sobolev space of divergence free, mean free
vector fields (see e.g. \cite{temam1995navier}) defined as
\begin{align}
  H^s_{div} := \left\{ \bq : \TT^2 \to \RR^2 : \bq(\bx) :=\sum_{ \bk \in \ZZ^2} q_\bk
  \frac{\bk^\perp}{\|\bk\|_2} e^{2\pi i \bk \cdot \bx},
  \bar{q}_\bk = - q_{-\bk}, \|\bq\|_{H^s} < \infty \right\}
\end{align}
where $q_\bk$ are complex numbers with $\bar{q}_\bk$ representing
the complex conjugate and 
\begin{align}\label{eq:ad:unk}
  \|\bq\|_{H^s}^2 := \sum_{ \bk \in \ZZ^2} \| \bk\|^{2s} |q_\bk|^2.
\end{align}
The Sobolev spaces $H^s$ of mean free scalar fields are defined
analogously.  When $s > 1$ we have that
$\Sol(\bq, \theta_0) = \theta(\bq, \theta_0)$, the solution of
\eqref{eq:ad:eqn} corresponding to any $\bq \in H^s_{div}$ and
$\theta_0 \in H^s$, is an element in $C([0,T], H^s)$ i.e. the
continuous functions from $[0,T]$ taking values in $H^s$.  Note that
this solution map $\Sol$ depends continuously on $\bq, \theta_0$ in the
standard topologies; see \cite{borggaard2020consistency} for a proof
of these well-posedness claims.

Regarding the prior $\mu_0$, we consider a centered Gaussian with covariance
operator defined as (see \cref{sec:mpCN} above):
\begin{align}
  \cC \bq
  = \sum_{ \bk \in \ZZ^2} E_\bk\left(
  \int_{\TT^2} \bee_\bk(\bx) \cdot \bq(\bx) d\bx \right) \bee_\bk
  \quad
  \text{ where }
  E_\bk = \frac{E(\| \bk\|)}{2\pi \| \bk\|}.
  \label{eq:cov:Op:AD}
\end{align}
Here $\bee_{\bk} (\bx) := \frac{ \bk^\perp}{\|\bk\|} \exp(2 \pi i \bk
\cdot \bx)$ and
\begin{align}
  E(k) = E_0 \sum_{j = 0}^N \left( \frac{k}{2^{j/2}} \right)^4
  \exp\left( - \frac{3 k^2}{2^{i+1}} \right) 2^{- j\xi/2}.
  \label{eq:Ks:turb:mdl}
\end{align}
The form of $E$ in \eqref{eq:Ks:turb:mdl} determines the `energy
spectrum' we are placing on the prior for the unknown velocity field
$\bq$ and was inspired by a turbulence model due to Kraichnan; see
\cite{kraichnan1968small, kraichnan1991stochastic}.  The model is
specified by the three parameters, $E_0$, $\xi$, $N$ where $E_0$ is
related to the overall energy of turbulent flow, $N$ determines the
range of scales $k \in [0, 2^N]$ in an `inertial range' where $\xi$ is
the exponent in a power law decay over this range.  In our example we
specify
\begin{align}
  E_0 = 1, \quad N = 20, \quad \xi = 3/2.
  \label{eq:cov:param}
\end{align}
To make the sampling computationally tractable, we truncate the unknown \eqref{eq:ad:unk} to $0 < \|\bk\| \le 8$, yielding a 196-dimensional sample space.
\arxiv{See \cref{fig:prior:AD:model} below in
  \cref{sec:appndx:sup:mat} for a visualization of a typical draw from
  this prior.}

Regarding the observational data specifying our posterior measure we
exploit symmetries in the parameter to solution maps specified by
\eqref{eq:ad:eqn} to obtain a posterior with a complex, highly
non-Gaussian structure\arxiv{; see \cref{fig:ad:hist2dtrue} below
  and \cite{borggaard2020bayesian} for further details}.  We consider
\begin{align*}
  \theta_0(x,y) = \frac{1}{2} - \frac{1}{4}\cos(2\pi x) - \frac{1}{4}
  \cos(2\pi y),
  \quad \bq^*(x,y) = [8 \cos( 2\pi y), 8 \cos(2 \pi x)],
\end{align*}
and notice that
\begin{align*}
  \theta(t, \bx_i, \bq^*,\theta_0) = \theta(t, \bx_i, -\bq^*,\theta_0)
  \quad \text{ for any } t > 0 \text{ and } i = 1,2 \text{ where }
  \bx_1 = [0,0],  \bx_2 = [\tfrac{1}{2}, \tfrac{1}{2}].
\end{align*}
With this in mind we take as our data
\begin{align*}
  \mathcal{Y} := \{ \theta(t_k, \bx_i, \bq^*,\theta_0)  \}_{i = 1,2, \, t_k =
  0.001 k, \, k =1,\ldots, 50}
\end{align*}
Regarding the observation noise we consider the situation at which
each data point is subject to an independent and identically distributed
Gaussian error $N(0, \sigma_\noise)$ with $\sigma_\noise = 2^{-3}$.

\smallskip
\noindent\textbf{Numerical results}
\smallskip

\begin{figure}[h]
    \centering
    \includegraphics[width=0.48\textwidth]{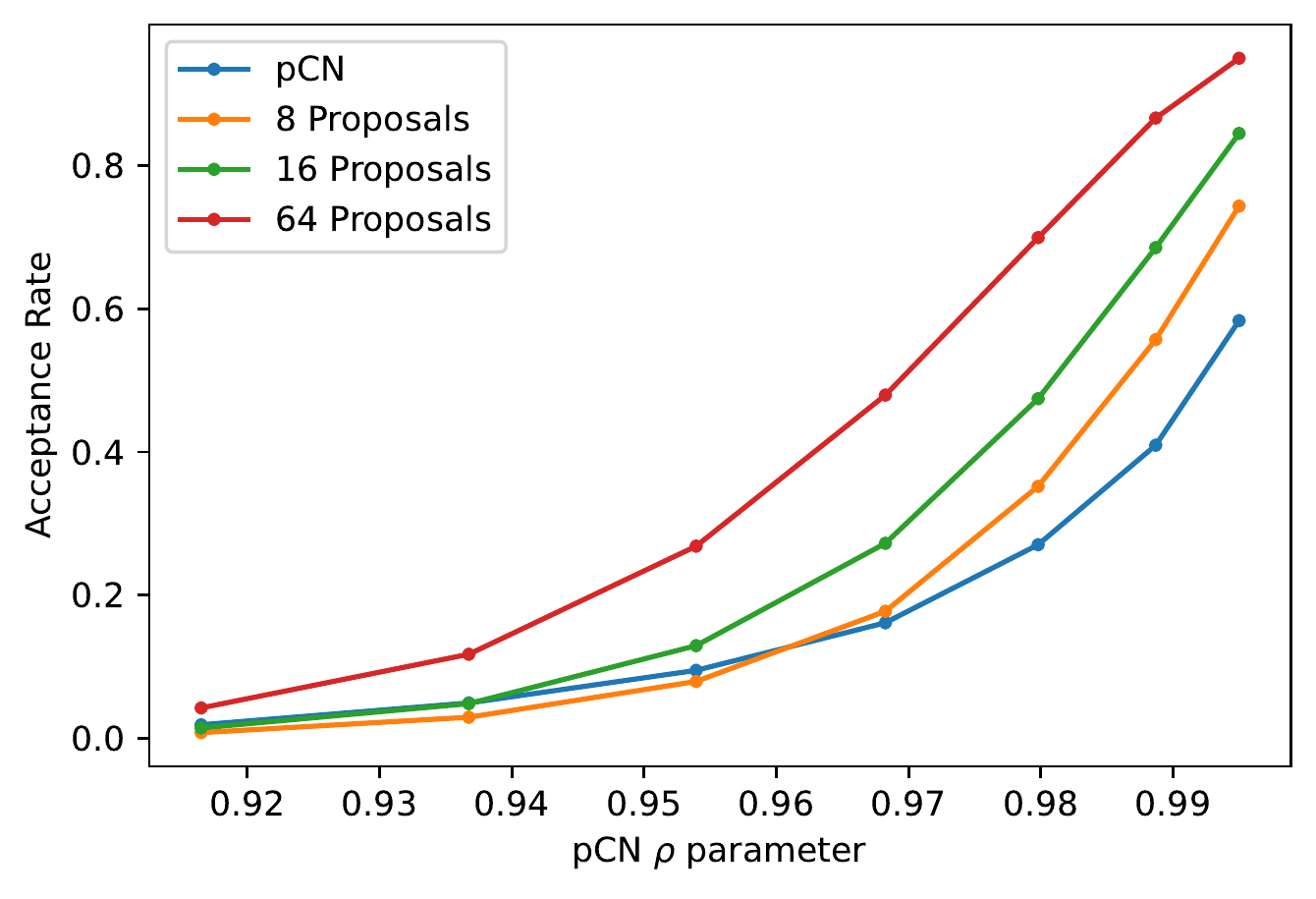}
    \hfill
    \includegraphics[width=0.48\textwidth]{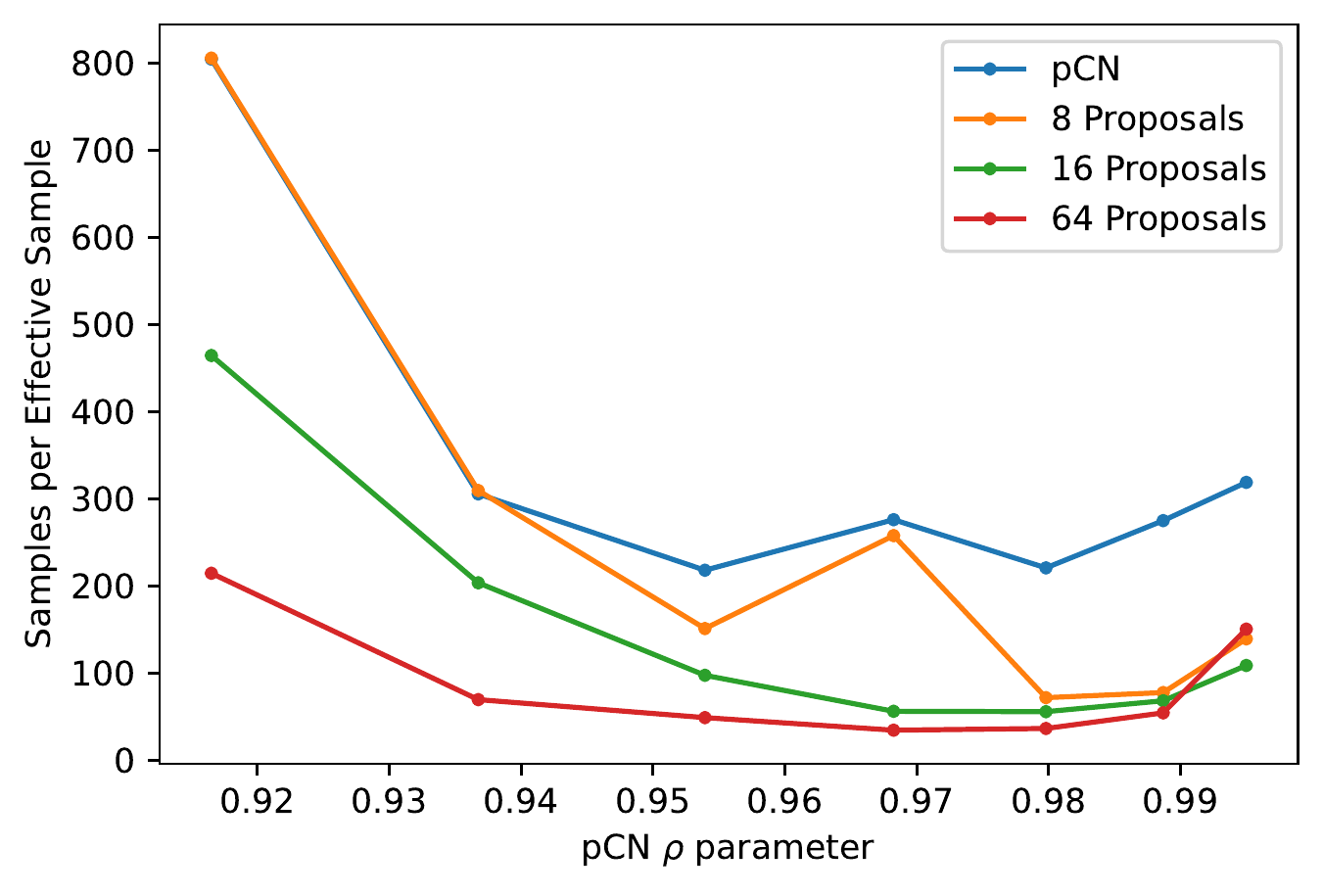}
    \caption{Sweep of tuning parameters $\rho$ and $p$ for the
      advection-diffusion problem with 250,000 samples (average values across two chains) generated using multiproposal preconditioned Crank-Nicolson  (mpCN, \cref{alg:mpCN}) and pCN. Left:
      Acceptance rate. Right: Samples per effective sample of the
      (unnormalized) posterior density (lower is better).}
    \label{fig:ad:ar:ess}
\end{figure}

\cref{fig:ad:ar:ess} shows the results of a parameter study
for mpCN, \cref{alg:mpCN}, in $\rho$ and in $p$ the number of
proposals. Two $250,000$-sample chains were run for several $\rho$
values ranging from very aggressive step sizes (low $\rho$) to very
conservative step sizes (high $\rho$) for the original ``vanilla'' pCN
algorithm and for multiproposal pCN with 8, 16, and 64 proposals. The
figure shows the average across these two chains for each case. The
plot on the left shows the acceptance rate for the chains; we see that
adding proposals of course increases the acceptance rate for most of
the range of $\rho$ values, although the effect diminishes as the
number of proposals grows large.

It is perhaps worth noting that the acceptance rate for $p=8$ was
lower than for vanilla pCN when $\rho$ was set to be very low
(overly-aggressive); likely this is because use of the Barkerized
acceptance ratio, \eqref{def:alphaj:pCN}, rather than the traditional
Metropolis-Hastings ratio increases the chance of rejection of a given
proposal. The right-hand plot shows the number of samples required to
achieve an effective sample -- i.e., the number of samples divided by
the effective sample size (ESS), where ESS is calculated according to
\cite[Section 11.5]{gelman2014bayesian} -- computed with the
unnormalized posterior density (the log-prior plus the log-likelihood)
as the estimand of interest and using the final $100,000$ samples from
each chain. The results show that using multiple proposals reduced the
number of samples required to generate a statistically independent
sample by nearly an order of magnitude, from roughly 200 samples in
the best cases for pCN to approximately 50 samples for 16 proposals
and 35 samples for 64 proposals, respectively. Also, when the step
size was well-tuned to an acceptance rate of $30-50\%$ even $8$ or
$16$ proposals seemed to be enough to substantially improve
performance according to this metric.

\begin{figure}[h]
    \centering
    \includegraphics[width=0.24\textwidth]{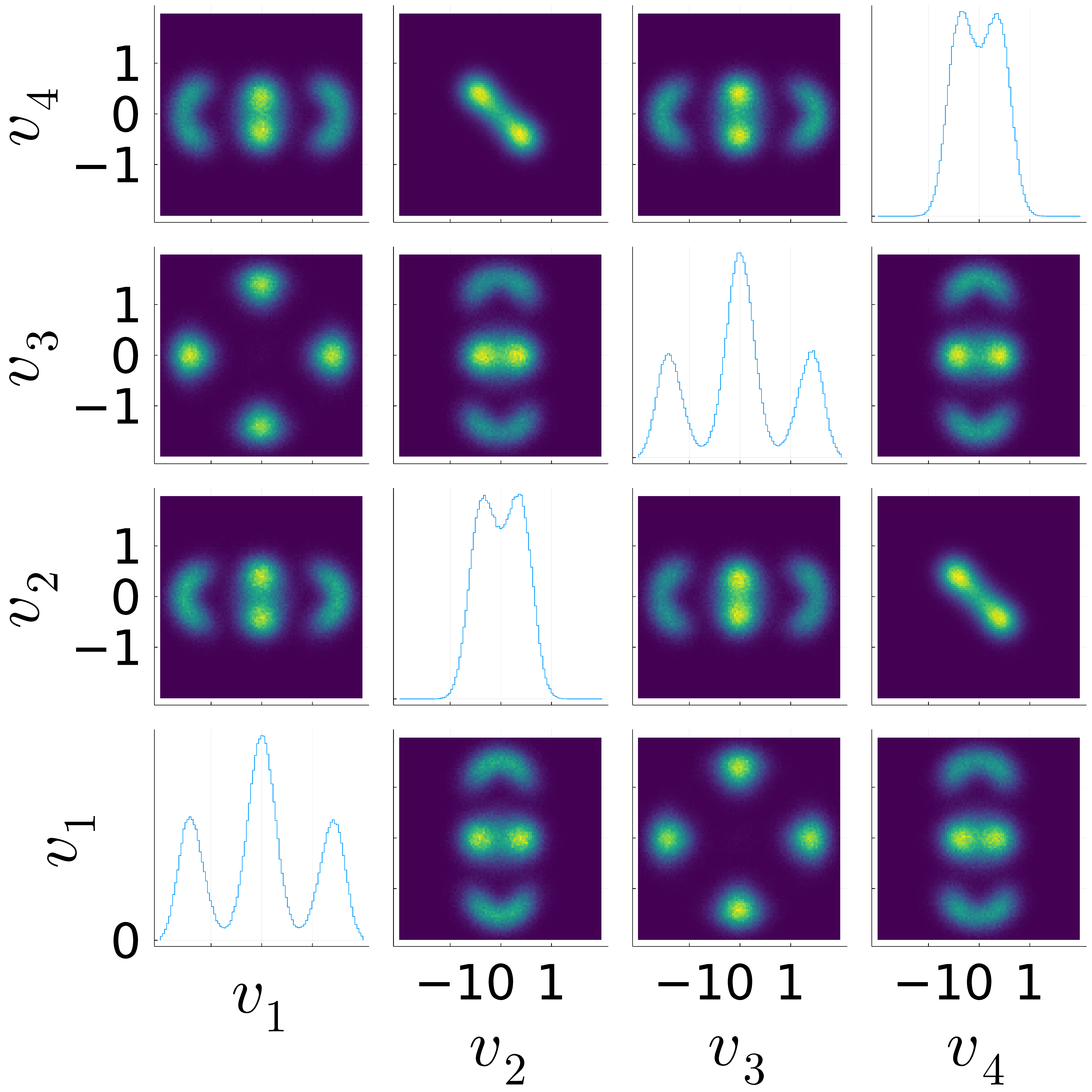}
    \hfill
    \includegraphics[width=0.24\textwidth]{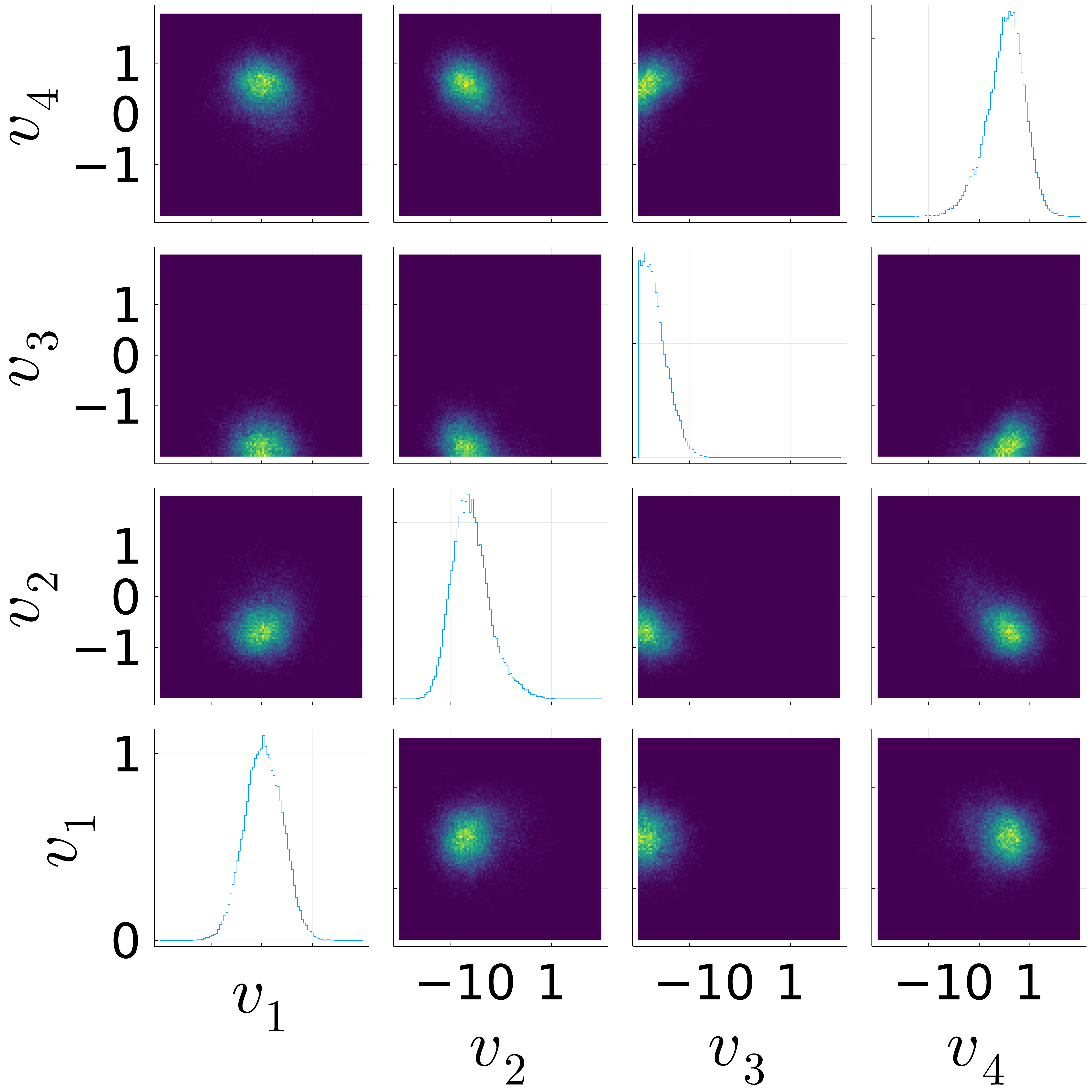}
    \hfill
    \includegraphics[width=0.24\textwidth]{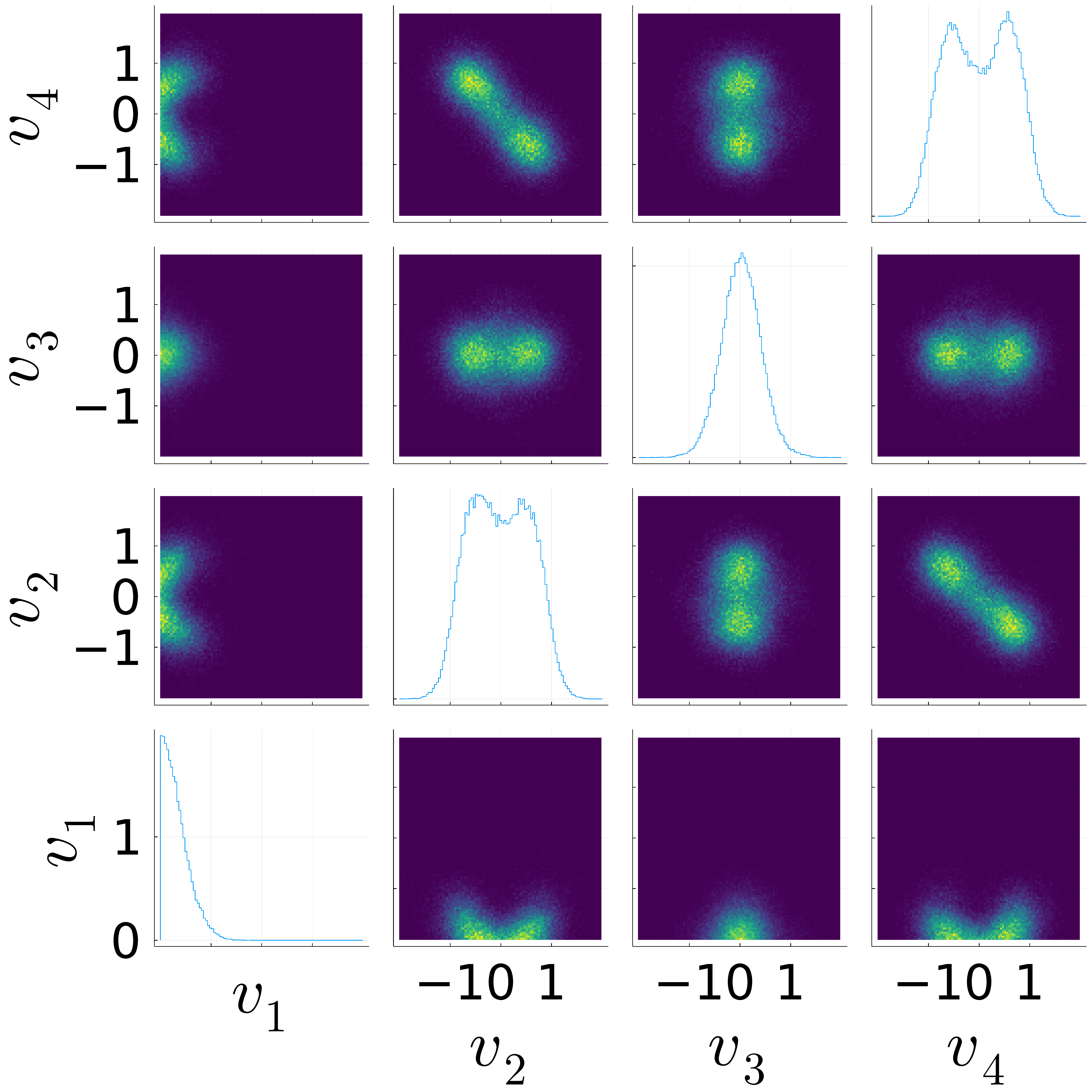}
    \hfill
    \includegraphics[width=0.24\textwidth]{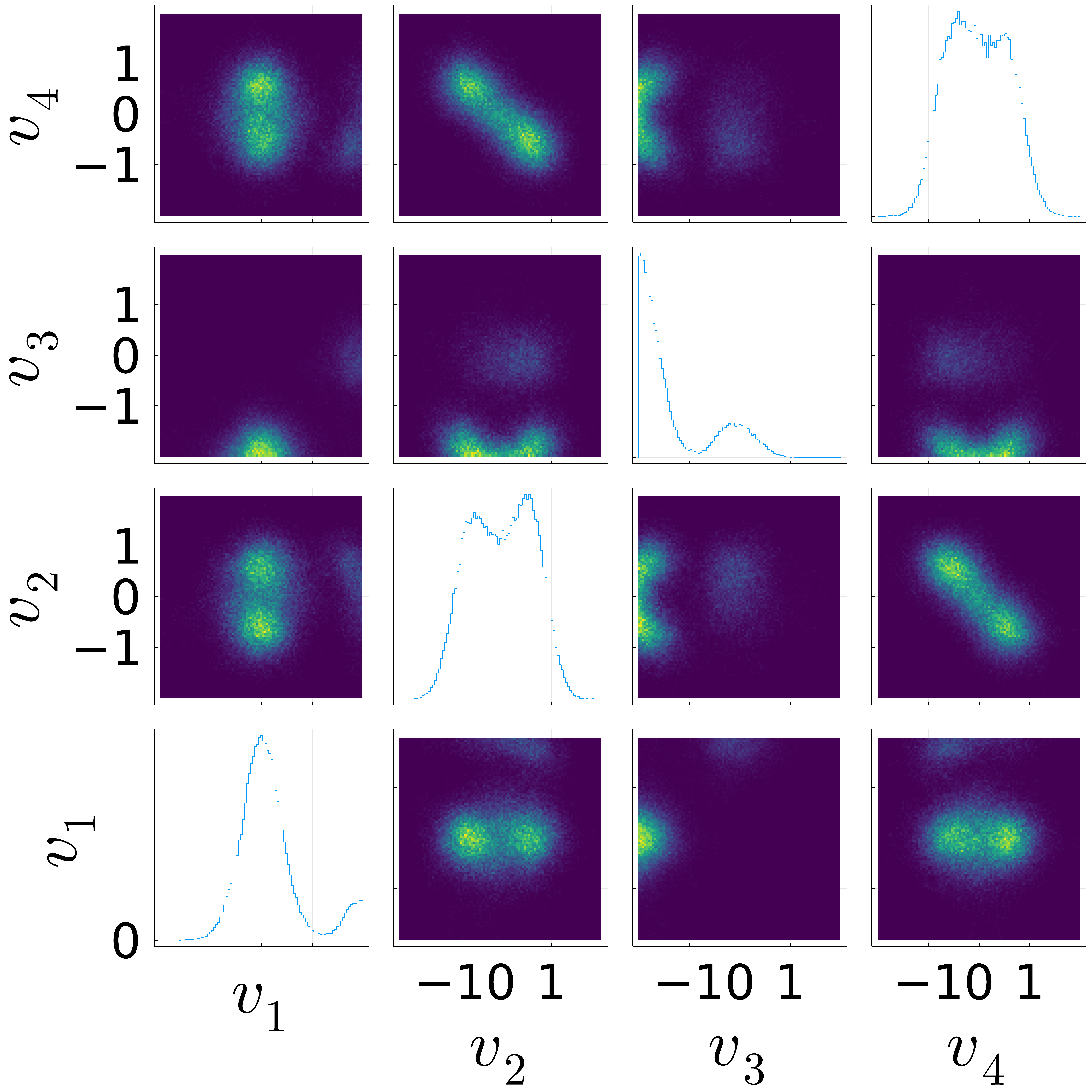}
    \caption{Posterior two-dimensional histograms for the first few
      components of the advection-diffusion problem. From left: (1)
      ``Truth'' from \cite{borggaard2020bayesian}; (2) preconditioned Crank-Nicolson (pCN)
      ($\rho=0.980$) after 250,000 samples; (3) and (4) two chains of
      multiproposal pCN (\cref{alg:mpCN}) ($\rho=0.968$, $64$ proposals) after 250,000
      samples.}
    \label{fig:ad:hist2d}
\end{figure}

Based on the results shown in \cref{fig:ad:ar:ess}, we selected
``optimal'' $\rho$ values of $0.980$ for vanilla pCN and $0.968$ for
multiproposal pCN with 64 proposals. \cref{fig:ad:hist2d} compares the
results of the two-dimensional histograms for the first few components
of the posterior measure for these $\rho$ values compared with the
``true'' posterior (computed via many long chains) from
\cite{borggaard2020bayesian}, which is shown in the plot on the
left. The result a pCN chain is show in the second figure from the
left and two 64-proposal multiproposal pCN chains are shown in the
figures on the right. We see that through 250,000 samples the pCN
chain has yet to explore the multimodal structure that characterizes
the posterior; a second pCN chain was run for this $\rho$ value and
yielded the same result. The two multiproposal pCN chains, by
contrast, appear to have fully explored the ``gentle'' multimodality
in the second and fourth components; one of the two chains has also
made the larger jump between modes in the first and third components,
although the balance between modes has not yet been achieved. These
results show the promise of multiproposal methods for exploring
complex posterior measures in high-dimensional spaces.

\subsubsection{Estimation of boundary shape characteristics in a
  rotating Stokes flow}
\label{sec:rot:Stokes}

Our final statistical inversion problem is a shape estimation problem
from \cite{borggaard2023statistical} wherein the unknown
parameter $\bq$ specifies the shape of a domain upon which a system of
PDEs is defined. The forward map therefore requires solving the PDEs
on an irregular domain, making the derivation of an appropriate
adjoint method a difficult task.  Furthermore, typical implementation
will involve the use of a meshing algorithm, often third-party and
black box to re-mesh the domain at each iteration step, ruling out use
of automatic differentiation algorithms. This problem is therefore a
natural fit for exploration of gradient-free methods such as
\cref{alg:mpCN}.

\smallskip
\noindent\textbf{Problem specification}
\smallskip
  
We specify our shape estimation problem in the statistical inversion
framework, \eqref{eq:stat:inv}, as follows. Start with the forward map
$\G$. Consider a steady (i.e. time-independent) Stokes flow coupled to
an advection-diffusion equation describing the equilibrium
concentration of solute sitting passively in this flow.  The governing
PDEs for this situation are
\begin{align}
  \nu \Delta \bU = \nabla p, \quad \nabla \cdot \bU = 0,
  \quad
    \bU \cdot\nabla \pS =  \kappa \Delta \pS + \src
  \quad \text{ in } \DD_\bq.
  \label{eq:AD:Stokesbulk}
\end{align}
The unknown parameter $\bq$ determines the shape of an annular domain
$\DD_\bq \subset \RR^2$ with a circular outer boundary on
$\Gamma_{\bq}^o$ and unknown inner boundary $\Gamma_{\bq}^i$. Here
$\bU: \DD_\bq \to \RR^2$, $p: \DD_\bq \to \RR$ represent the fluid
velocity and pressure and $\pS: \DD_\bq \to \RR$ is the concentration
of the solute.  The physical constants $\nu,\kappa >0$ are the
viscosity and diffusivity while $\src$ is a source continuously
injecting solute into the system.  Regarding boundary conditions for
our problem, we posit
\begin{align}
  \bU = \rtrt \bx^\perp \text{ on } \Gamma^o_\bq
  \quad  \text{ and } \quad 
  \bU = 0 \text{ on } \Gamma^i_\bq, \quad
   \pS = 0 \text{ on } \text{ on } \Gamma^o_\bq
  \quad  \text{ and } \quad 
  \nabla  \pS \cdot \bnh = 0 \text{ on } \Gamma^i_\bq,
  \label{eq:AD:bc}
\end{align}
that is the fluid is being rotated at its outer boundary with a rate
$\rtrt \in \RR$.  Here $\bnh$ is the outward normal to $\Gamma^i_\bq$
so that $\nabla \pS \cdot \bnh = 0$ represents an insulation condition
if we interpret $\theta$ as the temperature of the fluid.  The precise
values of the various physical parameters $\nu, \kappa, \rtrt$ as well
as the form of $\src$ which we used for our test example are given
in \cref{tab:stokes:parameters} below.

We specify our fluid domain $\DD_\bq$ as a function of $\bq$ as
follows.  We suppose that the parameter space $\spq$ is given by the
collection of $2\pi$ periodic functions with Sobolev regularity
$s > 0$ namely
\begin{align}
  \spq  = H^s := \left\{ \bq: [0,2\pi] \to \RR :
  \bq(x) :=
  \sum_{k=1}^\infty (a_{2k-1} \cos (k x) + a_{2k} \sin
  (k x)),
  \|\bq\|_{H^s} < \infty \right\}
  \label{eq:sob:sp}
\end{align}
where
\begin{align}
  \| \bq\|_{H^s}^2 := 
  \frac{1}{2}\sum_{k=1}^{\infty} k^{2s} (a_{2k-1}^2+a_{2k}^2).
    \label{eq:sob:sp:norm}
\end{align}
Here $s$ will determine the `degree of smoothness' of the inner
boundary $\Gamma^i_\bq$.  To define the map associating each
$\bq \in \spq$ its associated $\DD_\bq$, we let
\begin{align}
  \DD := \{\bx \in \RR^2 \,:\,  \ir < |\bx| \leq \oR\}.
  \label{eq:com:domain}
\end{align}
Our domains $\DD_\bq$ are defined so that they always contain the
subdomain
\begin{align}
  \DD^0 := \{\bx \in \RR^2 \,:\,  \mr \leq |\bx| \leq \oR\},
  \label{eq:com:sub:domain}
\end{align}
for some $0< \ir < \mr < \oR$.  We denote by
$(\pth, \pr) : \DD \to [0,2\pi) \times [\ir,\oR]$ the conversion to
polar coordinates.  To ensure that the inner radius does not decrease
below $\ir$ or extend too far into the domain, we next fix a
`clamping' function $\lnk: \RR \to (\ir, \mr)$ to be a smooth strictly
increasing function with
\begin{align*}
  \lim_{t \to -\infty} \lnk(t) = \ir,
  \quad
  \lim_{t \to \infty}\lnk(t) = \mr.
\end{align*}
To limit the effect of clamping on the shapes, we used quadratic
interpolation to develop a smooth $\lnk$ that changes the radius only
within a given range $\epsilon$ of either boundary $\ir$ or $\mr$; the
precise form of this interpolant is given in \cite[Section
3.2.2]{borggaard2023statistical}.  With $\lnk$ in place,
the radius of the inner boundary for a given angle $\pth$ is then
given by $\lnk(\bq(\pth))$. Then given any $\bq:\RR \to \RR$ which is
sufficiently smooth and $2\pi$ periodic, we consider the domain
\begin{align}
  \DD_{\bq} := \{\bx \in \RR^2 \,:\,  \lnk(a_0+\bq(\pth(\bx))) \leq |\bx| \leq \oR\},
  \label{eq:b:dom}
\end{align}
where $a_0$ is a fixed mean radius and and denote the boundary as
\begin{align*}
  \Gamma^i_{\bq} := \{\bx \in \RR^2 \,:\, \lnk(a_0+\bq(\pth(\bx))) = |\bx| \},
  \quad
  \Gamma^o := \{ \bx \in \RR^2 \,:\, |\bx| = \oR\}.
\end{align*}
\arxiv{This procedure to construct annular domains $\DD_{\bq}$ is
  visualized in \cref{fig:Domain:Trans}.}

Having specified in \eqref{eq:AD:Stokesbulk}, \eqref{eq:AD:bc} and
\eqref{eq:b:dom} the `parameter to solution map' $\Sol$ portion of the
forward map $\G$ in \eqref{eq:forward:map:decomp}, it remains to
describe our observation procedure $\Obs$.  In fact there are a number
of different physically interesting possibilities $\Obs$ which yield
interesting statistics; see
\cite{borggaard2023statistical} for further details.  For
our purposes here we restrict our attention to a procedure based on
the observation of \emph{scalar variance} by quadrant.  Scalar
variance is an important metric for mixing; low scalar variance
indicates that $\pS$ has been well mixed while high scalar variance
indicates that $\pS$ has been ``trapped'' somewhere in the region.

To make this precise define $\pth_j = j\pi/2$ for $j=0,\dots,4$ and
define the quadrants
\begin{align*}
  \qS_j(\DD_\bq) = \{ \bx \in \DD_\bq \,:\, \pth_{j -1} < \pth(\bx)  \leq \pth_{j} \}
\end{align*}
for $j=1,\dots,4$. Then we define the four observations to be the
average scalar variance for each of the four quadrants, i.e.,
\begin{align}
  \Obs_j(\bU, \pS, \DD_\bq) := \frac{1}{|\DD_\bq|} \int\limits_{\qS_j(\DD_\bq)}
  \biggl( \pS(\bx;\bU) -  \frac{1}{|\DD_\bq|}
     \!\int\limits_{\DD_\bq} \pS(\by;\bU) d \by \biggr)^2 d\bx.
 \label{eq:scalar:sect:var:msr}
\end{align}
where $|\DD_\bq|$ is the area of the domain $\DD_\bq$ so that the
observations sum to the average scalar variance across the whole
domain.

For the data, $\Data = (\data_1, \data_2, \data_3, \data_4)$ in
\eqref{eq:stat:inv}, we choose high scalar variance (indicating that
the scalar should be trapped) in the first and third quadrants and low
scalar variance (indicating that the scalar should be mixed) in the
second and fourth quadrants.  For the observational noise $\eta$ in
\eqref{eq:stat:inv} we assume that each sectorial observation is
perturbed by an independent mean zero Gaussian noise. See
\cref{tab:stokes:parameters} for precise parameter values for $\Data$
and the distribution of $\eta$.

To parameterize the inner boundary, we 

let the components 
$\{a_k\}_{k=1}^\infty$ be our unknown parameters. For the prior
measure on these parameters, we choose $\mu_0 = N(0,C)$ where $C$ is a
diagonal operator such that components $a_k$, $k\ge 1$, are mutually
independent and distributed as $a_{2k-1},a_{2k} \sim
N(0,k^{-2s-1})$. Using \eqref{eq:sob:sp:norm}, this ensures that draws
from the prior are almost surely elements of $H^{s'}$ for any $s' < s$:
\begin{align}\label{eq:stokes:prior}
    \Exp \| \bq\|_{H^s}^2 
    = \Exp \left[
    \frac{1}{2}\sum_{k=1}^\infty k^{2s} (a_{2k-1}^2+a_{2k}^2) \right]
    = 
    \frac{1}{2}\sum_{k=1}^\infty k^{2s} \Exp (a_{2k-1}^2+a_{2k}^2) 
    = 
    \sum_{k=1}^\infty k^{-1 - 2(s-s')} < \infty.
\end{align}

To compute the forward map $\G(\bq)$ numerically from a given (finite)
array of components $\{a_k\}_{k=1}^K$, we (1) compute the Fourier
expansion \eqref{eq:sob:sp}, (2) compute the optimal B-spline
approximation to the Fourier expansion, (3) clamp via $\lnk$ to get a
B-spline representation of the inner boundary $\ib$, (4) mesh the
interior of the domain $\DD_\bq$ with Gmsh \cite{geuzaine2009gmsh}, a
popular meshing software, (5) solve the Stokes and advection-diffusion
PDEs \eqref{eq:AD:Stokesbulk}, \eqref{eq:AD:bc} via a finite element
solver, and (6) compute the observations by computing the integrals in
\eqref{eq:scalar:sect:var:msr}. Here step (2) is included because a
B-spline representation is required by Gmsh. Additional details are
available in \cite{borggaard2023statistical}; code
implementing the full solver and MCMC routines are publicly available
at \url{https://github.com/jborggaard/BayesianShape}. \arxiv{
  \cref{fig:stokes:solve} shows plots of the solution to the coupled
  Stokes and advection-diffusion equations; the example was taken from
  a sample from one of the multiproposal pCN chains described below.}

The remaining problem parameters are summarized in
\cref{tab:stokes:parameters}.
\begin{table}[htbp]
  {\footnotesize
  \caption{Parameter choices for the Stokes Problem.} \label{tab:stokes:parameters}
  \centering
  \begin{tabular}{|>{\raggedright}p{0.25\textwidth}|p{0.3\textwidth}|>{\raggedright}p{0.25\textwidth}|p{0.075\textwidth}|} \hline
    Parameter & Value &
    Parameter & Value \\
    \hline\hline
    Observations, $\Obs$ \eqref{eq:forward:map:decomp} & Sectoral Scalar Variance \eqref{eq:scalar:sect:var:msr} &
    Sampling dimension, $K$ & 320 \\\hline
    Data, $\Data = (\data_1, \data_2, \data_3, \data_4)$ \eqref{eq:stat:inv} & $\data_1=\data_3=0.4,\, \data_2=\data_4=0$ &
    Angular Velocity, $\rtrt$ \eqref{eq:AD:bc} & 10 \\\hline
    Prior, $\mu_{0}$ \eqref{eq:stat:inv:sol} & $a_{2k-1},a_{2k} \sim N(0,k^{-2s-1})$ &
    Mean radius, $a_0$ \eqref{eq:b:dom} & $1.0$ \\\hline
    Noise, $\noise$ \eqref{eq:stat:inv} & $N(0,\sigma_{\noise}^2 I)$, $\sigma_\noise=0.05$ &
    Diffusion, $\kappa$ \eqref{eq:AD:Stokesbulk} & $1.0$ \\\hline
    Radius Constraints \eqref{eq:com:domain} & $\ir=0.5,\,\mr=1.5,\,\oR=2$ &
    Number of B-splines & 160 \\\hline
    Source, $\src(\bx)$ \eqref{eq:AD:Stokesbulk} & $4 \exp[-(\bx-\bx_0)^2/100]$, $\bx_0=(1.5,1)$ &
    Range for $\lnk$, $\epsilon$ & $0.1$ \\\hline
  \end{tabular}
  }
\end{table}

\smallskip
\noindent\textbf{Numerical results}
\smallskip

\begin{figure}[h]
    \centering
    \includegraphics[width=0.48\textwidth]{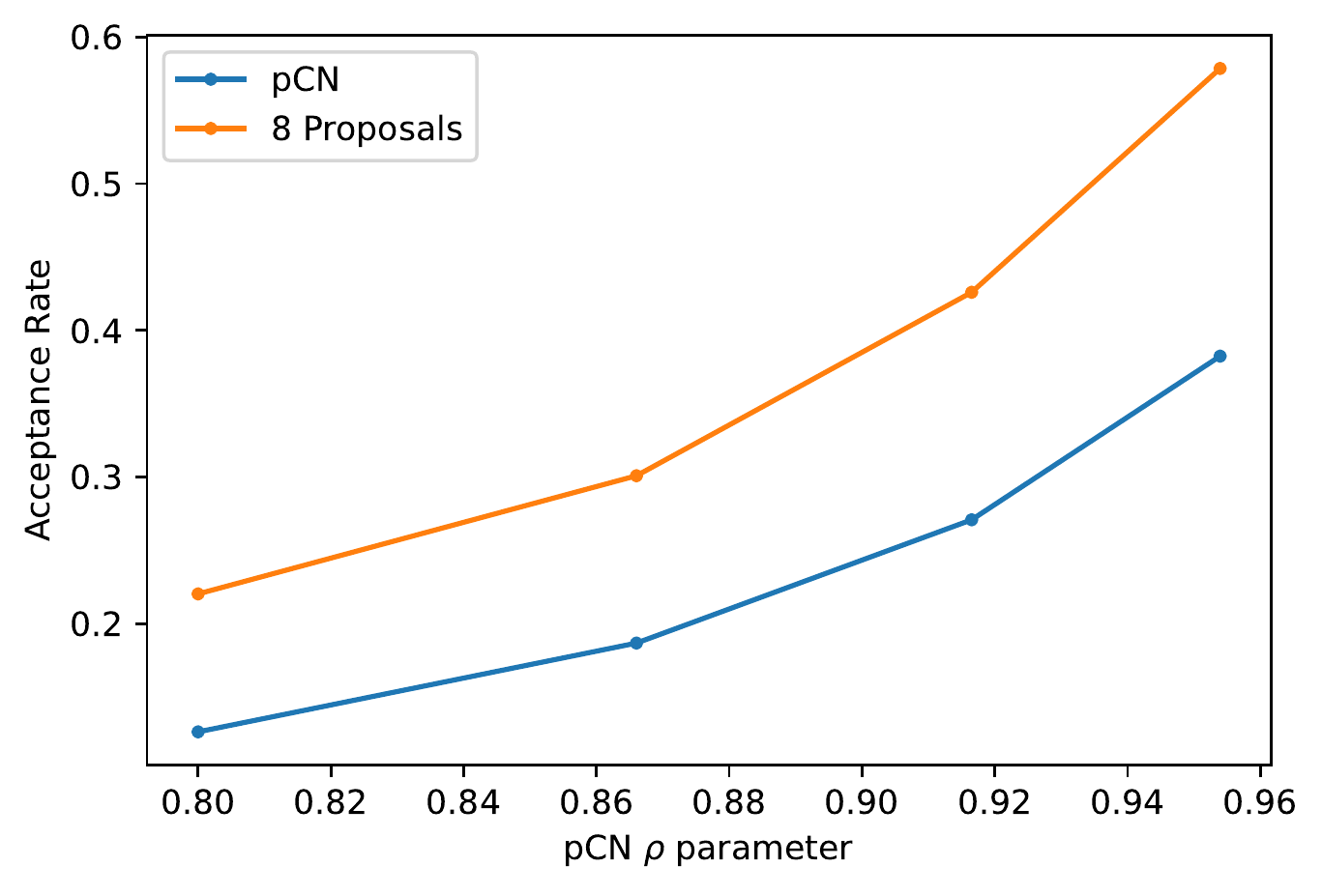}
    \hfill
    \includegraphics[width=0.48\textwidth]{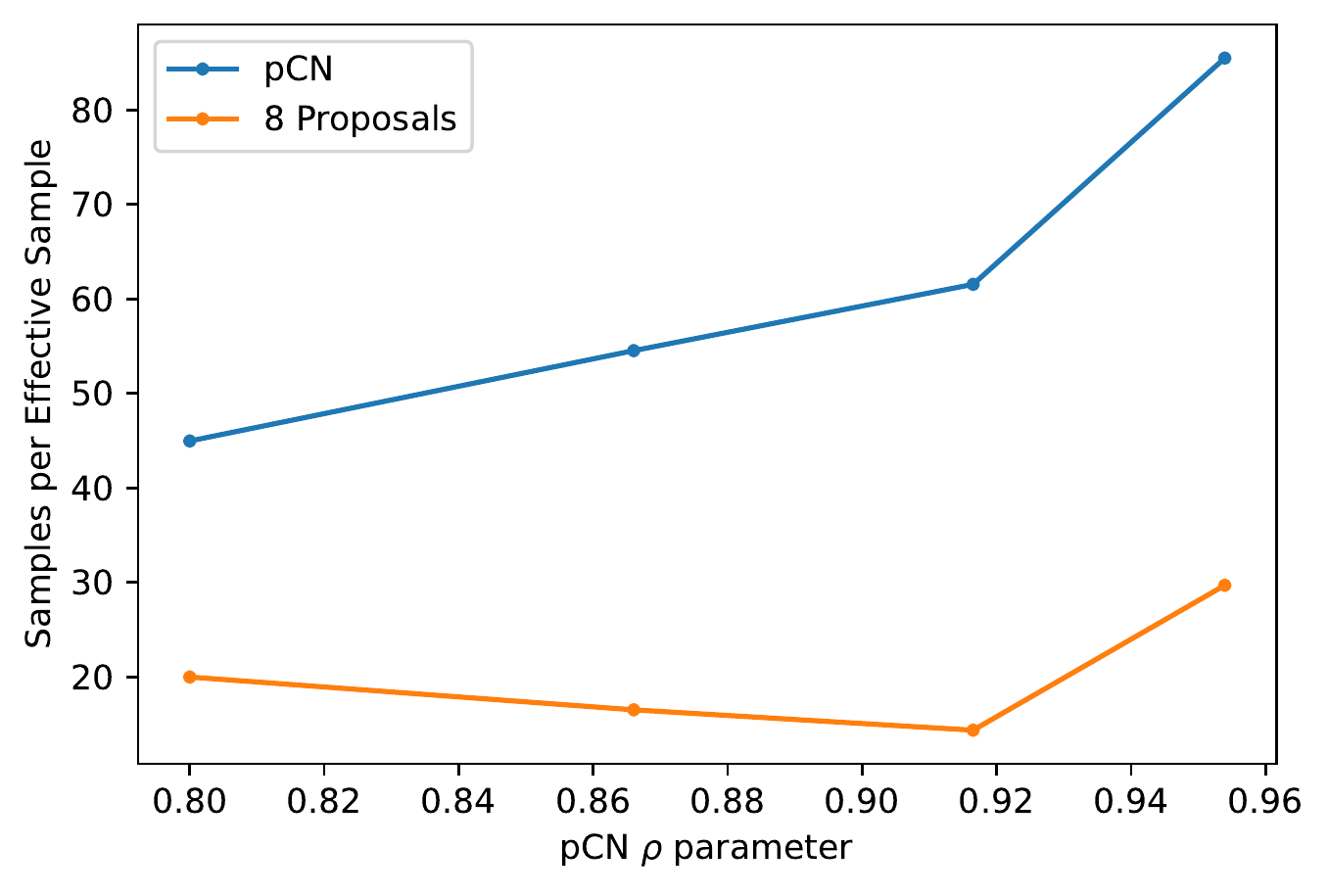}
    \caption{Sweep of tuning parameters $\rho$ and $p$ for the Stokes
      problem with 10,000 samples generated using multiproposal preconditioned Crank-Nicolson  (mpCN, \cref{alg:mpCN}) and pCN. Left: Acceptance rate. Right:
      Samples per effective sample of the (unnormalized) posterior
      density (lower is better).}
    \label{fig:stokes:ar:ess}
\end{figure}

PDE solves for the Stokes problem
\eqref{eq:AD:Stokesbulk}--\eqref{eq:AD:bc} are, due to the irregular
boundary conditions that change for each choice of parameter, quite a
bit more expensive than for the advection-diffusion example,
\eqref{eq:ad:eqn} covered in \cref{sec:AD:JUQ}. For this reason, we
limit the number of proposals to eight and the chains to $10,000$
samples each. We begin by presenting the acceptance rate and samples
per effective sample by MCMC method and pCN $\rho$ parameter, which
are shown in \cref{fig:stokes:ar:ess}. Each value shown is the average
across two chains. To minimize the effects of burn in, the effective
sample size in this case was computed on the final $5,000$ samples of
the $10,000$-sample chains. As in the analogous figure for the
advection-diffusion example (\cref{fig:ad:ar:ess}), this shows that
multiproposal pCN yields higher acceptance rates and requires fewer
samples to achieve an effective sample for a given choice of $\rho$.

\begin{figure}[h]
    \centering
    \includegraphics[width=0.48\textwidth]{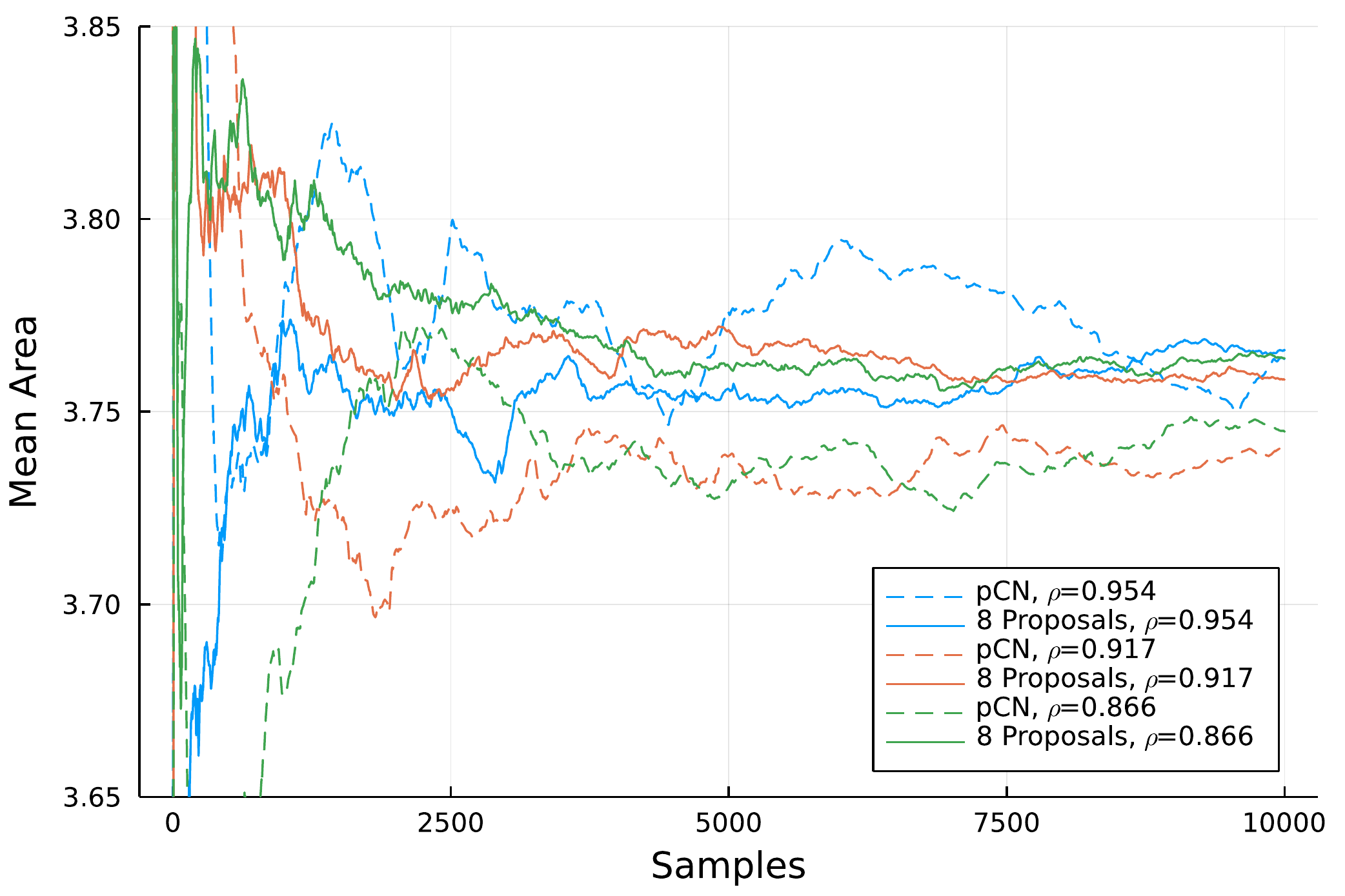}
    \hfill
    \includegraphics[width=0.48\textwidth]{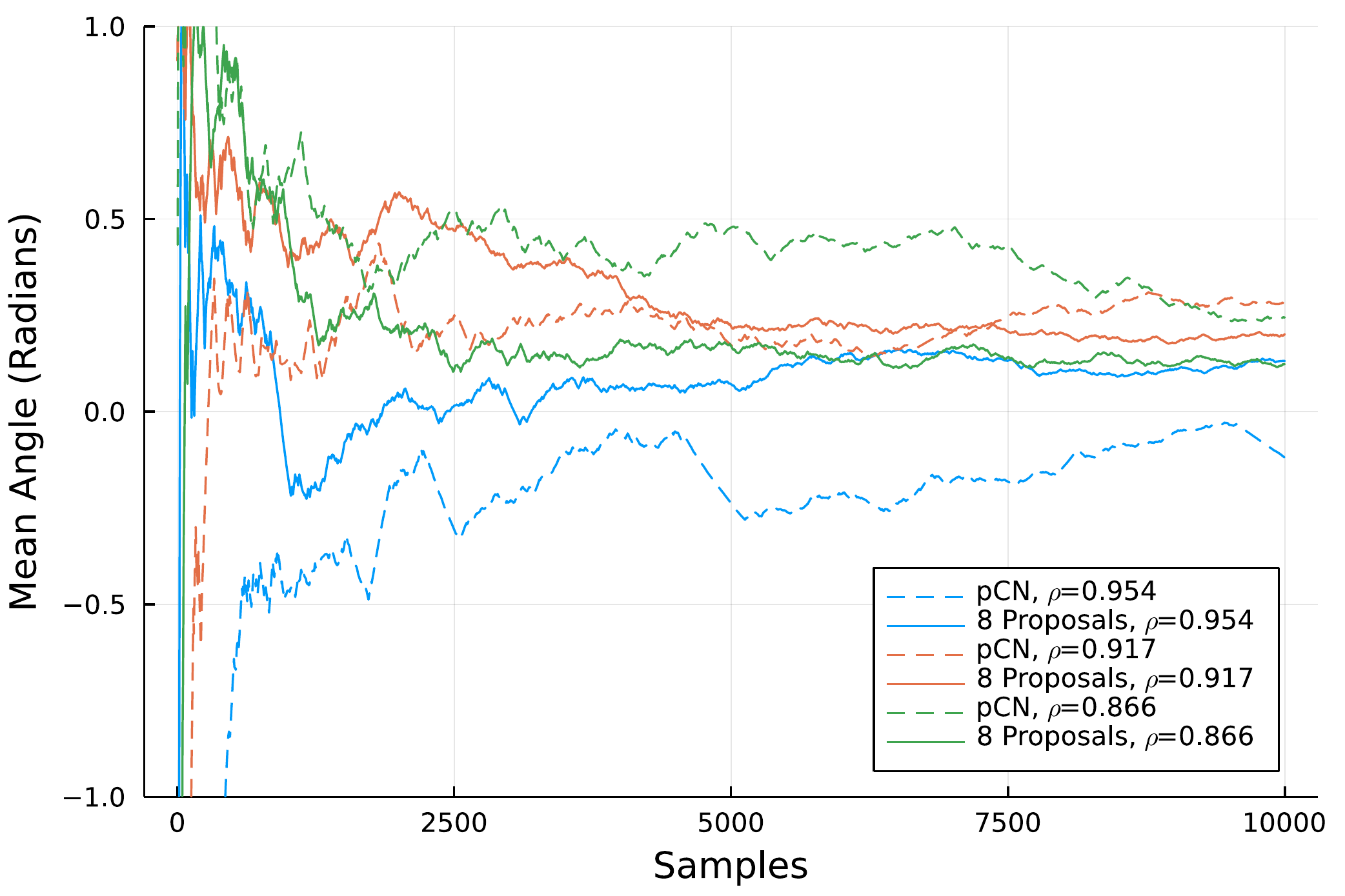}
    \caption{Running averages of inner boundary enclosed area (left)
      and primary angle (right) by $\rho$ (color) and MCMC method
       (preconditioned Crank-Nicolson (pCN) dashed and multiproposal pCN (\cref{alg:mpCN}) solid).}
    \label{fig:stokes:running:avg}
\end{figure}

We now consider convergence of two quantities of interest that provide
key information about the boundaries that make up the posterior. The
first is the area enclosed by the inner boundary, which describes the
size of the domain $\DD_{\bq}$. The second is the principal angle
associated with the inner boundary, which is given by the orientation
of its first Fourier component, i.e., $\arctan( Im(z) / Real(z) )$
where $z = \int_0^{2\pi} e^{i \pth} \lnk(b(x)) \,d\pth$. This scalar
quantity describes the polar orientation where the radius of the inner
boundary tends to be largest.  \cref{fig:stokes:running:avg} shows
running averages for these two scalar quantities; the results are
given for both pCN and multiproposal pCN with eight proposals, with
two chains run for each of three values of pCN's $\rho$ parameter. The
results in each case show that the multiproposal chains have, for the
most part, converged to a steady state about halfway through the
10,000 sample chains; the pCN chains, meanwhile, still deviate quite a
bit from each other and what appears to be the true value.

\begin{figure}[h]
    \centering
    \includegraphics[width=0.48\textwidth]{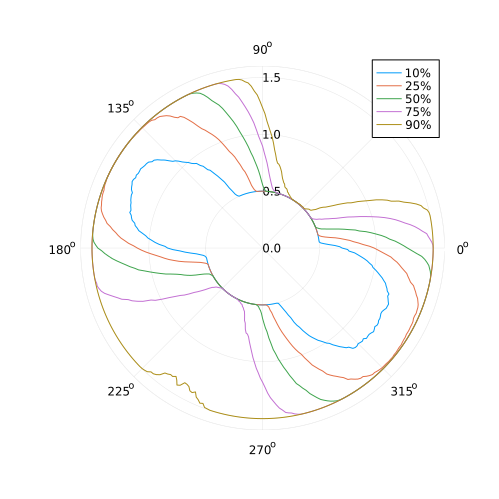}
    \hfill
    \includegraphics[width=0.48\textwidth]{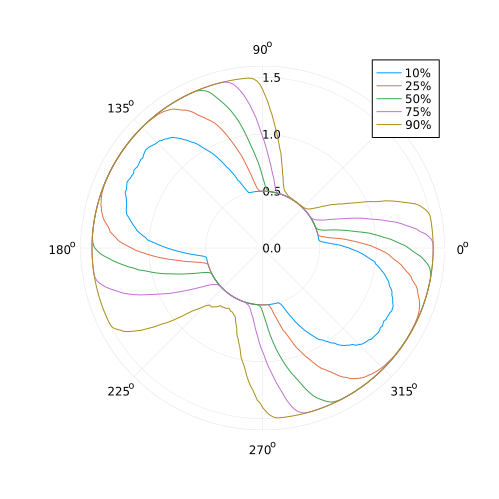}
    \caption{Radius quantiles for 10,000 samples with $\rho=0.917$. Left: preconditioned Crank-Nicolson (pCN). Right: multiproposal pCN (\cref{alg:mpCN}) with $8$ proposals.}
    \label{fig:stokes:radii:quantiles}
\end{figure}

Finally, we can consider the computed distributions on the shapes for
each MCMC method. \cref{fig:stokes:radii:quantiles} shows radius
quantiles by angle for pCN and multiproposal ($p=8$) pCN chains; to
generate the figure, we compute the radius of the inner boundary at
each angle for each sample and then tabulate the quantiles of the set
of radii for that angle. The bowtie shape of the boundaries are a
result of the desire to ``trap'' the scalar $\pS$ in the first and
third quadrants as dictated by the data. The two quantile plots are
largely similar, although the pCN (left) plot shows significant
divergence in the tails (e.g., the $90$\textsuperscript{th} percentile
in the third quadrant) as a result of the slower convergence.

\section{Outlook}
\label{sec:outlook}

Moving forward, the present work suggests a number of avenues for
future research.  First, certain foundational
questions remain to be addressed stemming from our analysis in
\cref{sec:framework} and \cref{sec:ext:Phase}.  It is clear that these
two formalisms, namely \cref{alg:main} and \cref{alg:tj:gen}, do not
entirely coincide while nevertheless demonstrate a significant
overlapping scope.  This overlapping scope includes conditionally
independent proposal structures encompassing important examples,
i.e., \cref{alg:TJ:cor:FD} and \cref{alg:mpCN}.  See
\cref{rmk:overlap}.  In any case, a systematic account of the
relationship between the formulations leading to \cref{alg:main},
\cref{alg:tj:gen} remain elusive at the time of writing.  Note also
that \cref{thm:tj} only addresses invariance (i.e., lack of bias) with
respect to the target whereas \cref{thm:main} and the other results in
\cref{sec:framework} provide comprehensive conditions for
reversibility.  Thus, this question of reversibility for the framework
in \cref{sec:ext:Phase} and its significance at the level of the
extended (indexed) phase space remain to be addressed.

Leaving aside these general considerations, we would like to emphasize
that the scope for novel concrete pMCMC algorithm design and
development from our formalisms (and other potential generalizations
of these formalisms) which we initiated above in \cref{sec:Ex:alg} is
far from exhausted.  For example, mpCN (\cref{alg:mpCN},
\cref{alg:mpCN:TJ}) suggests that a number of other such `infinite-dimensional' sampling methods, namely $\infty$HMC and $\infty$MALA,
\cite{cotter2013mcmc,Beskosetal2011}, can be fruitfully extended to
the multiproposal setting. We would also like to emphasize that
\cref{sec:HMC} only scratches the surface as far as the range of
possible formulations for multiproposal variants of the Hamiltonian
(Hybrid) Monte Carlo paradigm.

Note that our high-performance computing studies suggest the potential
of using GPUs to power pMCMC implementations with massive numbers of
proposals, and \cite{holbrook2023quantum} shows that quantum computing
also provides speedups for pMCMC. From an engineering perspective, our
extended phase space formulation of pMCMC may motivate further
high-performance computing methods that support gradient based pMCMC
such as that of \cref{sec:HMC}. From a theoretical perspective, the
same extended phase space formulation may extend to other pMCMC
algorithms motivated by novel computational paradigms.  For example,
it has been demonstrated that parallel, asynchronous updates of
different parameters within Gibbs sampling is theoretically valid so
long as one can bound latency between updates
\cite{terenin2020asynchronous}.  If each multiproposal constitutes a
(many-pronged) fork in the road, we are interested in pMCMC algorithms
that employ a branching structure, accepting multiple proposals at
once and conditionally assigning multiple individual iterations to new
nodes. Such a viral program could prove useful on large computing
resources with nodes constantly going on- and off-line.

Another outstanding question is that of optimal scaling.  Here, one
has additional degrees of freedom to consider beyond those of the
traditional random walk Monte Carlo analysis \cite{gelman1997weak}.
Optimal scaling should depend on the number of proposals, and the
optimal number of proposals should depend on the specific form of the
acceptance probabilities (e.g., \eqref{def:alphaj} or
\eqref{def:alphaj:metr:1}).  A useful analysis should account for
multimodal structure of the target distribution, i.e., the number of
modes and their mutual distances relative to individual scales.

Related to this question of optimal scaling is the establishment of
mixing rates and geometric ergodicity for various pMCMC algorithms
applied to specific target distributions.  Ultimately one would like
to rigorously quantify the improvements in mixing rates on a per
sample basis in comparison to traditional single proposal methods.
Here, one might suspect that mixing improves monotonically with the
number of proposals but any rigorous justification of this conjecture
appears to be far from reach at present. More tractably than simply establishing mixing rates for, e.g., \cref{alg:mpCN} one may
appeal to the strategies from \cite{hairer2014spectral,
  glatt2021mixing} as a fruitful starting point.  These works use a
weak Harris theorem and a generalized coupling argument to establish
geometric ergodicity for, respectively, pCN and HMC in
infinite-dimensional Hilbert spaces.

Finally, rigorous investigation is outside the scope of this manuscript, but intuition suggests that the truncation of acceptance probabilities in MH-type mechanism 
\eqref{def:alpha:simple:MH} might result in poor performance in the large proposal limit. An in depth study of the relative benefits of the MH-type \eqref{def:alpha:simple:MH} and Barker-type \eqref{def:alphaj:bb} acceptance probabilities remains for future work.

\section*{Acknowledgements}

Our efforts are supported under the grants NSF-DMS-1816551,
NSF-DMS-2108790 (NEGH), NSF-DMS-2108791 (JAK), NIH-K25-AI153816,
NSF-DMS-2152774, NSF-DMS-2236854 (AJH) and NSF-DMS-2009859, NSF-DMS-2239325 (CFM).  The authors thank
J. Borggaard for help in formulating and preparing the code for the
Stokes example presented in \cref{sec:rot:Stokes}. We would also like
to express our appreciation to H. Riggs for clarifying some issues
related to macrodata refinement.  The authors acknowledge Advanced
Research Computing at Virginia Tech (\url{https://arc.vt.edu}) for
providing computational resources that have contributed to the results
reported within this paper.

%% References

\begin{footnotesize}
\addcontentsline{toc}{section}{References}
\bibliographystyle{alpha}
\bibliography{refs}
\end{footnotesize}

\begin{multicols}{2}
\noindent
Nathan E. Glatt-Holtz\\ {\footnotesize
	Department of Statistics\\
	Indiana University\\
	AND Department of Mathematics\\
	Tulane University\\
	Web: \url{https://negh.pages.iu.edu}\\
	Email: \url{negh@iu.edu}} \\[.2cm]

\noindent
Andrew J. Holbrook\\  {\footnotesize
Department of Biostatistics\\
University of California, Los Angeles\\
Web: \url{https://andrewjholbrook.github.io}\\
Email: \url{aholbroo@g.ucla.edu}}

\columnbreak

\noindent Justin Krometis\\
{\footnotesize
National Security Institute\\
AND Department of Mathematics\\
Virginia Tech\\
Web: \url{https://krometis.github.io/}\\
Email: \href{mailto:jkrometis@vt.edu}{\nolinkurl{jkrometis@vt.edu}}} \\[.2cm]

\noindent Cecilia F. Mondaini \\
{\footnotesize
  Department of Mathematics\\
  Drexel University\\
  Web: \url{https://www.math.drexel.edu/~cf823}\\
  Email: \url{cf823@drexel.edu}}\\[.2cm]
 \end{multicols}

\newpage

%% Appendix

\appendix
%\addcontentsline{toc}{section}{Appendix}

%\counterwithin{figure}{section}
%\counterwithin{table}{section}
%\renewcommand\thefigure{\thesection\arabic{figure}}
%\renewcommand\thetable{\thesection\arabic{table}}

\section{Appendix: Some measure-theoretic tools}
\label{subsec:meas:th}

This first appendix gathers together for the convenience of the reader
some measure theoretic elements used extensively throughout the
manuscript. We refer to
e.g. \cite{Bogachev2007,Folland1999,Aliprantis2013}
for further general background.

Let $(\cX, \Sigma_{\cX})$ and $(\cY, \Sigma_{\cY})$ be measurable
spaces. Given a measurable function $\phi: \cX \to \cY$ and a measure
$\nu$ on $(\cX, \Sigma_{\cX})$, the \emph{pushforward} of $\nu$ by
$\phi$, denoted $\phi^* \nu$, is the measure on $\cY$ defined as
\begin{align}\label{def:pushfwd}
  \phi^*\nu(E) \coloneqq \nu (\phi^{-1}(E)) \quad
  \mbox{ for all } E \in \Sigma_\cY.
\end{align}
In a statistical context, we notice that if $\bw$ is a random variable
with probability distribution given by a measure $\nu$ then
$\phi(\bw)$ is another random variable that is distributed as
$\phi^* \nu$.

For any measurable functions $\phi_1, \phi_2$ defined on appropriate
spaces so that the composition $\phi_1 \circ \phi_2$ makes sense, it
follows immediately from \eqref{def:pushfwd} that
\begin{align*}
	(\phi_1 \circ \phi_2)^* \nu = \phi_1^*(\phi_2^* \nu).
\end{align*}
We also recall the following change of variables formula regarding
pushforward measures. Namely, given a $(\phi^* \nu)$-integrable
function $\psi: \cY \to \RR$, i.e. $\psi \in L^1(\phi^* \nu)$, it
follows that the composition $\psi \circ \phi: \cX \to \RR$ belongs to
$L^1(\nu)$ and
\begin{align}
  \int_\cY \psi(\bw) \phi^*\nu(d\bw)
  = \int_\cX \psi(\phi(\bw))  \nu(d\bw).
  \label{def:pushfwd:obs}
\end{align}

In the particular case when $\cX = \RR^N$ and $\nu$ is any Borel
measure on $\RR^N$ which is absolutely continuous with respect to
Lebesgue measure, namely
\begin{align*}
	\nu(d\bw) = \pi(\bw) d\bw
\end{align*}
for some density function $\pi: \RR^N \to \RR$, then for any
diffeomorphism $\phi: \RR^N \to \RR$, we have
\begin{align}\label{pfwd:diff:fd}
	\phi^* \nu(d\bw) = \pi(\phi^{-1}(\bw)) |\det \nabla \phi^{-1}(\bw)| d\bw.
\end{align}

Further, recall that a measure $\nu$ on $(\cX, \Sigma_\cX)$ is
\emph{absolutely continuous} with respect to another measure $\rho$ on
$(\cX, \Sigma_\cX)$, denoted $\nu \ll \rho$, if $\nu(E) = 0$ whenever
$\rho(E) = 0$, for $E \in \Sigma_\cX$. If $\nu$ and $\rho$ are two
sigma-finite measures on $(\cX, \Sigma_\cX)$ such that $\nu \ll \rho$
then there exists a $\rho$-almost unique function
$d \nu/d \rho \in L^1(\rho)$ such that
\begin{align*}
	\nu(E) = \int_E \frac{d\nu}{d\rho}(\bw) \rho(d\bw), \quad E \in \Sigma_\cX,
\end{align*}
called the \emph{Radon-Nikodym derivative} of $\nu$ with respect to
$\rho$. Moreover, given sigma-finite measures $\nu$, $\rho$ and
$\gamma$ on $(\cX, \Sigma_\cX)$ such that $\nu \ll \rho$ and
$\rho \ll \gamma$, it follows that $\nu \ll \gamma$ and
\begin{align*}
  \frac{d\nu}{d\gamma}(\bw)
  = \frac{d\nu}{d\rho}(\bw)\frac{d\rho}{d\gamma}(\bw)
  \quad \mbox{ for $\gamma$-a.e. } \bw \in \cX.
\end{align*} 
In particular, if $\nu_1$, $\nu_2$ and $\rho$ are sigma-finite
measures on $(\cX, \Sigma_\cX)$ with $\nu_1 \ll \rho$ and
$\nu_2 \ll \rho$, so that
\begin{align*}
  \nu_1(d\bw)
  = \phi_1(\bw) \rho(d\bw),
  \quad
  \nu_2(d\bw) = \phi_2(\bw) \rho(d\bw),
\end{align*}
where $\phi_1 = d \nu_1/d\rho$ and $\phi_2 = d\nu_2/d\rho$, and if
$\phi_2 > 0$ $\rho$-a.e., then $\nu_1 \ll \nu_2$ and
\begin{align}\label{RN:nu1:nu2}
  \frac{d\nu_1}{d\nu_2}(\bw) = \frac{\phi_1(\bw)}{\phi_2(\bw)}
  \quad \mbox{ for $\rho$-a.e. } \bw \in \cX.
\end{align}

Finally, under the above definitions the following identities can be
easily verified, see \cite[Section 2.1]{glatt2020accept}. Firstly,
given a measurable and invertible mapping $\phi: \cX \to \cX$ with
measurable inverse $\phi^{-1}: \cX \to \cX$, and sigma-finite measures
$\nu$ and $\rho$ on $(\cX, \Sigma_\cX)$ with $\nu \ll \rho$, it
follows that $\phi^*\nu \ll \phi^* \rho$ and
\begin{align}\label{RN:phi:nu:rho}
	\frac{d \phi^*\nu}{d\phi^*\rho}(\bw) =
  \frac{d\nu}{d\rho}(\phi^{-1}(\bw))
  \quad \mbox{ for $(\phi^*\rho)$-a.e. } \bw \in \cX.
\end{align}
Secondly, if $\nu$ is a sigma-finite measure on $(\cX, \Sigma_\cX)$,
and $\phi_i: \cX \to \cX$, $i=1,\ldots,n$, a sequence of measurable
and invertible functions with measurable inverses
$\phi_i^{-1}: \cX \to \cX$ such that $\phi_i^*\nu \ll \nu$ for all
$i=1,\ldots,n$, then $(\fm_n \circ \cdots \circ \fm_1)^* \mu \ll \mu$
and
\begin{align}\label{RN:push:comp}
\frac{d (\fm_n \circ \cdots \circ \fm_1)^*\mu}{d \mu} (\bw)
= \frac{d\fm_n^* \mu}{d\mu}(\bw) \prod_{i=1}^{n-1}
\frac{d \fm_i^* \mu}{d \mu}((\fm_n\circ \cdots \circ \fm_{i+1})^{-1}(\bw))
\quad \mbox{ for $\mu$-a.e. } \bw \in \cX.
\end{align}

\section{Appendix: Rigorous Proofs}\label{appx:proofs}

This appendix gathers together the rigorous proofs for all the claims
made above in \cref{sec:framework}, \cref{sec:ext:Phase} and
\cref{sec:Ex:alg}.

\subsection{Proof of \cref{thm:main}}\label{appx:thm:main}

To show that \ref{P1} and \ref{P2} imply
$\mu \Pker^{\alpha, S, \Vker} = \mu$, notice that for any bounded and
measurable function $\varphi: \spq \to \RR$ we have
\begin{align*}
\int_\spq \varphi(\bq) \mu \Pker^{\alpha, S, \Vker} (d\bq) 
&= \int_\spq \varphi(\bq) \int_\spq \Pker^{\alpha,S,\Vker}(\btq, d \bq) \mu(d \btq) \\
&= \sum_{j=0}^p \int_\spq \int_\spq \int_\spv \varphi(\bq) \alpha_j(\btq, \bv) \delta_{\Pi_1 S_j (\btq, \bv)} (d\bq) \Vker(\btq, d \bv) \mu(d \btq) \\
&= \sum_{j=0}^p \int_\spq \int_\spv \varphi(\Pi_1 S_j(\btq, \bv)) \alpha_j(\btq, \bv) \cM(d\btq, d\bv) \\
&= \sum_{j=0}^p \int_\spq \int_\spv  \varphi(\btq) \alpha_j(S_j(\btq, \bv)) S_j^* \cM(d\btq, d\bv) 
= \int_\spq \varphi(\btq) \mu(d\btq).
\end{align*}
Since $\varphi$ is arbitrary, this implies $\mu \Pker^{\alpha, S, \Vker} = \mu$ as desired. 

For the second part of the statement, the fact that \ref{P3} implies
\ref{P2} follows immediately upon summing \eqref{eq:det:bal:cond} over
$j=0,\ldots,p$, taking the integral over $\spv$ and invoking
\eqref{sum:alpha:1}. For \eqref{det:bal}, it suffices to show that for
every bounded and measurable function
$\psi: \spq \times \spq \to \mathbb{R}$, we have
\begin{align}\label{eq:phi}
\sum_{j=0}^p \int_{\spq} \int_{\spv}
\psi(\bq,  \Proj \circ S_j(\bq, \bv)) \alpha_j( \bq, \bv) \cM( d\bq, d \bv)
= \sum_{j=0}^p \int_{\spq} \int_{\spv}
\psi(\Proj \circ S_j(\btq, \bv), \btq) \alpha_j( \btq, \bv)
\cM( d\btq, d \bv).
\end{align}
Invoking \ref{P1} and changing variables, cf. \eqref{def:pushfwd:obs}, we obtain that 
\begin{align}\label{eq:phi:1}
\sum_{j=0}^p \int_{\spq} &\int_{\spv}
\psi(\bq,  \Proj \circ S_j(\bq, \bv)) \alpha_j( \bq, \bv) \cM( d\bq, d \bv) \notag\\
=&\sum_{j=0}^p \int_{\spq} \int_{\spv}
\psi(\Proj \circ S_j \circ S_j (\bq, \bv),  \Pi_1 \circ S_j(\bq, \bv)) 
\alpha_j( S_j \circ S_j (\bq, \bv)) \cM( d\bq, d \bv) \notag\\
=& \sum_{j=0}^p \int_{\spq} \int_{\spv}  \psi(\Proj \circ S_j(\bq, \bv),  \bq) 
\alpha_j( S_j(\bq, \bv)) S^*_j\cM( d\bq, d \bv).
\end{align}
Now from \ref{P3} we deduce \eqref{eq:phi}, completing the proof.

 \subsection{Proof of \cref{cor:ar:Barker}}\label{appx:cor:ar:Barker}

  We first notice that $S_j^*\cM \ll (S_0^*\cM + \cdots + S_p^*\cM)$,
  for each $j = 0, \ldots, p$, so that the Radon-Nikodym derivative in
  \eqref{def:alphaj} is well-defined. Moreover, clearly
\begin{align}\label{sum:alphaj:1}
	\sum_{j=0}^p \frac{d S_j^*\cM}{d(S_0^*\cM + \cdots + S_p^*\cM)}(\bq, \bv)
	= \frac{d(S_0^*\cM + \cdots + S_p^*\cM)}{d(S_0^*\cM + \cdots + S_p^*\cM)}(\bq, \bv) = 1
\end{align}
for $\left(\sum_{j=0}^p S_j^*\cM\right)$-a.e.
$(\bq,\bv) \in \spq\times \spv$, which implies that $\alpha_j$,
$j=0,\ldots,p$, are well-defined, and condition \eqref{sum:alpha:1}
from \ref{P3} in \cref{thm:main} holds.

It remains to verify \eqref{eq:det:bal:cond}. Since
$S_j \circ S_j = I$, it follows from \eqref{RN:phi:nu:rho} that
$\cM \ll S_j^*(S_0^*\cM + \cdots + S_p^*\cM)$ and
\begin{align*}
	\frac{d \cM}{d S_j^*(S_0^*\cM + \cdots + S_p^*\cM)} (\bq, \bv)
	= \frac{d S_j^*\cM}{d(S_0^*\cM + \cdots + S_p^*\cM)}(S_j(\bq, \bv))
	= \alpha_j(S_j(\bq, \bv))
\end{align*}
for $S_j^* \left(\sum_{k=0}^p S_k^*\cM\right)$-a.e.
$(\bq, \bv) \in \spq \times \spv$. But from our assumption
\eqref{sum:cond}, we have
\begin{align*}
	S_j^*(S_0^*\cM + \cdots + S_p^*\cM)
	= (S_j \circ S_0)^*\cM + \cdots + (S_j \circ S_p)^*\cM
	= S_0^*\cM + \cdots + S_p^*\cM.
\end{align*}
Hence, $\cM \ll S_0^*\cM + \cdots + S_p^*\cM$ and
\begin{align*}
  \frac{d\cM}{d (S_0^* \cM + \cdots + S_p^*\cM)} (\bq, \bv)
     = \alpha_j(S_j(\bq, \bv))
  \quad \mbox{ for $\left(\sum_{j=0}^p S_j^*\cM\right)$-a.e. }
  (\bq, \bv) \in \spq \times \spv.
\end{align*}
We thus obtain
\begin{align*}
	&\alpha_j(S_j(\bq, \bv)) S_j^*\cM (d\bq, d\bv) 
	= \frac{d\cM}{d (S_0^* \cM + \cdots + S_p^*\cM)} (\bq, \bv) \, S_j^*\cM (d\bq, d\bv)  \\
	&\qquad\qquad\qquad= \frac{d\cM}{d (S_0^* \cM + \cdots + S_p^*\cM)} (\bq, \bv)
   \frac{d S_j^*\cM}{d (S_0^* \cM + \cdots + S_p^*\cM)} (\bq, \bv) \, (S_0^* \cM + \cdots + S_p^*\cM) (d\bq, d \bv) \\
	&\qquad\qquad\qquad=  \frac{d S_j^*\cM}{d (S_0^* \cM + \cdots + S_p^*\cM)} (\bq, \bv) \, \cM(d\bq, d \bv)
	= \alpha_j(\bq, \bv) \cM(d\bq, d \bv),
\end{align*}
so that \eqref{eq:det:bal:cond} also holds. Therefore, condition
\ref{P3} of \cref{thm:main} is satisfied. This concludes the proof.

\subsection{Proof of \cref{cor:wedge:alpha}}
\label{appx:cor:wedge:alpha}
Clearly, condition \eqref{sum:alpha:1} follows immediately from the
definition of $\alpha_0$. Regarding \eqref{eq:det:bal:cond}, the case
$j=0$ is readily verified since $S_0 = I$. For $j=1,\ldots, p$,
\begin{align*}
	\alpha_j(S_j(\bq, \bv)) S_j^* \cM (d\bq, d \bv) 
	&= \overline{\alpha}_j \left( 1 \wedge \frac{d S_j^*\cM}{d \cM} (S_j(\bq, \bv)) \right)  \frac{d S_j^* \cM}{d \cM}(\bq, \bv) \cM(d\bq, d \bv) \\
	&= \overline{\alpha}_j \left[\frac{d S_j^* \cM}{d \cM}(\bq, \bv) \wedge \left( \frac{d S_j^*\cM}{d \cM} (S_j(\bq, \bv)) \frac{d S_j^* \cM}{d \cM}(\bq, \bv) \right) \right] \cM(d\bq, d \bv).
\end{align*}
From \eqref{RN:push:comp} and since $S_j \circ S_j = I$, we deduce that
\begin{align*}
	\alpha_j(S_j(\bq, \bv)) S_j^* \cM (d\bq, d \bv) 
	&= \overline{\alpha}_j \left[ 1 \wedge \frac{d S_j^*\cM}{d\cM} (S_j(\bq,\bv)) \right] \frac{d S_j^*\cM}{d\cM}(\bq, \bv) \cM(d\bq, d \bv) \\
	&= \overline{\alpha}_j \left[\frac{d S_j^* \cM}{d \cM}(\bq, \bv) \wedge 1 \right]  \cM(d\bq, d \bv) = \alpha_j(\bq, \bv) \cM(d\bq, d \bv),
\end{align*}
so that \eqref{eq:det:bal:cond} also holds in this case. This concludes the proof. 

\subsection{Proof of \cref{thm:multi:ind:Tjmd}}
\label{sec:thm:multi:ind:Tjmd}

Recall that $\sigma$-finite measures on $(\spq^p, \Sigma_{\spq^p})$
are uniquely determined by their evaluation on cylinder sets of the
form $E = A_0 \times  \cdots \times A_p$ where $A_j \in \Sigma_\spq$.  
Thus, it suffices to verify that
\begin{align}\label{eq:Sj:Sk:phi}
\sum_{k=0}^p \int_{\spq^{p+1}} \varphi(\bq_0, \bv) (S_j \circ S_k)^*\cM (d\bq_0, d \bv) 
= \sum_{k=0}^p \int_{\spq^{p+1}} \varphi(\bq_0, \bv) S_k^*\cM (d\bq_0, d \bv),
\end{align}
for every $j=0,1,\ldots,p$, and all $\varphi: \spq^{p+1} \to \RR$ of the form
\begin{align*}
  \varphi(\bq_0, \bv) = \prod_{l=0}^p \varphi_l(\bq_l),
  \quad (\bq_0, \bv) = (\bq_0, \bq_1, \ldots, \bq_p) \in \spq^{p+1},
\end{align*}
with bounded and measurable functions $\varphi_l: \spq \to \spq$,
$l=0,1,\ldots,p$.

Clearly, \eqref{eq:Sj:Sk:phi} holds for $j=0$, so we may assume from
now on that $j \in \{1,\ldots, p\}$.  Now for any such $j$,
noting that $S_k$ is an involution for each $k = 0, 1, \ldots, p$ we
infer that \eqref{eq:Sj:Sk:phi} reduces to showing that
\begin{align}
  \sum_{\stackrel{k=1}{k\neq j}}^p \int_{\spq^{p+1}}
  \varphi(\bq_0, \bv) (S_j \circ S_k)^*\cM (d\bq_0, d \bv) 
  = \sum_{\stackrel{k=1}{k\neq j}}^p \int_{\spq^{p+1}}
  \varphi(\bq_0, \bv)S_k^*\cM (d\bq_0, d \bv).
  \label{eq:Sj:Sk:phi:red}
\end{align}
Regarding the right-hand side of
\eqref{eq:Sj:Sk:phi:red}, we obtain by changing variables,
\eqref{def:pushfwd:obs}, and recalling the definition of $\cM$ that
for every $k=1,\ldots,p$
\begin{align}
\int_{\spq^{p+1}} \varphi(\bq_0, \bv) S_k^*\cM (d\bq_0, d \bv) 
&= \int_{\spq^{p+1}} \varphi(S_k(\bq_0, \bv)) \cM (d\bq_0, d \bv) \notag\\
  &= \int_{\spq^{p+1}} \varphi(\bq_k, \bq_1, \ldots, \bq_{k-1}, \bq_0,\bq_{k+1}, \ldots, \bq_p)
     \int_\spq \prod_{i=1}^p \Qker(\bq, d \bq_i) \bQker(\bq_0, d\bq) \mu(d \bq_0) \notag\\
  &= \int_{\spq^{p+2}} \varphi_0(\bq_k) \varphi_k(\bq_0)
    \prod_{\stackrel{l=1}{l\neq k}}^p \varphi_l(\bq_l)
    \prod_{i=1}^p \Qker(\bq, d \bq_i) \bQker(\bq_0, d \bq) \mu(d \bq_0) \\
&= \int_{\spq^2} \varphi_k(\bq_0) \left( \int_\spq \varphi_0(\bq_k) \Qker(\bq, d \bq_k) \right)
    \prod_{\stackrel{l=1}{l\neq k}}^p \left( \int_\spq\varphi_l(\bq_l) \Qker(\bq, d \bq_l)  \right)
                                                                             \bQker(\bq_0, d \bq) \mu(d\bq_0) \notag\\
  &= \int_{\spq^2} \varphi_k(\bq_0) \, \Qker \varphi_0(\bq)
    \prod_{\stackrel{l=1}{l\neq k}}^p \Qker \varphi_l(\bq)  \,  \bQker(\bq_0, d \bq) \mu(d\bq_0) \notag\\ 
&= \int_\spq \varphi_k(\bq_0) \int_\spq \prod_{\stackrel{l=0}{l \neq k}}^p
                   \Qker \varphi_l(\bq) \,  \bQker(\bq_0, d \bq) \mu(d\bq_0),
\label{eq:Sj:Sk:LHS}                                                                                                            
\end{align}
where we recall the notation
$\tilde{Q} \phi(\bq) = \int \phi(\bq') \tilde{Q}(\bq, d \bq')$ for the
action of a Markov kernel $\tilde{Q}$ on a measurable function $\phi$.

Turning to the left-hand side of \eqref{eq:Sj:Sk:phi:red}, it follows
once again by change of variables and the definition of $\cM$ that for
every $k = 1,\ldots, p$ with $k \neq j$,
\begin{align}
&\int_{\spq^{p+1}} \varphi(\bq_0, \bv) (S_j \circ S_k)^*\cM (d\bq_0, d \bv)
= \int_{\spq^{p+1}} \varphi(S_j \circ S_k(\bq_0, \bv)) \cM (d\bq_0, d \bv) \notag \\
&\qquad = \int_{\spq^{p+2}} \varphi_0(\bq_j) \varphi_j(\bq_k) \varphi_k(\bq_0) \prod_{\stackrel{l=1}{l\neq k,j}}^p \varphi_l(\bq_l) \prod_{i=1}^p \Qker(\bq, d \bq_i) \, \bQker(\bq_0, d \bq) \, \mu(d \bq_0) \notag \\
&\qquad = \int_{\spq^2} \varphi_k(\bq_0) \left( \int_\spq \varphi_0(\bq_j) \Qker(\bq,d \bq_j) \right) \left( \int_\spq \varphi_j(\bq_k) \Qker(\bq,d \bq_k) \right) \prod_{\stackrel{l=1}{l\neq k,j}}^p \left( \int_\spq \varphi_l(\bq_l) \Qker(\bq, d \bq_l) \right) \bQker(\bq_0, d\bq) \,\mu(d \bq_0) \notag \\
&\qquad = \int_\spq \varphi_k(\bq_0) \int_\spq \prod_{\stackrel{l=0}{l \neq k}}^p \Qker \varphi_l(\bq) \, \bQker(\bq_0, d \bq) \, \mu(d\bq_0).                                                                              \label{eq:Sj:Sk:M}
\end{align}
Comparing \eqref{eq:Sj:Sk:M} with \eqref{eq:Sj:Sk:LHS} and summing
yields \eqref{eq:Sj:Sk:phi:red} and hence \eqref{eq:Sj:Sk:phi}. This
concludes the proof of \eqref{eq:Sj:Sk:M:A}. The second part of the
statement follows immediately from \cref{cor:ar:Barker}.

\subsection{Proof of \cref{thm:cond:ind:mu:mu0}}
\label{sec:thm:cond:ind:mu:mu0}

We need to establish the absolute continuity $S_j^*\cM \ll \cM$ and
\eqref{RN:SjM:M} for $j = 0, \ldots, p$.  Since $S_0 = I$, clearly
this holds for $j=0$. Now we consider $j \in \{1,\ldots, p\}$.  Let
$\cM_0(d\bq_0, d \bv) \coloneqq \Vker(\bq_0, d \bv) \mu_0(d\bq_0)$. We
first claim that assumption \eqref{Q:mu0:det:bal} implies that
$S_j^* \cM_0 = \cM_0$. Indeed, take any measurable and bounded
function $\varphi: \spq^{p+1} \to \RR$. By change of variables, 
\eqref{def:pushfwd:obs}, and the
definitions of $\Vker$ and $S_j$, $j=1,\ldots,p$, in \eqref{def:Vker:multi:ind}-\eqref{def:Sj:cond:ind} it follows that
\begin{align*}
	\int_{\spq^{p+1}} \varphi(\bq_0, \bv) S_j^* \cM_0(d\bq_0, d \bv) 
	&= \int_{\spq^{p+1}} \varphi (S_j(\bq_0, \bv)) \cM_0(d\bq_0, d \bv) \\
	&= \int_{\spq^{p+1}} \varphi (\bq_j, \bq_1, \ldots, \bq_{j-1}, \bq_0, \bq_{j+1},\ldots,\bq_p) 
	      \int_\spq \prod_{i=1}^p \Qker(\bq, d \bq_i) \bQker(\bq_0, d\bq) \mu_0(d\bq_0).
\end{align*}
From \eqref{Q:mu0:det:bal} and Fubini's theorem, we thus obtain
\begin{align*}
	&\int_{\spq^{p+1}} \varphi(\bq_0, \bv) S_j^* \cM_0(d\bq_0, d \bv) 
	= \int_{\spq^{p+1}} \varphi (\bq_j, \bq_1, \ldots, \bq_{j-1}, \bq_0, \bq_{j+1},\ldots,\bq_p) \int_\spq \prod_{i=1}^p \Qker(\bq, d \bq_i) \Qker (\bq, d\bq_0) \mu_0(d\bq) \\
	&\quad = \int_{\spq^{p+2}} \varphi (\bq_j, \bq_1, \ldots, \bq_{j-1}, \bq_0, \bq_{j+1},\ldots,\bq_p) \,\, \Qker(\bq, \cdot)^{\otimes (p+1)} (d \bq_j, d \bq_1, \ldots, d \bq_{j-1}, d \bq_0, d \bq_{j+1}, \ldots, d\bq_p) \, \mu_0(d\bq) \\
	&\quad = \int_{\spq^{p+2}} \varphi (\bq_0, \bq_1, \ldots,\bq_p) \,\, \Qker(\bq, \cdot)^{\otimes (p+1)} (d \bq_0, d \bq_1, \ldots, d \bq_p) \, \mu_0(d\bq) \\
	&\quad = \int_{\spq^{p+2}} \varphi (\bq_0, \bq_1, \ldots,\bq_p)  \prod_{i=1}^p \Qker(\bq, d \bq_i) Q(\bq, d\bq_0) \mu_0(d\bq) \\
	&\quad = \int_{\spq^{p+2}} \varphi (\bq_0, \bq_1, \ldots,\bq_p)  \prod_{i=1}^p \Qker(\bq, d \bq_i) \bQker(\bq_0, d\bq) \mu_0(d\bq_0) 
	= \int_{\spq^{p+1}}  \varphi (\bq_0, \bv) \cM_0(d \bq_0, d \bv),
\end{align*}
where $\Qker(\bq, \cdot)^{\otimes (p+1)}$ denotes the $(p+1)$-fold
product of the probability measure $\Qker(\bq, \cdot)$. Since
$\varphi: \spq^{p+1} \to \RR$ is an arbitrary measurable and bounded
function, we deduce that $S_j^* \cM_0 = \cM_0$.

Next, again for any such function $\varphi$, we obtain by invoking the assumption $\mu \ll \mu_0$ that
\begin{align*}
	\int_{\spq^{p+1}} \varphi(\bq_0, \bv) S_j^* \cM(d\bq_0, d \bv) 
	&= \int_{\spq^{p+1}} \varphi(S_j(\bq_0, \bv)) \Vker(\bq_0, d \bv) \mu(d\bq_0) \\
	&= \int_{\spq^{p+1}} \varphi(S_j(\bq_0, \bv)) \frac{d\mu}{d\mu_0}(\bq_0) \Vker(\bq_0, d \bv) \mu_0(d\bq_0) \\
	&= \int_{\spq^{p+1}} \varphi(S_j(\bq_0, \bv)) \frac{d\mu}{d\mu_0}(\Pi_1(\bq_0,\bv)) \cM_0(d\bq_0, d \bv) \\
	&= \int_{\spq^{p+1}} \varphi(\bq_0, \bv) \frac{d\mu}{d\mu_0}(\Pi_1 \circ S_j(\bq_0,\bv)) S_j^*\cM_0(d\bq_0, d \bv).
\end{align*}
Since $S_j^*\cM_0 = \cM_0$, then
\begin{align*}
	\int_{\spq^{p+1}} \varphi(\bq_0, \bv) S_j^* \cM(d\bq_0, d \bv)
	&=  \int_{\spq^{p+1}} \varphi(\bq_0, \bv) \frac{d\mu}{d\mu_0}(\Pi_1 \circ S_j(\bq_0,\bv)) \cM_0(d\bq_0, d \bv) \\
	&= \int_{\spq^{p+1}} \varphi(\bq_0, \bv) \frac{d\mu}{d\mu_0}(\Pi_1 \circ S_j(\bq_0,\bv)) \left( \frac{d\mu}{d\mu_0}(\bq_0)\right)^{-1} \cM(d\bq_0, d \bv) .
\end{align*}
This implies that $S_j^*\cM \ll \cM$ and \eqref{RN:SjM:M} holds, which concludes the proof.
%
%The second part of the statement follows immediately from
%\cref{thm:multi:ind:Tjmd} referring to \cref{rmk:alphaj:ac},
%\eqref{def:alphaj:ac} for the form of the $\alpha_j$.  The proof is
%complete.

\subsection{Proof of \cref{thm:tj}}
\label{sec:thm:tj}

Let $\Phi : \ExtSp \to \RR$ be any bounded measurable function.
Regarding the first item, (i),
\begin{align*}
	\int_{\ExtSp} \Phi( \btz) \MN\KR( d \btz)
  =& \int_{\ExtSp} \Phi( \btq, \btv, \tk)
     \int_{\ExtSp}\KR(\bq, \bv, k, d\btq, d\btv, d\tk)  \MN( d\bq, d\bv, dk)\\
  =&  \int_{\ExtSp} \Phi( \btq, \btv, \tk) \int_{\ExtSp} S_k^*(\Vker_k( \Proj S_k( \bq, \bv), d \btv) \delta_{\Proj S_k( \bq,\bv)} (d \btq))
     \delta_{k}(d\tk)  \MN( d\bq, d\bv, dk)\\
    =&  \int_{\ExtSp} \Phi( S_k(\btq, \btv), \tk) \int_{\ExtSp} \Vker_k( \Proj S_k( \bq, \bv), d \btv) \delta_{\Proj S_k( \bq,\bv)} (d \btq)
     \delta_{k}(d\tk)  \MN( d\bq, d\bv, dk)\\
     =&  \int_{\ExtSp}  \int_{\spv}\Phi( S_k(\Proj S_k( \bq,\bv), \btv), k)
       \Vker_k( \Proj S_k( \bq, \bv), d \btv)  \MN( d\bq, d\bv, dk)\\
     =&  \sum_{j = 0}^p \frac{1}{p+1}\int_{\ExtSp} \int_{\spv}
     \Phi( S_k(\Proj S_k( \bq,\bv), \btv), k) \Vker_k( \Proj S_k( \bq, \bv), d \btv)  S_j^*\cM_j(d \bq, d \bv) \delta_{j}(d k)\\
  =&  \sum_{j = 0}^p \frac{1}{p+1} \int_{\spv}
     \int_{\spq \times \spv} \Phi( S_j(\Proj S_j( \bq,\bv), \btv), j)\Vker_j( \Proj S_j( \bq, \bv), d \btv)  S_j^*\cM_j(d \bq, d \bv)\\
  =&\sum_{j = 0}^p \frac{1}{p+1}  \int_{\spv}
     \int_{\spq \times \spv} \Phi( S_j(\Proj( \bq,\bv), \btv), j)\Vker_j( \Proj(\bq, \bv), d \btv) \cM_j(d \bq, d \bv)\\
  =&\sum_{j = 0}^p \frac{1}{p+1}  \int_{\spv}
     \int_{\spq \times \spv} \Phi( S_j(\bq, \btv), j)\Vker_j(\bq, d \btv) 
    \Vker_j(\bq, d \bv) \mu(d \bq)\\
  =&\sum_{j = 0}^p \frac{1}{p+1}
     \int_{\spq \times \spv} \Phi( S_j(\bq, \btv), j)\Vker_j(\bq, d \btv) \mu(d \bq)\\
   =&\sum_{j = 0}^p \frac{1}{p+1}
     \int_{\spq \times \spv} \Phi( \bq, \btv, j) S_j^*\cM_j(d \bq, d \btv) = \int_{\ExtSp} \Phi( \bq, \btv, k) \MN( d \bq, d\btv, dk),
\end{align*}
as desired.
Turning to the second item we have
\begin{align*}
	\int_{\ExtSp} \Phi( \btz) \MN\KA( d \btz)
	=& \int_{\ExtSp} \Phi( \btq, \btv, \tk) \int_{\ExtSp}\KA(\bq, \bv, k, d\btq_0, d\btv, d\tk)  \MN( d\bq, d\bv, dk)\\
	=& \sum_{j = 0}^p \int_{\ExtSp}  \Phi( \bq, \bv, j) \alpha_{k,j}(\bq,\bv) \MN( d\bq, d\bv, dk)\\
	=& \sum_{j = 0}^p \sum_{l = 0}^p \frac{1}{p+1} \int_{\spq \times \spv}  \Phi( \bq, \bv, j) \alpha_{l,j}(\bq,\bv)  S_l^*\cM_l(d \bq, d \bv) \\
	=& \sum_{j = 0}^p \frac{1}{p+1} \int_{\spq \times \spv}  \Phi( \bq, \bv, j) \sum_{l = 0}^p \alpha_{l,j}(\bq,\bv)  S_l^*\cM_l(d \bq, d \bv) \\
	=&\sum_{j = 0}^p \frac{1}{p+1} \int_{\spq \times \spv}  \Phi( \bq, \bv, j) S_j^*\cM_j(d \bq, d \bv),
\end{align*}	
where we used \eqref{eq:ext:bal:ac} for the final equality.  The third
claim follows immediately from the first and second one.

Turning to the final claim, (iv), we have, for any $\Phi: \spq \to \RR$ 
bounded and measurable,
\begin{align*}
	\int_{\spq} \Phi( \bq)  \ProjES^* \MN (d \bq) &= 
	\int_{\ExtSp} \Phi(  \ProjES(\bq, \bv, k))  \MN (d \bq, d\bv, d k) 
	 = \sum_{j = 0}^p \frac{1}{p+1} \int_{\spq \times \spv} \Phi( \ProjES(\bq, \bv, j))  S_j^*\cM_j(d \bq, d \bv) \\
	 &= \sum_{j = 0}^p \frac{1}{p+1} \int_{\spq \times \spv} \Phi(  \Proj S_j( \bq, \bv))  S_j^*\cM_j(d \bq, d \bv) \\
	 &= 	 \sum_{j = 0}^p \frac{1}{p+1} \int_{\spq \times \spv} \Phi(  \bq)  \Vker_j(\bq, d \bv)\mu(d \bq) 
	 =\int_{\spq} \Phi(  \bq)  \mu(d \bq).
\end{align*}
The proof is complete.

\subsection{Proof of \cref{prop:sum:cond:simpl}}
\label{sec:prop:sum:simpl}

Since $S_0 = I$ then \eqref{sum:cond:simpl} is clearly satisfied for $j=0$. Let us thus assume from now on that $j \in \{1,\ldots,p\}$. Since $S_j$ is an involution, namely $S_j^2 = I$, then \eqref{sum:cond:simpl} reduces to showing that
\begin{align}\label{sum:Sj:Sk:M}
	\sum_{\stackrel{k=1}{k \neq j}}^p (S_j \circ S_k)^* \cM = \sum_{\stackrel{k=1}{k \neq j}}^p S_k^* \cM.
\end{align}

Fix $j \in \{1,\ldots,p\}$ and $k \in \{1,\ldots,p\}$ with $k \neq j$. Let $\varphi: \RR^{\dimx(p+1)} \to \RR$ be any measurable and bounded function. It follows by change of variables, \eqref{def:pushfwd:obs}, that
\begin{align*}
	\int_{\RR^{\dimx(p+1)}} \varphi(\bq,\bv) (S_j \circ S_k)^* \cM (d\bq, d\bv) 
	&= \int_{\RR^{\dimx(p+1)}} \varphi(S_j \circ S_k(\bq,\bv)) \rrt(\bq,\cdot)^* (\ed \otimes \Hd)(d\bv) \mu(d\bq) \\
	=& \int_{\RR^{\dimx(p+1)}} \varphi(S_j \circ S_k (\bq, \rrt(\bq, \lambda, Q))) \, \ed(d\lambda) \, \Hd(d Q) \, \mu(d \bq) \\
	=& \int_{\RR^{\dimx(p+1)}} \varphi(S_j \circ S_k (\bq, \bq + \lambda Q \vs_1, \ldots, \bq + \lambda Q \vs_p))  \ed(d\lambda) \Hd(d Q) \mu(d \bq) .
\end{align*}
Assuming without loss of generality that $j \geq k$, it thus follows from the definition of $(S_0,\ldots,S_p)$ in \eqref{def:Sj:simpl} that
\begin{align}\label{phi:Sj:Sk:M}
	&\int_{\RR^{\dimx(p+1)}} \varphi(\bq,\bv) (S_j \circ S_k)^* \cM (d\bq, d\bv) \notag \\
	&= \int_{\RR^{\dimx(p+1)}}  \varphi(\bq + \lambda Q \vs_j, \bq + \lambda Q \vs_1, \ldots, \bq + \lambda Q \vs_{k-1}, \bq, \bq + \lambda Q \vs_{k+1}, \ldots, \notag\\
	& \qquad \qquad \qquad \quad \bq + \lambda Q \vs_{j-1}, \bq + \lambda Q \vs_k, \bq + \lambda Q \vs_{j+1}, \ldots, \bq + \lambda Q \vs_p) \, \ed(d\lambda) \, \Hd(d Q) \, \mu(d \bq). 
\end{align}

Now let $A \in \Od$ be the reflection matrix between the vertices $\vs_j$ and $\vs_k$, so that $A \vs_j = \vs_k$, $A \vs_k = \vs_j$, and $A = A^T$. Concretely, $A$ can be written as $I - 2 (\vs_j - \vs_k) \otimes (\vs_j - \vs_k)$, namely $A \bw = \bw - 2 \langle \bw, \vs_j - \vs_k \rangle  (\vs_j - \vs_k)$ for all $\bw \in \RR^\dimx$, where $\langle \cdot, \cdot \rangle$ denotes the Euclidean inner product. From the assumed simplex structure in \eqref{equidist}, it follows immediately that $A$ leaves all other vertices invariant, namely $A \vs_l = \vs_l$ for all $l \in \{1,\ldots, p\}$ with $l \neq j$ and $l \neq k$. For any fixed $Q \in \Od$, we then define $\tQ := Q A Q^T$. Notice that $\tQ \in \Od$, and from the left-invariant property of the Haar measure $\Hd$, \eqref{left:inv:Haar}, we have $(\tQ^T)^*\Hd = \Hd$. Using this fact in \eqref{phi:Sj:Sk:M}, we obtain by changing variables and invoking the properties of $A$ that
\begin{align*}
&\int_{\RR^{\dimx(p+1)}} \varphi(\bq,\bv) (S_j \circ S_k)^* \cM (d\bq, d\bv) \\
&= \int_{\RR^{\dimx(p+1)}}  \varphi(\bq +  \lambda \tQ Q \vs_j, \bq + \lambda \tQ Q \vs_1, \ldots, \bq + \lambda \tQ Q \vs_{k-1}, \bq, \bq + \lambda \tQ Q \vs_{k+1}, \ldots, \\
& \qquad \qquad \qquad \quad \bq + \lambda \tQ Q \vs_{j-1}, \bq + \lambda \tQ Q \vs_k, \bq + \lambda \tQ Q \vs_{j+1}, \ldots, \bq + \lambda \tQ Q \vs_p) \, \ed(d\lambda) \, \Hd(d Q) \, \mu(d \bq) \\
&= \int_{\RR^{\dimx(p+1)}}  \varphi(\bq +  \lambda Q A \vs_j, \bq + \lambda Q A \vs_1, \ldots, \bq + \lambda Q A \vs_{k-1}, \bq, \bq + \lambda Q A \vs_{k+1}, \ldots, \\
& \qquad \qquad \qquad \quad \bq + \lambda Q A \vs_{j-1}, \bq + \lambda Q A \vs_k, \bq + \lambda Q A \vs_{j+1}, \ldots, \bq + \lambda Q A \vs_p) \, \ed(d\lambda) \, \Hd(d Q) \, \mu(d \bq) \\
&= \int_{\RR^{\dimx(p+1)}}  \varphi(\bq +  \lambda Q \vs_k, \bq + \lambda Q \vs_1, \ldots, \bq + \lambda Q \vs_{k-1}, \bq, \bq + \lambda Q \vs_{k+1}, \ldots, \\
& \qquad \qquad \qquad \quad \bq + \lambda Q \vs_{j-1}, \bq + \lambda Q \vs_j, \bq + \lambda Q \vs_{j+1}, \ldots, \bq + \lambda Q \vs_p) \, \ed(d\lambda) \, \Hd(d Q) \, \mu(d \bq) \\
&= \int_{\RR^{\dimx(p+1)}}  \varphi(S_k(\bq, \rrt(\bq, \lambda, Q))) \, \ed(d\lambda) \, \Hd(d Q) \, \mu(d \bq) \\
&= \int_{\RR^{\dimx(p+1)}}  \varphi(S_k(\bq, \bv)) \rrt(\bq, \cdot)^*(\ed \otimes \Hd)(d\bv) \, \mu(d\bq)
= \int_{\RR^{\dimx(p+1)}}  \varphi(\bq, \bv) S_k^*\cM(d\bq, d \bv).
\end{align*}
Since $\varphi$ is arbitrary, we deduce that $(S_j \circ S_k)^* \cM = S_k^* \cM$, which shows \eqref{sum:Sj:Sk:M} and concludes the proof.

\subsection{Proof of \cref{prop:SjM:M:simpl}}
\label{sec:prop:SjM:M:simpl}

We first notice that since $S_0 = I$, then \eqref{RN:SjM:M:simpl} is readily satisfied when $j=0$. Let us now fix $j \in \{1,\ldots,p\}$. For any bounded and measurable function $\varphi: \RR^{\dimx(p+1)} \to \RR$, we have
\begin{align*}
	&\int_{\RR^{\dimx(p+1)}} \varphi(\bq,\bv) S_j^* \cM (d\bq, d\bv)
	= \int_{\RR^{\dimx(p+1)}} \varphi(S_j(\bq,\bv)) \dmu(\bq) \, \rrt(\bq, \cdot)^*(\ed \otimes \Hd)(d\bv) \, d\bq \\
	&= \int_{\RR^{\dimx(p+1)}} \varphi(S_j(\bq,\rrt(\bq, \lambda, Q))) \dmu(\bq) \, \ed(d\lambda) \, \Hd(d Q) \, d \bq \\
	&= \int_{\RR^{\dimx(p+1)}} \varphi( \bq + \lambda Q \vs_j, \bq + \lambda Q \vs_1, \ldots, \bq + \lambda Q \vs_{j-1}, \bq, \bq + \lambda Q \vs_{j+1}, \ldots, \bq + \lambda Q \vs_p) \dmu(\bq) \, \ed(d\lambda) \, \Hd(d Q) \, d \bq.
\end{align*}
With the change of variables $\btq = \bq + \lambda Q \vs_j$, it follows that
\begin{align}\label{int:phi:btq}
	&\int_{\RR^{\dimx(p+1)}} \varphi(\bq,\bv) S_j^* \cM (d\bq, d\bv) \notag \\
	&= \int_{\RR^{\dimx(p+1)}} \varphi(\btq, \btq + \lambda Q (\vs_1 - \vs_j), \ldots,  \btq + \lambda Q (\vs_{j-1} - \vs_j),  \btq - \lambda Q \vs_j,  \btq + \lambda Q (\vs_{j+1} - \vs_j), \ldots, \notag \\
	&\qquad\qquad \qquad \quad \btq + \lambda Q (\vs_p - \vs_j)) \dmu(\btq - \lambda Q \vs_j)  \, \ed(d\lambda) \, \Hd(d Q) \, d \btq.
\end{align}

Let us now consider the reflection matrix $A$ of $\vs_j$ about the origin, so that $A \vs_j = - \vs_j$ and $A = A^T$. This can be written explicitly as $A = I - 2 \vs_j \otimes \vs_j$, namely $A \bw = \bw - 2 \langle \bw, \vs_j \rangle \vs_j$ for all $\bw \in \RR^\dimx$, where $\langle \cdot, \cdot \rangle$ denotes the Euclidean inner product. Then, assumption \eqref{equidist} regarding the simplex $(\vs_1,\ldots,\vs_p,\bzero)$ implies that $A\vs_l = \vs_l - \vs_j$ for all $l \in \{1,\ldots,p\}$ with $l \neq j$.

Next, similarly as in the proof of \cref{prop:sum:cond:simpl}, we define $\tQ = Q A Q^T \in \Od$ and notice that $(\tQ^T)^*\Hd = \Hd$ by the left-invariance property of $\Hd$, \eqref{left:inv:Haar}. From \eqref{int:phi:btq}, we thus obtain after change of variables and the properties of $A$ that
\begin{align*}
	&\int_{\RR^{\dimx(p+1)}} \varphi(\bq,\bv) S_j^* \cM (d\bq, d\bv) \\
	&= \int_{\RR^{\dimx(p+1)}} \varphi(\btq, \btq + \lambda \tQ Q (\vs_1 - \vs_j), \ldots,  \btq + \lambda \tQ Q (\vs_{j-1} - \vs_j),  \btq - \lambda \tQ Q \vs_j,  \btq + \lambda \tQ Q (\vs_{j+1} - \vs_j), \ldots, \notag \\
	&\qquad\qquad \qquad \quad \btq + \lambda \tQ Q (\vs_p - \vs_j)) \dmu(\btq - \lambda \tQ Q \vs_j)  \, \ed(d\lambda) \, \Hd(d Q) \, d \btq \\
	&= \int_{\RR^{\dimx(p+1)}} \varphi(\btq, \btq + \lambda Q A (\vs_1 - \vs_j), \ldots,  \btq + \lambda Q A (\vs_{j-1} - \vs_j),  \btq - \lambda Q A \vs_j,  \btq + \lambda Q A (\vs_{j+1} - \vs_j), \ldots, \notag \\
	&\qquad\qquad \qquad \quad \btq + \lambda Q A (\vs_p - \vs_j)) \dmu(\btq - \lambda Q A \vs_j)  \, \ed(d\lambda) \, \Hd(d Q) \, d \btq \\
	&= \int_{\RR^{\dimx(p+1)}} \varphi(\btq, \btq + \lambda Q \vs_1 , \ldots,  \btq + \lambda Q \vs_{j-1} ,  \btq + \lambda Q \vs_j,  \btq + \lambda Q \vs_{j+1}, \ldots, \btq + \lambda Q \vs_p)) \, \dmu(\btq + \lambda Q\vs_j) \\
	&\hspace{12.5cm} \ed(d\lambda) \, \Hd(d Q) \, d \btq \\
	&= \int_{\RR^{\dimx(p+1)}} \varphi(\btq, \rrt(\btq, \lambda, Q)) \, \dmu(\Proj S_j (\btq , \rrt(\btq, \lambda, Q))) \, \ed(d\lambda) \, \Hd(d Q) \, d \btq \\
	&= \int_{\RR^{\dimx(p+1)}} \varphi( \btq, \bv) \, \dmu(\Proj S_j(\btq, \bv)) \rrt(\btq, \cdot)^*(\ed \otimes \Hd)(d\bv) d \btq \\
	&= \int_{\RR^{\dimx(p+1)}} \varphi(\btq, \bv) \frac{\dmu(\Proj S_j(\btq, \bv))}{ \dmu(\btq)} \, \cM(d\btq, d \bv).
\end{align*}
Since $\varphi$ is arbitrary, we conclude that 
\begin{align*}
	\frac{d S_j^* \cM}{d \cM}(\bq, \bv) =  \frac{\dmu(\Proj S_j(\bq, \bv))}{ \dmu(\bq)} \quad \mbox{ for $\cM$-a.e. } (\bq, \bv) \in \RR^{\dimx(p+1)}.
\end{align*}
This shows \eqref{RN:SjM:M:simpl} and completes the proof. 

 \arxiv{
   
% \newpage

%  \setcounter{equation}{0}
% \appendix
%\addcontentsline{toc}{section}{Appendix}

\section{Appendix: Supplementary Figures}
\label{sec:appndx:sup:mat}

This appendix collects various additional materials from our  
case studies described in \cref{sec:case:Study}.

\begin{figure}[!htp]
    \centering
    \includegraphics[width=0.7\textwidth]{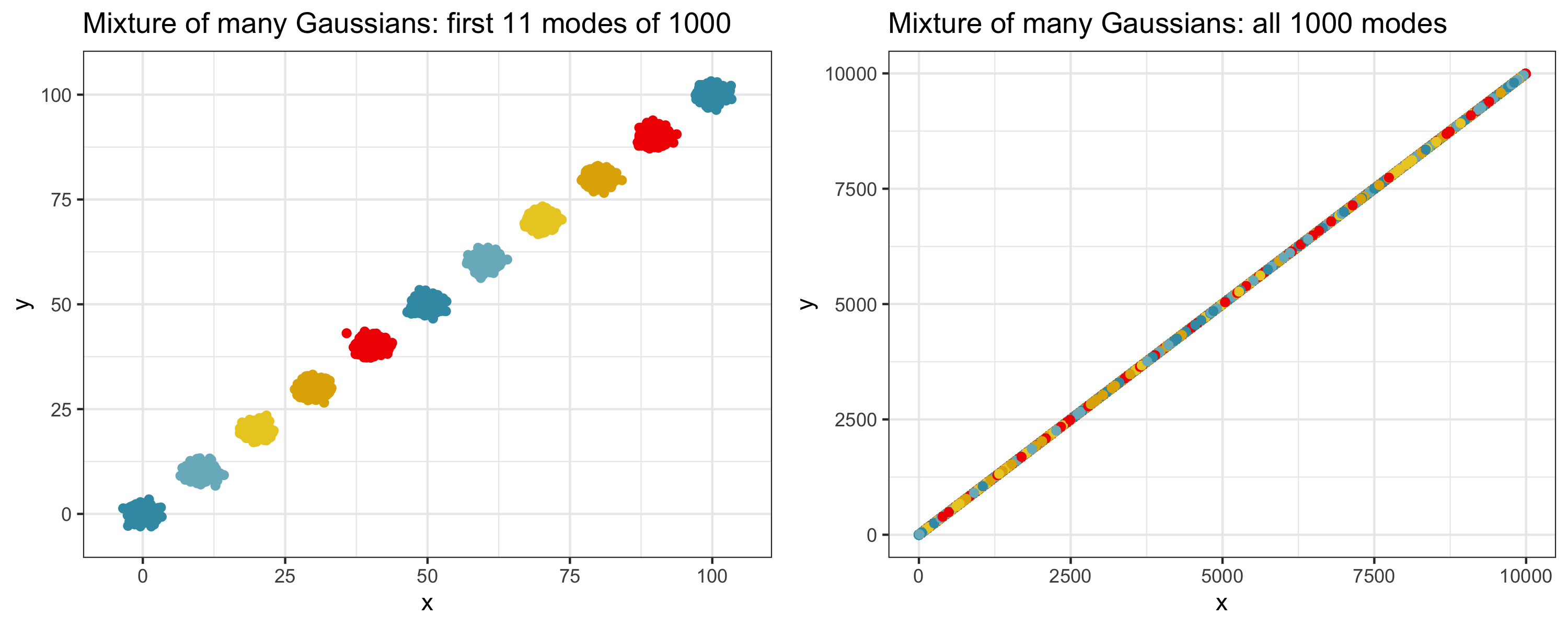}
    \caption{Death mixture}
    \label{fig:deathMixture}
\end{figure}

\begin{figure}[!htb]
	\centering
	\includegraphics[width=\linewidth]{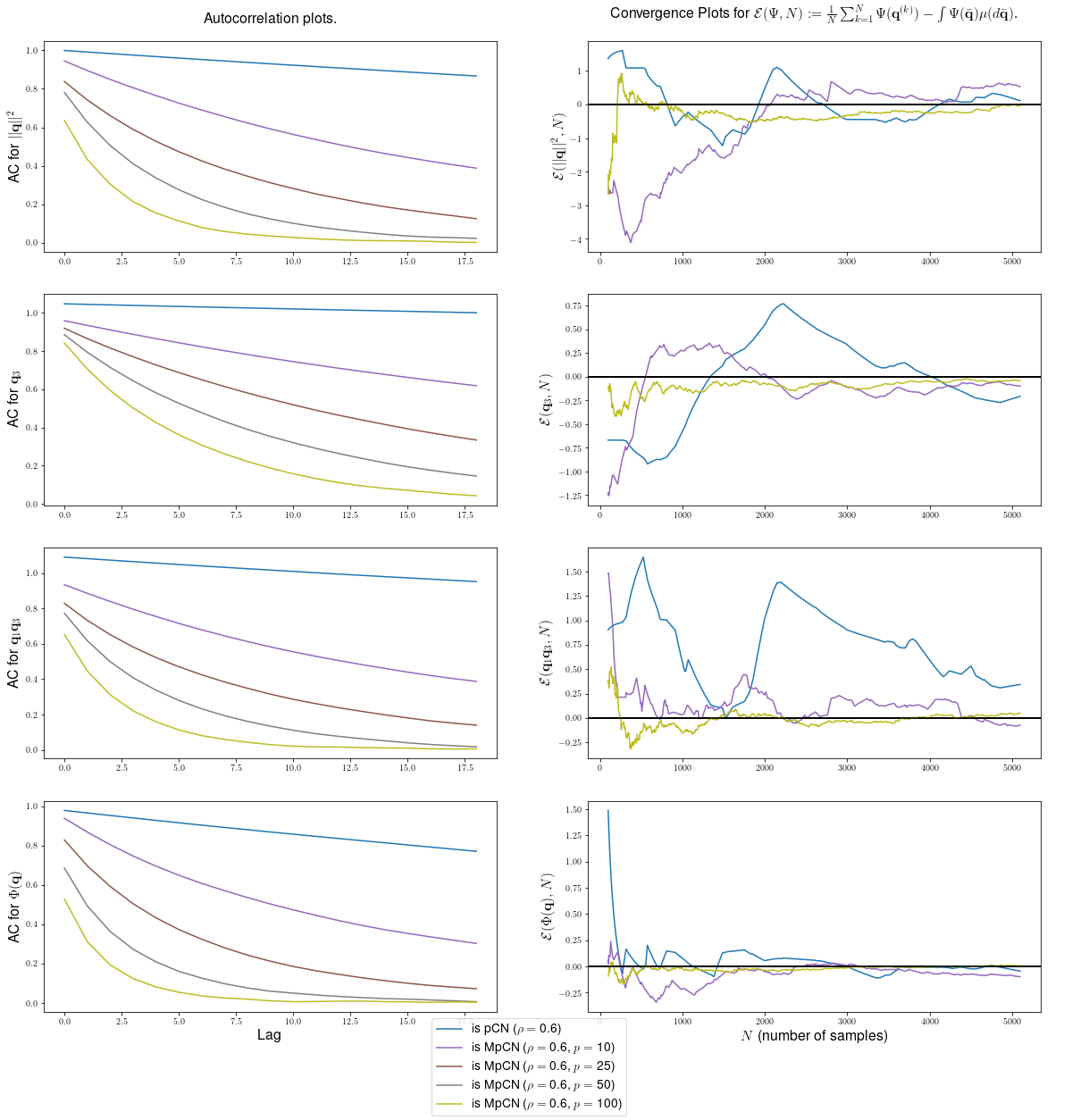}
	\caption{Autocorrelation and convergence plots for pCN and
          mpCN for increasing values of $p$ over
          various observables.}\label{fig:AD:conv:plt:EXT}
\end{figure}
\begin{figure}[!htb]
	\centering
	\includegraphics[width=\linewidth]{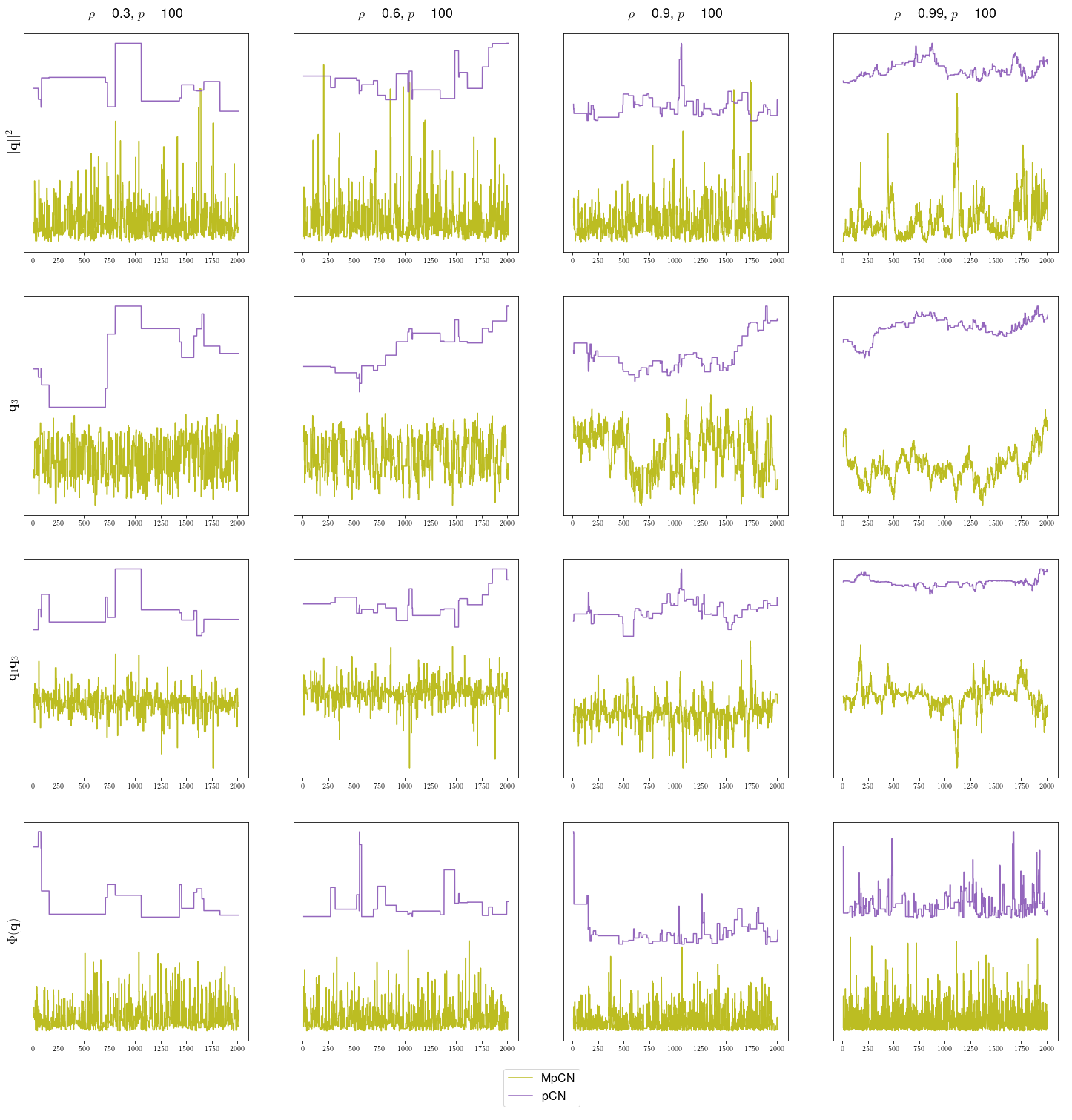}
	\caption{Time series plots comparing mpCN with $p = 100$ and
          pCN for various values of $\rho$ for different
          observables.}\label{fig:AD:TS:Plots:EXT}
\end{figure}
\begin{figure}[!htb]
	\centering
	\includegraphics[width=\linewidth]{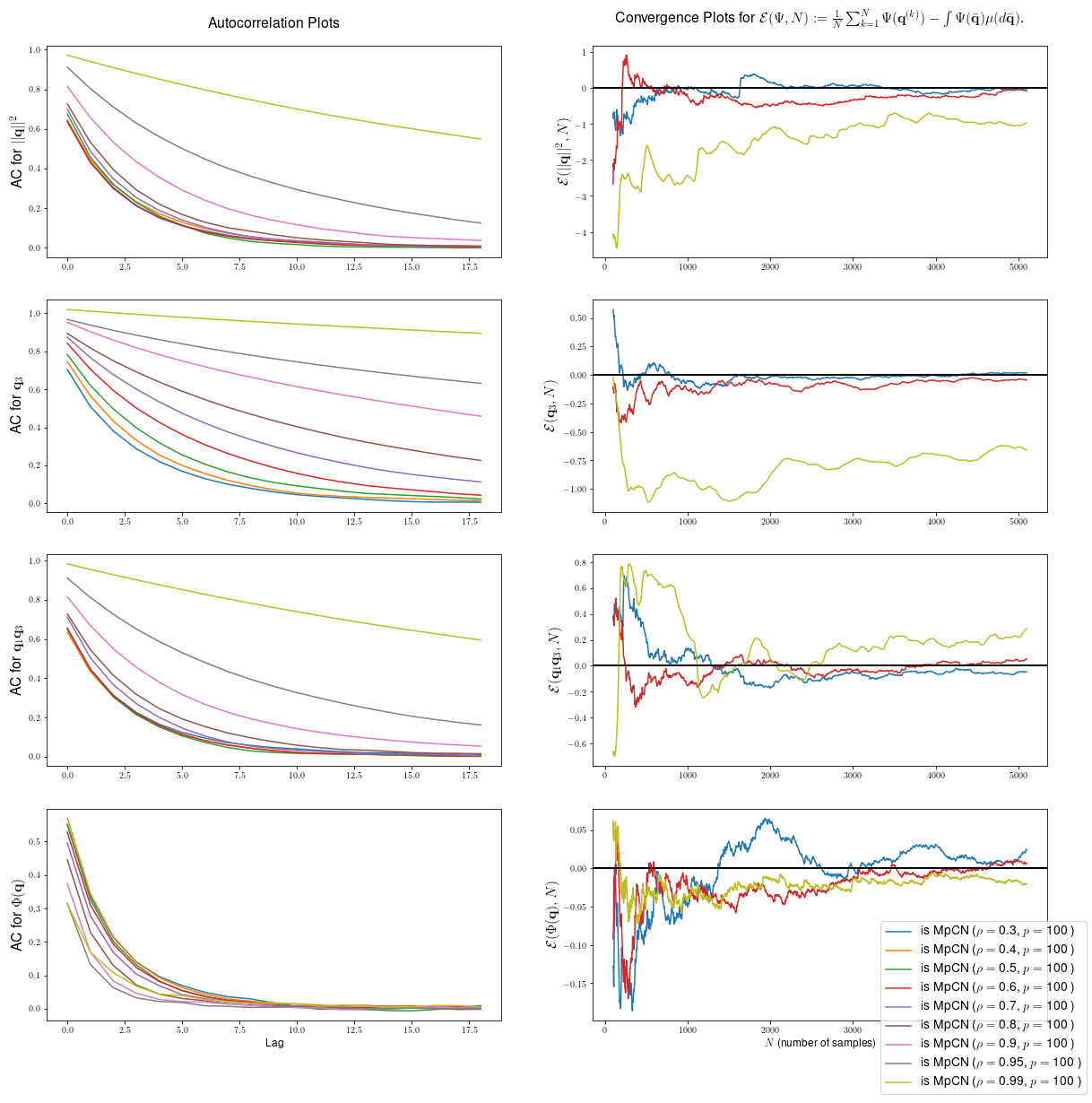}
	\caption{Autocorrelation and convergence plots for ergodic
          averages plots comparing mpCN with $p = 100$ over various
          values of $\rho$ for different
          observables.}\label{fig:AC:Conv:rho:for:p100}
\end{figure}

\begin{figure}[!htb]
	\centering
	\includegraphics[width=.45\linewidth]{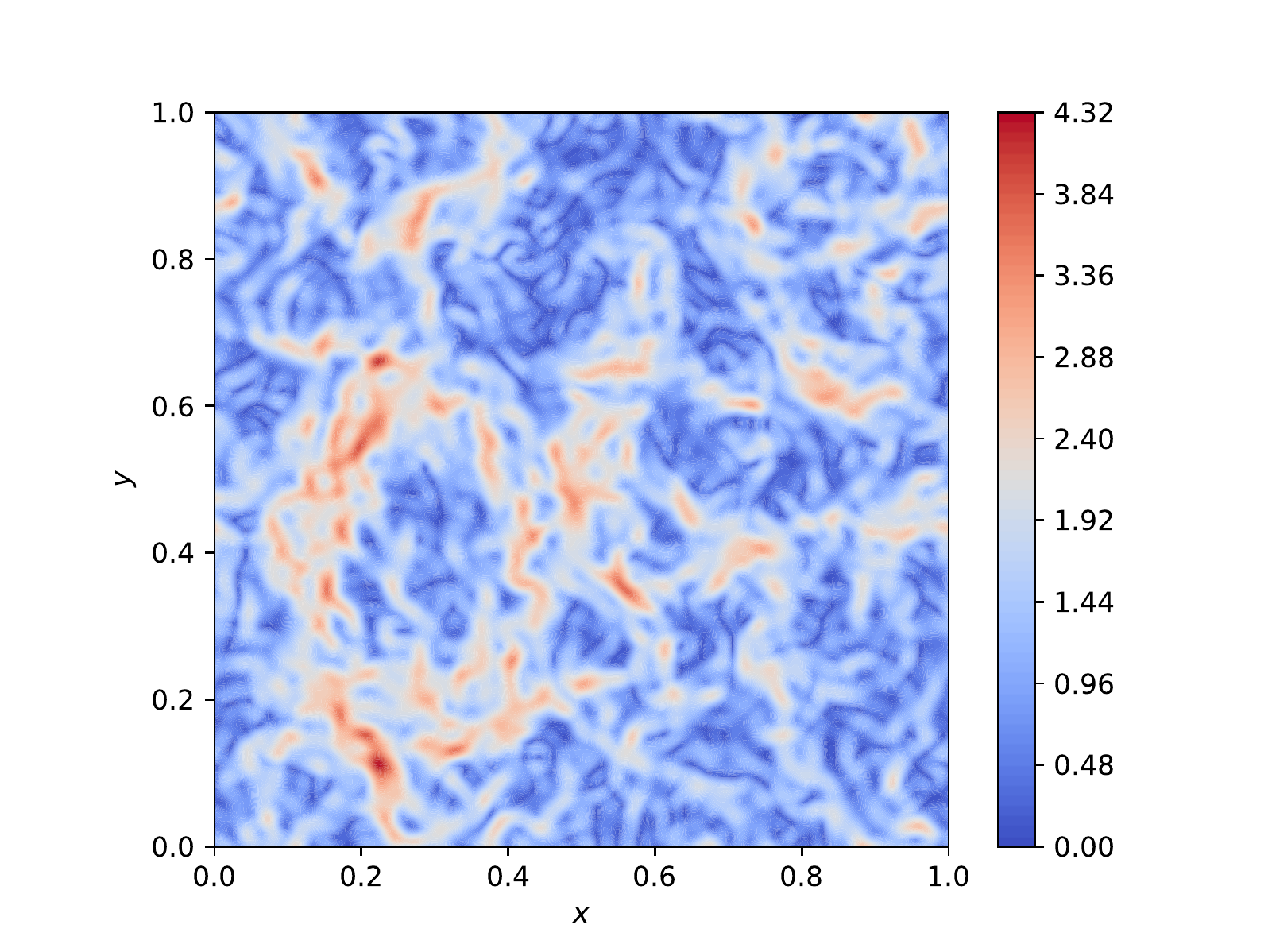}
	\includegraphics[width=.45\linewidth]{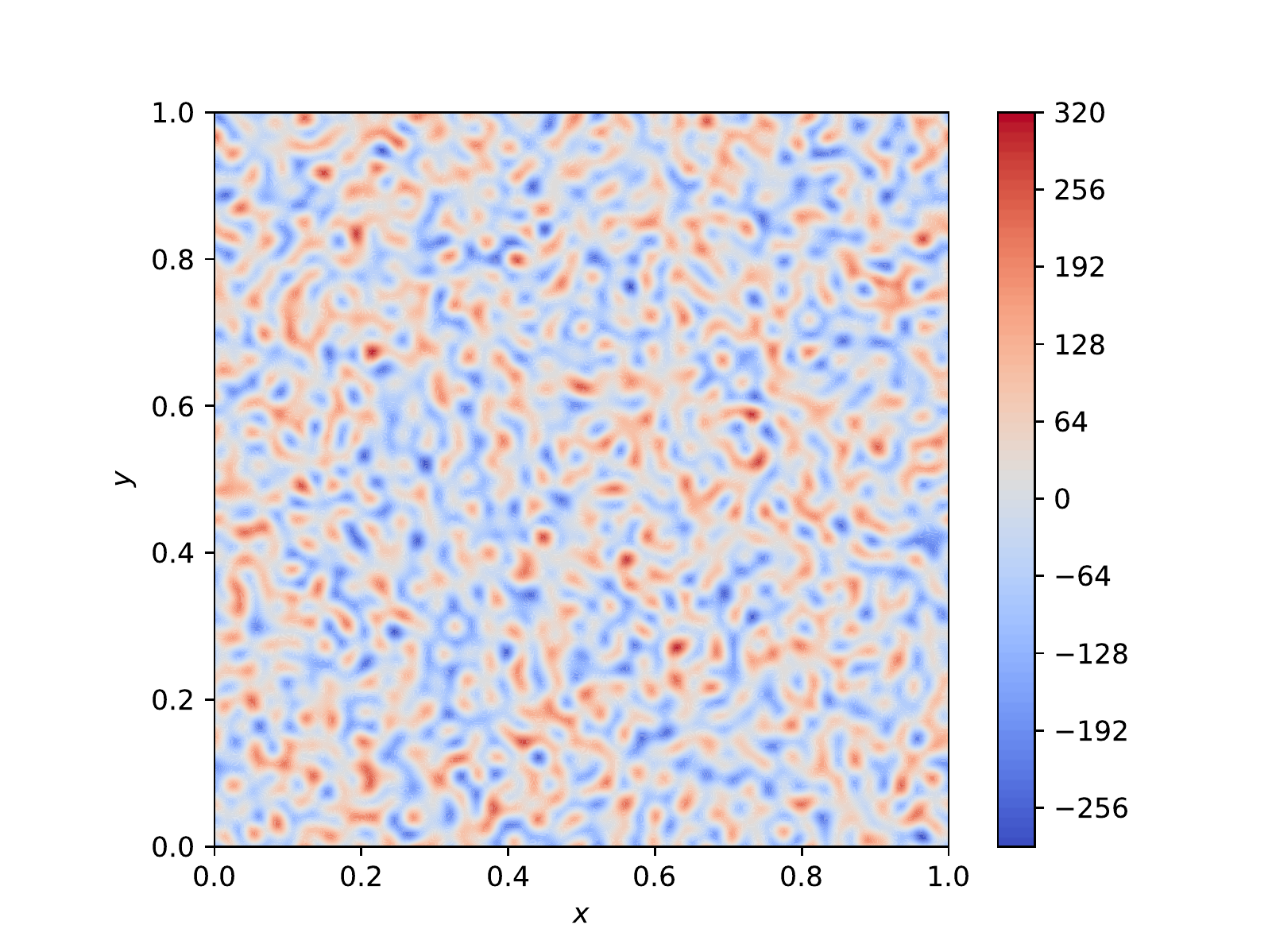}
	\caption{Typical draw from the prior $\mu_0 = N(0,\cC)$ with
          $\cC$ defined by \eqref{eq:cov:Op:AD}.  The right panel is 
          $\mbox{curl}(\bq(\bx)) = (\partial_{x_2} \bq_1 - \partial_{x_1} \bq_2)(\bx)$
          while the left panel is $\| \bq(\bx)\|$.}\label{fig:prior:AD:model}
\end{figure}
\begin{figure}[!htb]
	\centering
	\includegraphics[width=\linewidth]{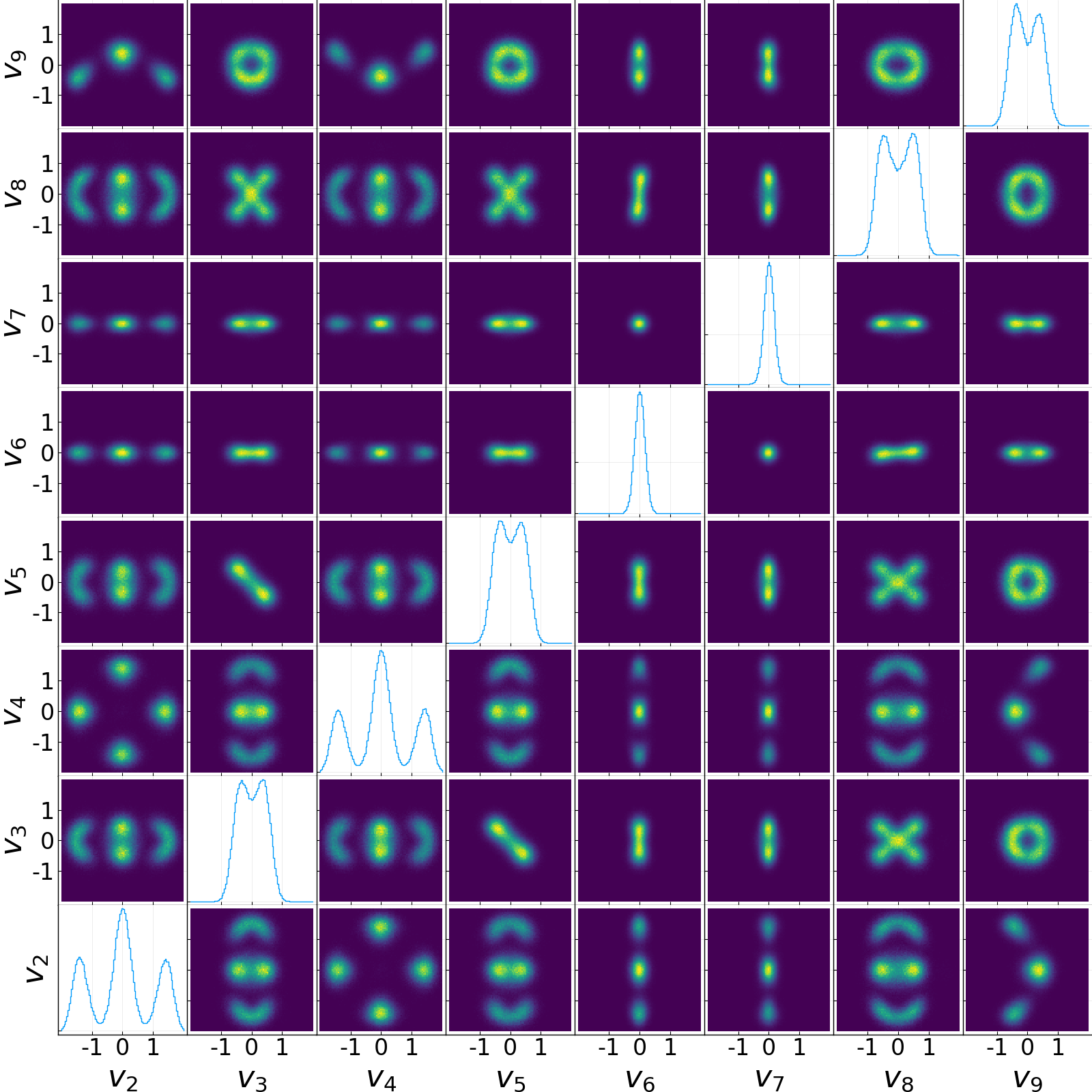}
	\caption{A grid of histogram of the (fully resolved) two
          dimensional marginal onto various low frequencies of the
          posterior measure.  More precisely the $i-j$th off diagonal
          is $F_{i,j}^\#\mu$ where
          $
          F_{i,j}(\bq) = [ \int_{\TT^2}
          \bff_i(\bx) \cdot  \bq(\bx) d\bx,\int_{\TT^2}
          \bff_j(\bx) \cdot  \bq(\bx) d\bx]
         $
         with
         $
         \bff_2(x,y) = [0, \cos(2\pi y)], %\quad
            \bff_3(x,y) = [0, -\sin(2\pi y)], %\quad
            \bff_4(x,y) = [\cos(2\pi x), 0], %\quad
            \bff_5(x,y) = [-\sin(2 \pi x), 0],
            $
            $
            \bff_6(x,y) = [0, \cos(4\pi y)],  %\quad
            \bff_7(x,y) = [0, -\sin(4\pi y)], %\quad
            \bff_8(x,y) = [\cos(2\pi x), \cos(2\pi y)], \%quad
            \bff_9(x,y) =  [-\sin(2\pi x), \sin(2\pi y)].
         $
        }\label{fig:ad:hist2dtrue}
\end{figure}

\begin{figure}[ht!]
	\centering
	\includegraphics[width=0.7\linewidth]{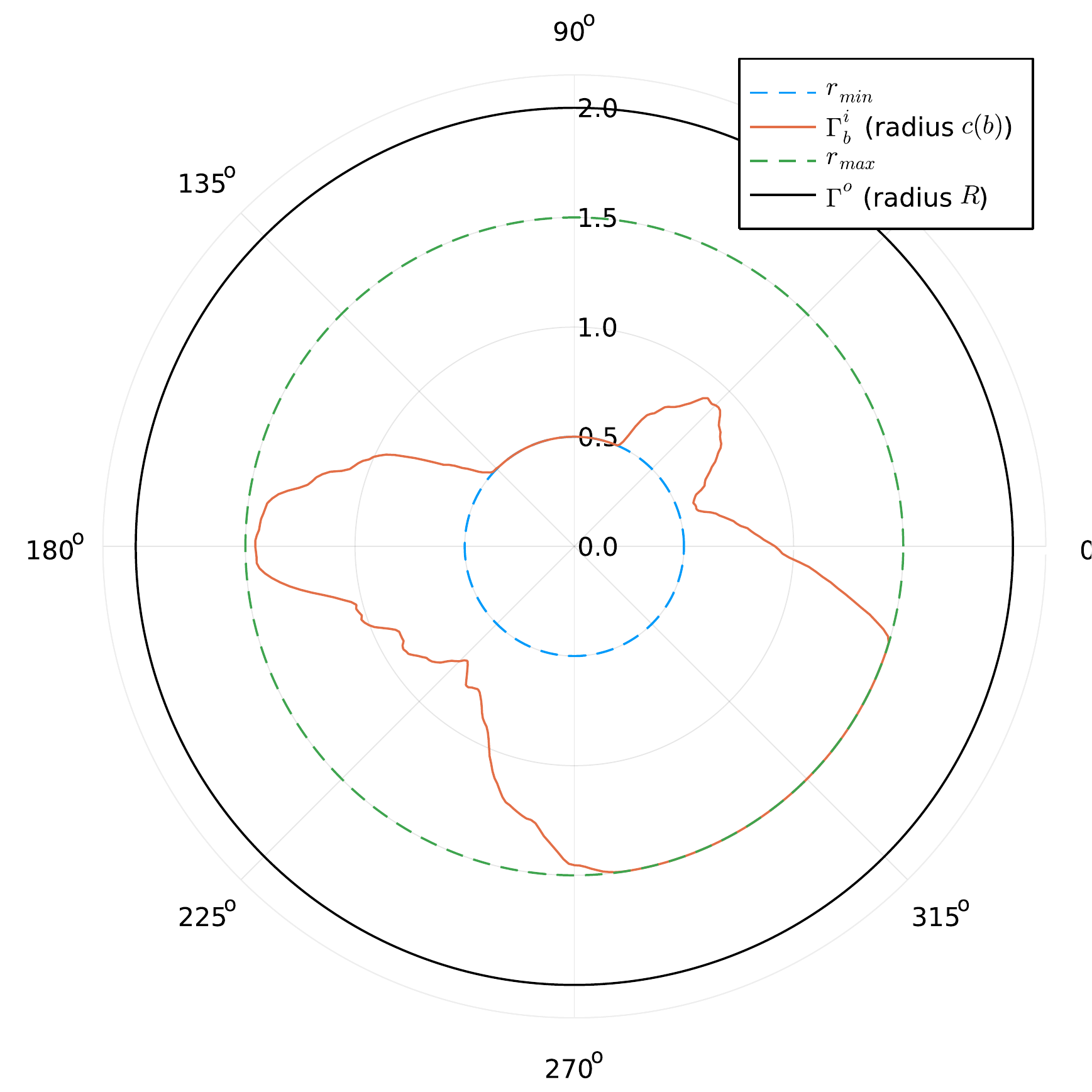}
	\caption{Diagram of the radius constraints involved in constructing the domain $\DD_\bq$.}\label{fig:Domain:Trans}
\end{figure}

\begin{figure}[ht!]
	\centering
	\includegraphics[width=0.45\linewidth]{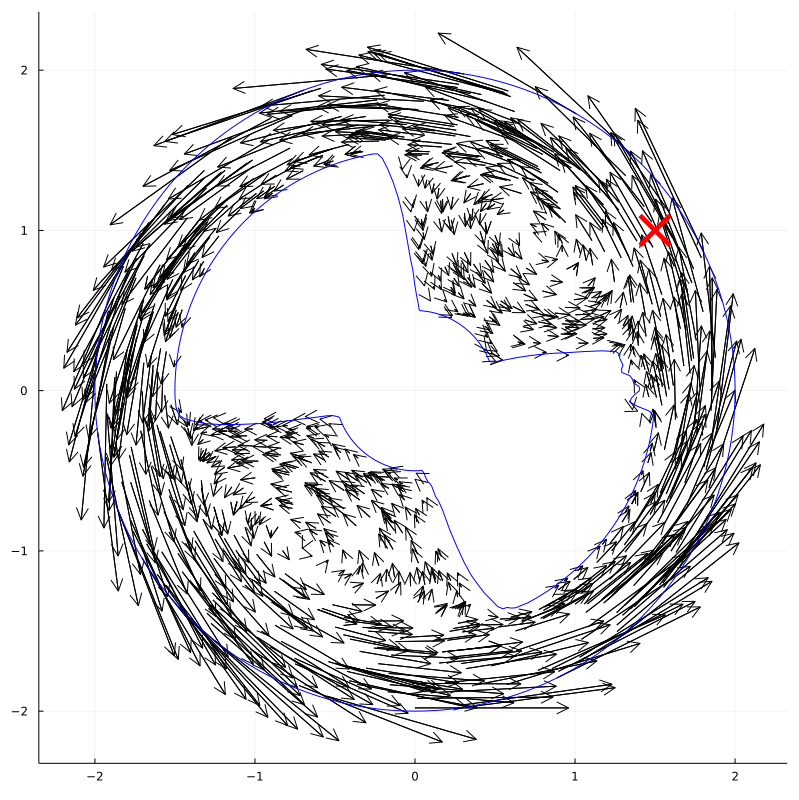}
	\hfill
	\includegraphics[width=0.51\linewidth]{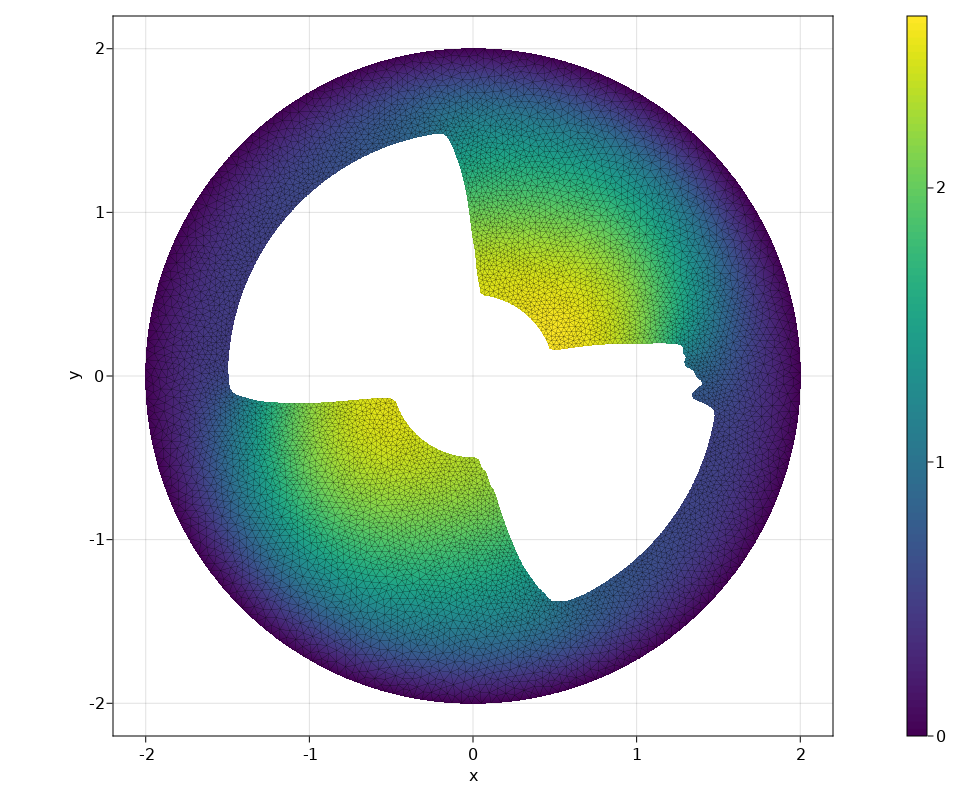}
	\caption{Solutions of the coupled Stokes and
          advection-diffusion PDEs. Left: Quiver plot of the solution
          to the Stokes PDE. The red X marks the point at which the
          scalar is injected into the system. Right: Contour plot of
          the solution to the advection-diffusion equation associated
          with the Stokes flow in the left-hand
          plot.}\label{fig:stokes:solve}
\end{figure}
      
}

\end{document}